\RequirePackage[right]{lineno}
\documentclass[aps,prl,preprint,nopacs,superscriptaddress]{revtex4}
\usepackage{graphicx}
\usepackage{verbatim}
\usepackage{relsize}
\usepackage{mathrsfs}
\usepackage{color}
\usepackage{ulem}
\usepackage{amsmath,amsfonts,amssymb}
\usepackage{graphicx}
\usepackage{braket}
\definecolor{mygray}{gray}{0.6}

\usepackage{bbding}
\usepackage{pifont}
\usepackage{wasysym}
\usepackage{amssymb}
\usepackage{multirow}
\usepackage{wrapfig}
\usepackage{braket}
\usepackage{bbold}
\usepackage{multirow}

\def \beq {\begin{equation}}
\def \eeq {\end{equation}}
\begin{document}


\title{
Observation of the  Axion quasiparticle in 2D MnBi$_2$Te$_4$

}

\author{\footnotesize Jian-Xiang Qiu}\affiliation{\footnotesize Department of Chemistry and Chemical Biology, Harvard University, Massachusetts 02138, USA}

\author{\footnotesize Barun Ghosh}\affiliation{\footnotesize Department of Physics, Northeastern University, Boston, MA 02115, USA}
\affiliation{\footnotesize Quantum Materials and Sensing Institute, Northeastern University, Burlington, MA 01803, USA.}
\affiliation{\footnotesize Department of Condensed Matter and Materials Physics
S. N. Bose National Center for Basic Sciences, Kolkata-700106, India}

\author{\footnotesize Jan Sch\"utte-Engel}\affiliation{\footnotesize Department of Physics, University of California, Berkeley, CA 94720, USA}\affiliation{\footnotesize RIKEN iTHEMS, Wako, Saitama 351-0198, Japan}
\author{\footnotesize Tiema Qian}\affiliation{\footnotesize Department of Physics and Astronomy and California NanoSystems Institute, University of California, Los Angeles, Los Angeles, CA 90095, USA.}

\author{\footnotesize Michael Smith}\affiliation{\footnotesize Materials Science Division, Argonne National Laboratory, Lemont, Illinois 60439, USA}

\author{\footnotesize Yueh-Ting Yao}\affiliation{\footnotesize Department of Physics, National Cheng Kung University, Tainan 70101, Taiwan}

\author{\footnotesize Junyeong Ahn}\affiliation{\footnotesize Department of Physics, Harvard University, Cambridge, MA 02138, USA}

\author{\footnotesize Yu-Fei Liu}\affiliation{\footnotesize Department of Chemistry and Chemical Biology, Harvard University, Massachusetts 02138, USA}\affiliation{\footnotesize Department of Physics, Harvard University, Cambridge, MA 02138, USA}

\author{\footnotesize Anyuan Gao}\affiliation{\footnotesize Department of Chemistry and Chemical Biology, Harvard University, Massachusetts 02138, USA}

\author{\footnotesize Christian Tzschaschel}\affiliation{\footnotesize Department of Chemistry and Chemical Biology, Harvard University, Massachusetts 02138, USA}
\affiliation{\footnotesize Department of Physics, University of Zurich, Zurich, Switzerland} 

\author{\footnotesize Houchen Li}\affiliation{\footnotesize Department of Chemistry and Chemical Biology, Harvard University, Massachusetts 02138, USA}

\author{\footnotesize Ioannis Petrides}\affiliation{\footnotesize  College  of  Letters  and  Science,  University  of  California,  Los  Angeles, USA}

\author{\footnotesize Damien B\'erub\'e}\affiliation{\footnotesize Department of Chemistry and Chemical Biology, Harvard University, Massachusetts 02138, USA}

\author{\footnotesize Thao Dinh}\affiliation{\footnotesize Department of Chemistry and Chemical Biology, Harvard University, Massachusetts 02138, USA}\affiliation{\footnotesize Department of Physics, Harvard University, Cambridge, MA 02138, USA}

\author{\footnotesize Tianye Huang}\affiliation{\footnotesize Department of Chemistry and Chemical Biology, Harvard University, Massachusetts 02138, USA}

\author{\footnotesize Olivia Liebman}\affiliation{\footnotesize  College  of  Letters  and  Science,  University  of  California,  Los  Angeles, USA}\affiliation{\footnotesize  Department of Materials Science and Engineering, University of California, Los Angeles, California 90095, USA}

\author{\footnotesize Emily M. Been}\affiliation{\footnotesize  College  of  Letters  and  Science,  University  of  California,  Los  Angeles, USA}

\author{\footnotesize Joanna M. Blawat}\affiliation{\footnotesize National High Magnetic Field Laboratory, Los Alamos National Laboratory, Los Alamos, New Mexico 87545, USA}

\author{\footnotesize Kenji Watanabe}\affiliation{\footnotesize National Institute for Materials Science, 1-1 Namiki, Tsukuba 305-0044, Japan}


\author{\footnotesize Takashi Taniguchi}\affiliation{\footnotesize National Institute for Materials Science,  1-1 Namiki, Tsukuba 305-0044, Japan}

\author{\footnotesize Kin Chung Fong}\affiliation{\footnotesize Department of Physics, Northeastern University, Boston, MA 02115, USA}
\affiliation{\footnotesize Quantum Materials and Sensing Institute, Northeastern University, Burlington, MA 01803, USA.}
\affiliation{\footnotesize Department of Electrical and Computer Engineering, Northeastern University, MA 02115, USA}

\author{\footnotesize Hsin Lin}
\affiliation{\footnotesize Institute of Physics, Academia Sinica, Taipei 115201, Taiwan}

\author{\footnotesize Peter P. Orth}\affiliation{\footnotesize Department of Physics and Astronomy, Iowa State University, Ames, Iowa 50011, USA}\affiliation{\footnotesize Ames National Laboratory, Ames, Iowa 50011, USA}\affiliation{\footnotesize Department of Physics, Saarland University, 66123 Saarbr\"ucken, Germany}

\author{\footnotesize Prineha Narang}\affiliation{\footnotesize  College  of  Letters  and  Science,  University  of  California,  Los  Angeles, USA}\affiliation{\footnotesize  Electrical and Computer Engineering Department, University of California, Los Angeles, California 90095, USA}

\author{\footnotesize Claudia Felser}\affiliation{\footnotesize Max Planck Institute for Chemical Physics of Solids, Dresden, Germany.}
\author{\footnotesize Tay-Rong Chang}\affiliation{\footnotesize Department of Physics, National Cheng Kung University, Tainan 70101, Taiwan} \affiliation{\footnotesize Center for Quantum Frontiers of Research and Technology (QFort), Tainan 70101, Taiwan} \affiliation{\footnotesize Physics Division, National Center for Theoretical Sciences, Taipei 10617, Taiwan}

\author{\footnotesize Ross McDonald}\affiliation{\footnotesize National High Magnetic Field Laboratory, Los Alamos National Laboratory, Los Alamos, New Mexico 87545, USA}

\author{\footnotesize Robert J. McQueeney}\affiliation{\footnotesize Department of Physics and Astronomy, Iowa State University, Ames, Iowa 50011, USA}\affiliation{\footnotesize Ames National Laboratory, Ames, Iowa 50011, USA}

\author{\footnotesize Arun Bansil}\affiliation{\footnotesize Department of Physics, Northeastern University, Boston, MA 02115, USA}\affiliation{\footnotesize Quantum Materials and Sensing Institute, Northeastern University, Burlington, MA 01803, USA.}

\author{\footnotesize Ivar Martin}\affiliation{\footnotesize Materials Science Division, Argonne National Laboratory, Lemont, Illinois 60439, USA}

\author{\footnotesize Ni Ni}\affiliation{\footnotesize Department of Physics and Astronomy and California NanoSystems Institute, University of California, Los Angeles, Los Angeles, CA 90095, USA.}
\author{\footnotesize Qiong Ma}\affiliation{\footnotesize Department of Physics, Boston College, Chestnut Hill, MA, USA}\affiliation{\footnotesize The Schiller Institute for Integrated Science and Society, Boston College, Chestnut Hill, MA, USA}

\author{\footnotesize David J. E. Marsh}\affiliation{\footnotesize Theoretical Particle Physics and Cosmology, King's College London, UK}
\author{\footnotesize Ashvin Vishwanath}\affiliation{\footnotesize Department of Physics, Harvard University, Cambridge, MA 02138, USA}

\author{\footnotesize Su-Yang Xu\footnote{Corresponding author: suyangxu@fas.harvard.edu}}\affiliation{\footnotesize Department of Chemistry and Chemical Biology, Harvard University, Massachusetts 02138, USA}

\maketitle

\textbf{In 1978, Wilczek and Weinberg theoretically discovered a new boson - the Axion - which is the coherent oscillation of the $\theta$ field in QCD \cite{wilczek1978problem,weinberg1978new}. Its existence can solve multiple fundamental questions including the strong CP problem of QCD and the dark matter. However, its detection is challenging because it has almost no interaction with existing particles. Similar $\theta$ has been introduced to condensed matter and so far studied as a static, quantized value to characterize topology of materials \cite{ essin2009magnetoelectric, wu2016quantized, mogi2022experimental}. But the coherent oscillation of $\theta$ in condensed matter is proposed to lead to new physics directly analogous to the high-energy Axion particle, the dynamical Axion quasiparticle (DAQ) \cite{li2010dynamical, wang2019dynamic, zhang2020large,  wang2020dynamical, zhu2021tunable, roising2021axion, liu2022dynamical, lhachemi2024phononic, shiozaki2014dynamical, sekine2016chiral, sekine2017electric, taguchi2018electromagnetic, terccas2018axion, xiao2021nonlinear,   curtis2023finite,marsh2019proposal,  schutte2021axion, chigusa2021axion}. In this paper, we present the direct observation of the DAQ. By combining 2D electronic device with ultrafast pump-probe optics, we manage to measure the magnetoelectric coupling $\alpha$ ($\theta\propto\alpha$) of 2D MnBi$_2$Te$_4$ with sub-picosecond time-resolution. This allows us to directly observe the DAQ by seeing a coherent oscillation of $\theta$ at $\sim44$ GHz in real time, which is uniquely induced by the out-of-phase antiferromagnetic magnon. Interestingly, in 2D MnBi$_2$Te$_4$, the DAQ arises from the magnon-induced coherent modulation of Berry curvature. Such ultrafast control of quantum wavefunction can be generalized to manipulate Berry curvature and quantum metric of other materials in ultrafast time-scale.  Moreover, the DAQ enables novel quantum physics such as Axion polariton and electric control of ultrafast spin polarization \cite{li2010dynamical, sekine2016chiral, sekine2017electric, taguchi2018electromagnetic, terccas2018axion, xiao2021nonlinear,   curtis2023finite}, implying applications in unconventional light-matter interaction and coherent antiferromagnetic spintronics \cite{han2023coherent}. Beyond condensed matter, the DAQ can serve as a detector of the dark matter Axion particle \cite{marsh2019proposal,  schutte2021axion, chigusa2021axion}. We estimate the detection frequency range and sensitivity in the critically-lacking meV regime, contributing to one of the most challenging questions in fundamental physics.}

\vspace{0.5cm}
\textbf{Introduction}

In optically-driven quantum materials, phonons -- coherent modulation of lattice -- can drastically modify the electronic structure, leading to exotic non-equilibrium phenomena \cite{de2021colloquium,mitrano2016possible,sie2019ultrafast} such as light-induced superconductivity \cite{mitrano2016possible} and phonon-switch of Weyl fermions \cite{sie2019ultrafast}. Beyond phonons, quantum materials have many other collective excitations, suggesting novel handles to engineer electronic structure at ultrafast timescales. Magnets have coherent oscillation of spins -- magnons \cite{bae2022exciton, gao2024giant, kirilyuk2010ultrafast}. Specifically, in 2D MnBi$_2$Te$_4$, the antiferromagnetic Mn spins couple with the low-energy surface Dirac bands, giving rise to large Berry curvatures. A fundamentally interesting question is how the Berry curvature will be modulated by the magnons. Interestingly, the magnon-induced ultrafast Berry curvature modulation in 2D MnBi$_2$Te$_4$ also leads to a dynamical Axion quasiparticle (DAQ) that is directly analogous to the high energy Axion particle.

\begin{figure*}[t]
\includegraphics[width=15cm]{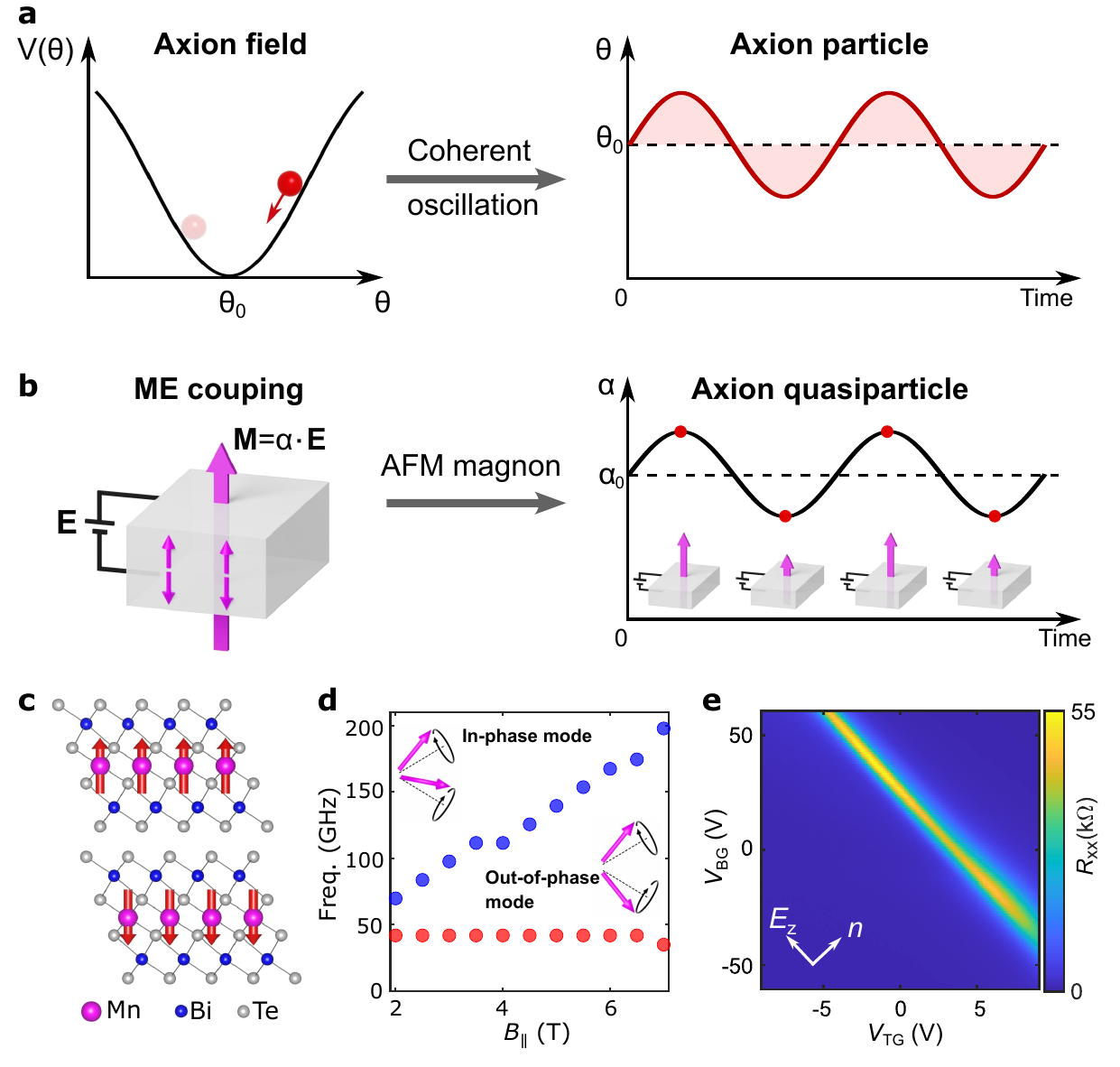}
\vspace{0cm}
\caption{{\bf Axion particle and quasiparticle and basic information of 2D MnBi$_2$Te$_4$.} \textbf{a,} The Axion particle is the coherent oscillation of the $\theta$ field in the QCD \cite{wilczek1978problem,weinberg1978new}. \textbf{b,} In condensed matter, the $\theta$ angle is directly proportional to the trace of the magnetoelectric coupling, $\theta=\pi\frac{2h}{3e^2}\sum_{i=x,y,z}\alpha_{ii}$. The DAQ is the coherent oscillation of the magnetoelectric coupling. \textbf{c,} MnBi$_2$Te$_4$ has a centrosymmetric rhombohedral lattice structure (the nonmagnetic group is $-3m$). It has a layered antiferromagnetic order, where Mn spins within a layer are ferromagnetic with an out-of-plane easy axis and Mn spins between two adjacent layers are anti-parallel (the magnetic group of an even-layer 2D MnBi$_2$Te$_4$ is $-3'm'$). \textbf{d,} Experimentally measured frequencies for the in-phase magnon and out-of-phase magnon modes as a function of the in-plane magnetic field $B_{\|}$ of 6-layer MnBi$_2$Te$_4$. See schematics in Extended Data Figs.~\ref{Magnon}\textbf{a-d} for a detailed visualization of the magnon modes. \textbf{e,} The dual gate resistance map for a 6-layer MnBi$_2$Te$_4$ electrical device.  }
\label{Fig1}
\end{figure*}

In particle physics, the Axion is a boson generated by the coherent oscillation of the $\theta$ field (Fig.~\ref{Fig1}\textbf{a}) \cite{wilczek1978problem,weinberg1978new}. This particle is of importance in QCD, cosmology and string theory \cite{wilczek1978problem,weinberg1978new}. However, due to the weak interaction with normal matter, the search for Axion has been a challenging problem. In condensed matter, similar $\theta$ was introduced \cite{ essin2009magnetoelectric} as a static value, which is proportional to a material's magnetoelectric coupling $\theta=\pi\frac{2h}{e^2}\alpha$. In the presence of time-reversal $\mathcal{T}$ or space-inversion $\mathcal{P}$ symmetry, $\theta$ is quantized in insulators and it describes the topological invariant. The experimental studies of the static $\theta$ have led to breakthroughs generating great interest \cite{wu2016quantized, allen2019visualization, mogi2022experimental}. Beyond the static $\theta$, the coherent oscillation of $\theta$ is proposed to lead to new physics directly analogous to the high-energy Axion particle, the DAQ. When $\mathcal{T}$ and $\mathcal{P}$ symmetries are both broken (e.g. in certain antiferromagnetic insulators), $\theta$ can coherently oscillate when coupled to certain magnetic fluctuations, giving rise to the DAQ.  Since the initial prediction by  \cite{li2010dynamical}, the DAQ has attracted great interest and has been anticipated by many theoretical works \cite{li2010dynamical, wang2019dynamic, zhang2020large,  wang2020dynamical, zhu2021tunable, roising2021axion, liu2022dynamical, lhachemi2024phononic, shiozaki2014dynamical, sekine2016chiral, sekine2017electric, taguchi2018electromagnetic, terccas2018axion, xiao2021nonlinear,   curtis2023finite,marsh2019proposal,  schutte2021axion, chigusa2021axion}: (1) The DAQ enables a wide range of novel quantum electromagnetic and spin phenomena \cite{li2010dynamical, sekine2016chiral, sekine2017electric, taguchi2018electromagnetic, terccas2018axion, xiao2021nonlinear,   curtis2023finite}. (2) The DAQ provides a condensed matter simulation of the high energy Axion particle. (3) Beyond a simulator, the DAQ can be used as a detector of the Axion particle \cite{marsh2019proposal,  schutte2021axion, chigusa2021axion}. 


In principle, any antiferromagnetic material that breaks $\mathcal{T}$ and $\mathcal{P}$ can host the DAQ. However, in most systems, both topologically nontrivial and trivial, the DAQ turns out to be very weak \cite{zhang2020large}: I.e., the coherent oscillation of $\theta$ is very weak even under a strong magnetic fluctuation.  Also, DAQ needs to be driven by specific magnetic fluctuations. Previous theoretical works considered the antiferromagnetic amplitude mode \cite{li2010dynamical, zhang2020large,  wang2020dynamical, zhu2021tunable}, which is difficult to excite experimentally. These challenges have hindered the experimental detection of DAQ. On a separate front, beyond antiferromagnetic insulators, the dynamical $\theta$ has also been proposed to arise from the phason (sliding mode) of CDW state in Weyl semimetals. Promising DC transport evidence has been reported in Ta$_2$Se$_8$I \cite{gooth2019axionic}, but definitive demonstration remains lacking \cite{sinchenko2022does}.  In this paper, we made critical conceptual advances to bridge theory and experiments including: (1) identifying an antiferromagnetic system that features large DAQ, (2) figuring out a magnetic fluctuation that can induce the DAQ but is much more experimentally accessible, and (3) developing an experimental scheme that can directly measure the $\theta$ oscillation. \color{black} In particular, previous works have investigated the optical detection as well as the ultrafast magnon dynamics in MnBi$_2$Te$_4$, CrI$_3$ and Cr$_2$Ge$_2$Te$_6$ \cite{bartram2022ultrafast, lujan2022magnons, bartram2023real, padmanabhan2022large, padmanabhan2022interlayer, cheng2024magnetic, Qiu2023axion}. In particular, Ref. \cite{bartram2023real} systematically studied the dynamics of the out-of-phase magnons in MnBi$_2$Te$_4$ and Ref. \cite{Qiu2023axion} reported the AFM Kerr effect in even-layer MnBi$_2$Te$_4$. These works provide important knowledge that guides our work here. \color{black} As such, we present the observation of the DAQ, driven by the out-of-phase antiferromagnetic magnons  under in-plane magnetic field  in 2D even-layer MnBi$_2$Te$_4$.

\vspace{0.5cm}
\textbf{Basic characterization}

Figure~\ref{Fig1}\textbf{c} shows a sideview of 2D even-layer MnBi$_2$Te$_4$ \cite{Otrokov2019a, Li2019intrinsic,Zhang2019topological,  Deng2020, Liu2020a, yang2021odd, Ovchinnikov2020, gao2021layer, cao2023switchable, li2024fabrication, chong2024intrinsic,bartram2022ultrafast, lujan2022magnons, padmanabhan2022large, padmanabhan2022interlayer, bartram2023real,cheng2024magnetic, fonseca2024picosecond}. The crystal structure is centrosymmetric, but the layered antiferromagnetic order breaks both time-reversal and space-inversion symmetries (the Mn spins have an out-of-plane easy axis). As such, 2D even-layer MnBi$_2$Te$_4$ is a possible platform to host the DAQ \cite{zhu2021tunable}. Note that symmetry only dictates the presence or absence of the DAQ. Why the DAQ is large in even-layer MnBi$_2$Te$_4$ will be discussed later.

We study the antiferromagnetic magnons in 2D even-layer MnBi$_2$Te$_4$, focusing on the scenario in the presence of an in-plane static magnetic field $B_{\|}$ as it is most relevant to our experiments. The $B_{\|}$ cants the Mn spins away from the out-of-plane easy axis, where the equilibrium spin directions under finite $B_{\|}$ are noted by the dashed lines in Fig.~\ref{Fig1}\textbf{d}. \color{black} SI.I.7 shows additional measurements to validate such evolution \color{black} The magnons are spin precessions away from the dashed lines.  Considering the two antiferromagnetic sublattices, there are naturally two magnon modes, where the spin precisions of the antiferromagnetic sublattices have the same phase (the in-phase mode) or opposite phase (the out-of-phase mode). Extended Data Figs.~\ref{Magnon}\textbf{a,b} show complete time evolution of the spin precessions. Moreover, it is informative to study the projection of the spins to the $\hat{z}$ axis ($S_z$). As shown in Extended Data Figs.~\ref{Magnon}\textbf{c,d}, for the out-of-phase mode, the net out-of-plane magnetization $M_z=S_{1z}+S_{2z}$ is zero. For the in-phase mode, by contrast, the net magnetization $M_z$ is nonzero and it oscillates as a function of time.  We have characterized their resonant frequencies as a function of $B_{\|}$ \cite{zhang2020gate} (see Methods). \color{black} As shown in Fig.~\ref{Fig1}\textbf{d}, the in-phase magnon frequency increases monotonically with $B_{\|}$; the out-of-phase magnon frequency remains roughly invariant and starts to slightly decrease at $7$ T. Strictly speaking, 6SL has six magnon modes. In SI, we present thorough discussion in connection to our experimental data. \color{black}


Next, we have made dual-gated electrical devices with 6-layer MnBi$_2$Te$_4$. The combination of the top and bottom gates allows us to control the charge density $n$ and the out-of-plane electric field $E_{z}$ independently.  \color{black} We always kept the Fermi level inside the gap at $n=0$ unless otherwise noted. Also, in the main text, we studied the $E$-field dependence by the lock-in method where $E$ was AC modulated (see Methods.4). \color{black} Figure~\ref{Fig1}\textbf{e} shows its resistance gate map, where the charge neutrality resistance peak is observed.

\vspace{0.5cm}
\textbf{Static $\theta$ measurements}

The DAQ is a coherent oscillation of $\theta(t)$, which directly manifests as a coherent oscillation of the magnetoelectric coupling $\alpha(t)$ because of their proportionality ($\theta=\pi\frac{2h}{e^2}\alpha$). Before searching for the time-dependent oscillation, we first study the static magnetoelectric coupling $\alpha$. The magnetoelectric coupling is defined as the electric field induced magnetization $\alpha=\frac{d M}{d E}$. Experimentally, \color{black} we measure the $\alpha$ by measuring the $E$-induced Kerr effect whiling keeping the Fermi level always at charge neutrality $n=0$ (unless otherwise noted)\color{black}. Specifically, out-of-plane electric field $E_{z}$ is applied by the dual gates;  We use the optical Kerr rotation under normal incidence to measure the net out-of-plane magnetization,  $\textrm{Kerr rotation}= M_{z}/\gamma $, $\gamma$ is a material specific conversion coefficient (\color{black} SI.I.3 presents additional studies including symmetry analysis, wavelength dependence and simultaneous Kerr and Faraday data that further strengthen the conclusion that our $E$-field induced Kerr rotation measures the $E$-field induced $M_z$ {black}).  Therefore, we measure $\alpha$ by measuring the slope of the Kerr rotation with respect to $E_{z}$, i.e., $\alpha =\gamma\frac{d \textrm{Kerr rotation}}{d E_z}$.\color{black}

\begin{figure*}[t]
\includegraphics[width=11.5cm]{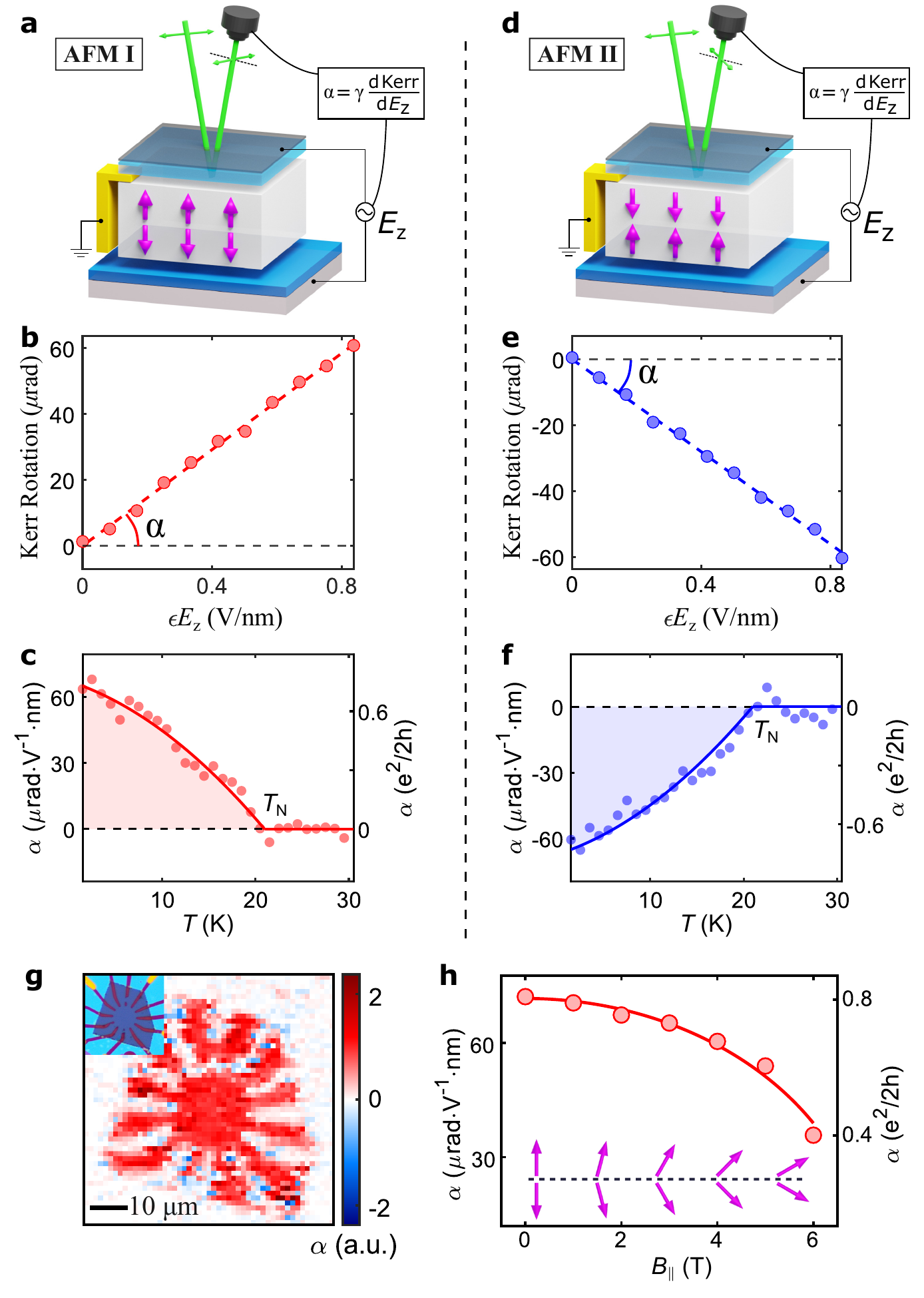}
\vspace{-0.5cm}
\caption{{\bf Probing the static magnetoelectric coupling in 2D MnBi$_2$Te$_4$.} \textbf{a,} Schematic experimental setup to probe the static magnetoelectric coupling. The out-of-plane electric field $E_{z}$ is applied by the top and bottom gates. The $E_{z}$-induced magnetization is measured by the optical Kerr rotation under normal incidence. Hence, $\alpha$ is measured by $\frac{d\textrm{Kerr rotation}}{d E_z}$. \textbf{b,} Measured Kerr rotation as a function of AC $E_{z}$ modulation for our 6-layer MnBi$_2$Te$_4$ device. The slope here is $\alpha$. \textbf{b,} Measured $\alpha$ as a function of temperature. We have converted $\alpha$ from the unit of $\mu\textrm{rad}\cdot V^{-1}\cdot \textrm{nm}$ (the left axis) to the unit of $\frac{e^2}{2h}$ (the right axis) (see Methods). \color{black} SI.I.4 shows the DC $E_{z}$ dependence, which yields consistent results of $\alpha$.  \color{black} \textbf{d-f,} The same as panels \textbf{a-c} but for the opposite antiferromagnetic domain. \textbf{g,} Spatial map of the measured $\alpha$ over the entire device. \textbf{h,} Measured $\alpha$ as a function of $B_{\|}$. \color{black} The probe wavelength was 515 nm. SI.I.3 shows the wavelength dependent data. The sample was cooled with parallel $\mathbf{E}\cdot\mathbf{B}$ field. Depending on the sign of $\mathbf{E}\cdot\mathbf{B}$, the sample was prepared into a specific AFM state. All measurements were performed at normal incidence. The small incidence angle in Figs. 2 and 3 are only an artistic choice. \color{black} }
\label{Fig2}
\end{figure*}

Indeed, our data in Fig.~\ref{Fig2}\textbf{b} shows that the measured Kerr rotation increases linearly with increasing $E_{z}$. Its slope is proportional to the magnetoelectric coupling $\alpha$. We then studied the temperature dependence of $\alpha$ (Fig.~\ref{Fig2}\textbf{c}): $\alpha$ is the largest at the base temperature and decreases to zero at the Neel temperature $T_{\textrm{N}}$. We also studied $\alpha$ in the opposite antiferromagnetic state (Figs.~\ref{Fig2}\textbf{e,f}), which shows the opposite sign. We further studied the spatial dependence. As shown in Fig.~\ref{Fig2}\textbf{g}, our data reveal homogeneous $\alpha$ across the entire device. As a side point, this also demonstrates a way to spatially resolve antiferromagnetic domains, which has been challenging. Moreover, we have converted $\alpha$ from the unit of $\mu\textrm{rad}\cdot V^{-1}\cdot \textrm{nm}$ (the left axis) to the unit of $\frac{e^2}{2h}$ (the right axis) (see Methods). We also studied the in-plane magnetic field $B_{\|}$ dependence (Fig.~\ref{Fig2}\textbf{h}). We found that $\alpha$ monotonically decreases as $B_{\|}$ cants the equilibrium spin directions away from the out-of-plane easy axis.

\vspace{0.5cm}
\textbf{Observing the dynamical Axion quasi-particle by ultrafast modulation of $\theta$}

\begin{figure*}[t]
\vspace{-0.5cm}
\includegraphics[width=13cm]{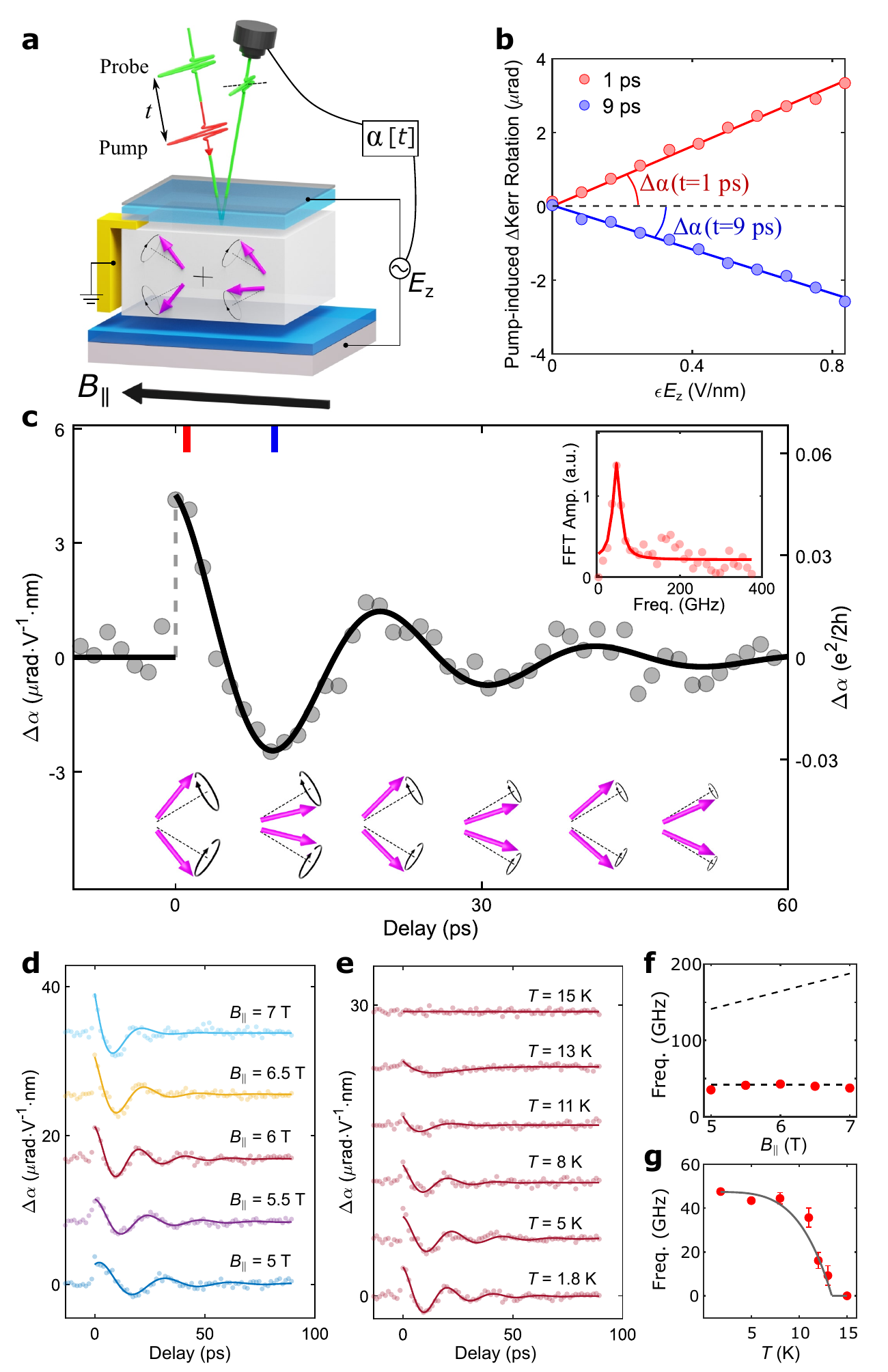}
\vspace{-0.5cm}
\caption{{\bf Observation of the dynamical Axion quasiparticle in 2D MnBi$_2$Te$_4$.} \textbf{a,} Schematic experimental setup to probe the time-dependent magnetoelectric coupling. The pump laser launches coherent magnons; The probe laser, combined with the dual gate $E_z$, measures $\frac{d \textrm{Kerr rotation}}{d E_z}$; By varying the delay time $t$, we can measure pump-induced $\Delta\alpha(t)$ with femtosecond resolution. \textbf{b,} Measured pump-induced Kerr rotation $\Delta \textrm{Kerr}(t)$ as a function of $E_z$ and $t$. \textbf{c,} Measured pump-induced $\Delta\alpha(t)$ at $B_{\|}=6$ T. A clear oscillation of $\alpha(t)$ provides the experimental observation of the DAQ. Inset: FFT of the data. \textbf{d,e,} $\Delta\alpha(t)$ at different $B_{\|}$ (\textbf{d}) and temperature (\textbf{e}) values. The curves are offset for clarity \textbf{f,g,} FFT of data in panels (\textbf{d,e}). }
\label{Fig3}
\end{figure*}

Now that we have individually characterized the magnons and the static $\theta$, we combine them to search for the ultrafast coherent oscillation of $\theta$. To this end, we construct an experimental setup that combines ultrafast pump-probe optics with 2D electronic devices. As shown in Fig.~\ref{Fig3}\textbf{a}, the pump laser launches coherent magnons (both the in-phase and out-of-phase modes); The probe detects the magnetoelectric coefficient $\alpha$, which is measured by $E_z$-induced Kerr rotation. By varying the delay time $t$ between the pump and probe, we can measure the time-dependent magnetoelectric coupling with femtosecond resolution $\alpha(t)$. Note that the time-dependent magnetoelectric coupling contains a static component and a pump-induced dynamical change, $\alpha(t)=\alpha_{\textrm{static}}+\Delta\alpha(t)$. The $\alpha_{\textrm{static}}$ has been studied above, so we focus on $\Delta\alpha(t)$.

Figure~\ref{Fig3}\textbf{c} shows the most essential dataset of our study here: The $x$ axis is the delay time $t$; the $y$ axis is the pump-induced dynamical change of the magnetoelectric coupling $\Delta\alpha(t)$; The in-plane magnetic field $B_{\|}$ is $6$ T.  From the data, we clearly observe a coherent oscillation of magnetoelectric coefficient $\alpha(t)$ ($\propto\theta(t)$), therefore demonstrating the DAQ by its definition. 

Interestingly, the magnitude of $\alpha$ oscillation is quite large: the $\Delta\alpha(t)$ oscillation reaches $0.05\frac{e^2}{2h}$ (right axis of Fig.~\ref{Fig3}\textbf{c}), which is $12\%$ of the static $\alpha$ at the corresponding magnetic field $B_{\|}=6$ T  (Fig.~\ref{Fig2}\textbf{h}). The FFT (inset of Fig.~\ref{Fig3}\textbf{c}) shows a distinct resonance at $\sim 44$ GHz, consistent with the out-of-phase magnon mode.  The underlying out-of-phase magnon oscillation (arrows in the Fig.~\ref{Fig3}\textbf{c}) are drawn synchronized with the observed $\alpha$ oscillation, which allows us to directly visualize how the coherent spin preccesion controls $\alpha$ at different time point.  Interestingly, although both the in-phase and out-of-phase magnons are launched by the pump laser, only the out-of-phase magnons directly couple to the $\theta$ field in MnBi$_2$Te$_4$, giving rise to a coherent oscillation of $\theta$, which will be explained in the next section.

To further substantiate our conclusion, we perform systematic measurements. First, it is important to check that we are really measuring the $\alpha$ oscillation in Fig.~\ref{Fig3}\textbf{c}. \color{black} In other words, each data point of $\Delta\alpha$ in Fig.~\ref{Fig3}\textbf{c} should correspond to the slope between pump-induced Kerr and electric field $E_z$ (see also Extended Data Fig.~\ref{Ultrafast_ME}) \color{black}.  This is explicitly checked in Fig.~\ref{Fig3}\textbf{b}, where the pump-induced $\Delta \textrm{Kerr}$ ($\propto\Delta M_z$) shows a linear dependence to $E_z$. Therefore, all $\Delta M_z$ observed in Fig.~\ref{Fig3}\textbf{b} is induced by electric field. Comparing Figs.~\ref{Fig3}\textbf{b,c} directly, the slope for $1$ ps and $9$ ps in Fig.~\ref{Fig3}\textbf{b} is opposite. Correspondingly, the $\Delta\alpha(t)$ at $1$ ps and $9$ ps in Fig.~\ref{Fig3}\textbf{c} indeed show opposite sign.   Second, we measure $\Delta\alpha(t)$ as a function of the in-plane magnetic field $B_{\|}$, where the raw data is shown in Fig.~\ref{Fig3}\textbf{d}. By performing FFT on every time trace, we found (Fig.~\ref{Fig3}\textbf{f}) that the resonant frequency show weak dependence on $B_{\|}$, again confirming that the observed DAQ is induced by the out-of-phase magnons. Third, we performed temperature dependent measurements. As shown in Figs.~\ref{Fig3}\textbf{e,g}, We found a critical slowdown behavior, i.e., the resonance frequency decreases to zero at the transition temperature. \color{black} In addition, it is crucial to check that the magnetic properties such as exchange coupling, anisotropy and magnon frequency of MnBi$_2$Te$_4$ are not significantly changed by gates. SI.I.3-5 presents additional systematic MOKE measurements including DC gate dependence and similar measurements in odd layer, which demonstrate that within the experimental $E$ and $n$ ranges studied here the magnetic properties are roughly invariant. We also present additional measurements to assess the sample quality following the processing of the electric contacts and to disentangle magnon and phonon origin for the observed DAQ (SI.I.3,6). These systematic studies further strengthen our experimental observation of the DAQ.
\color{black}

\vspace{0.5cm}
\textbf{Berry curvature origin}

We now study the microscopic origin of  $\alpha$.  We note that, independent of the microscopic origin, a coherent oscillation of $\theta$ always generates DAQ. In other material candidates, DAQ may arise from other mechanisms beyond Berry curvature (see discussion in the last section).   

\begin{figure*}[t]
\vspace{-0.5cm}
\includegraphics[width=14cm]{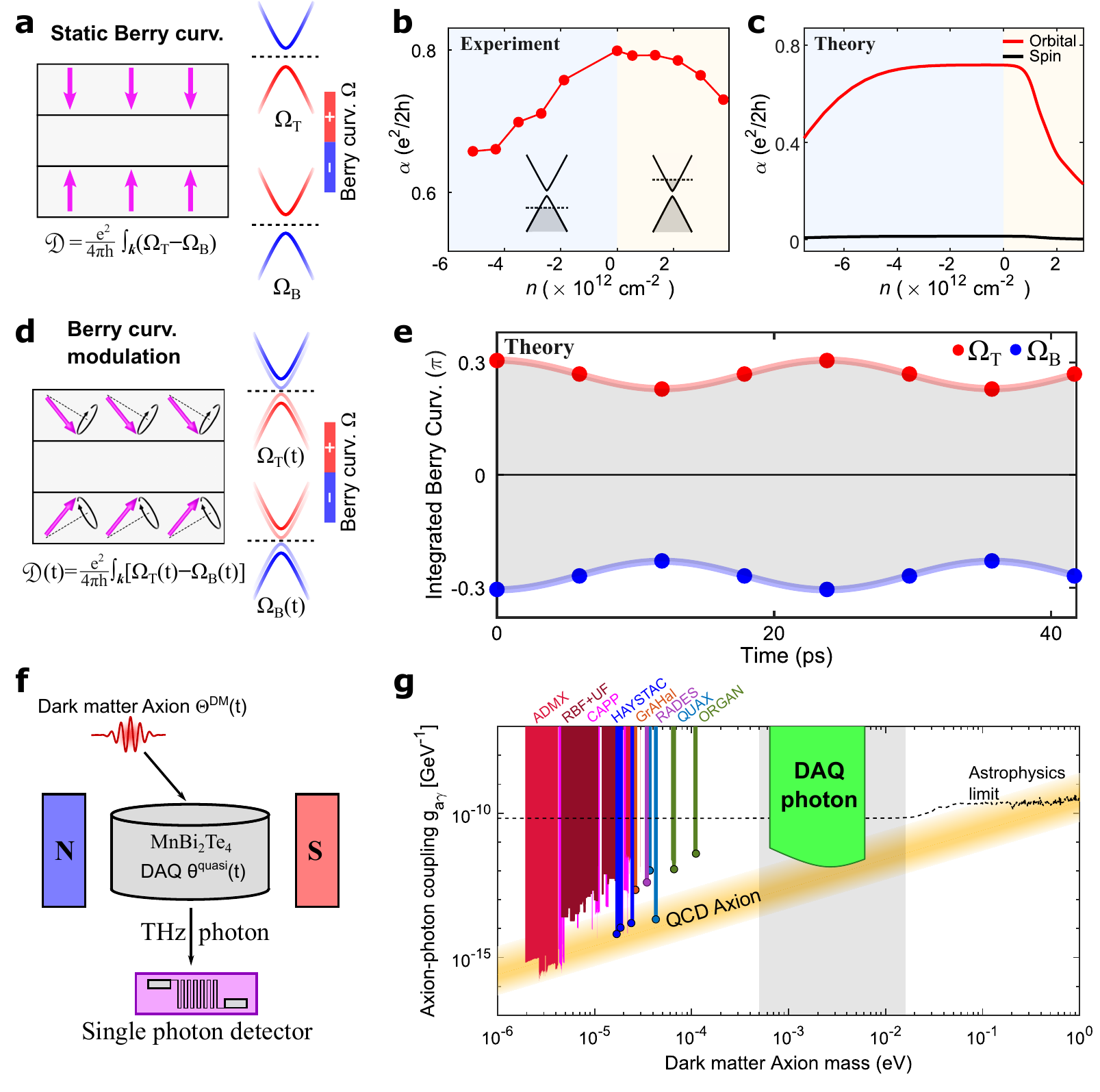}
\vspace{-0.5cm}
\caption{{\bf Ultrafast control of Berry curvature and dark matter detection sensitivity.} \textbf{a,} In 2D even-layer MnBi$_2$Te$_4$, the antiferromagnetic order generates opposite Berry curvatures on the top and bottom surfaces ($\Omega_{\textrm{T}}=-\Omega_{\textrm{B}}$), resulting in a Berry curvature real space dipole $\mathcal{D}$. \textbf{b,c,} Experimentally-measured (\textbf{b}) and theoretically-calculated (\textbf{c}) static $\alpha$ as a function of carrier density $n$ for 6-layer MnBi$_2$Te$_4$. For theory, the orbital and spin contributions are computed separately. \textbf{d,} The coherent precession of antiferromagnetic order potentially leads to an ultrafast control of Berry curvature. \textbf{e,} Calculated Berry curvature sum for the top and bottom surfaces at different spin angles for the out-of-phase magnons. The spin equilibrium angles are calculated based on the Heisenberg model (see SI.I.2.1). The grey area corresponds to the Berry curvature real space dipole $\mathcal{D}$. \textbf{f,} Schematic illustration of using DAQ to detect the dark matter Axions. \textbf{g,} Dark matter detection sensitivity ($g_{a\gamma}$) as a function of the Axion mass. The mission is to search for the Axion in the entire area that is below the horizontal dotted line (i.e., the astrophysics limit). The $\mu$eV regime has been extensively explored. Each vertical streak denotes a particular previous experiments. By contrast, the meV regime (the grey area) is particular promising according to cosmology and astrophysics \cite{gorghetto2021more, saikawa2024spectrum}, but there is no detector. The yellow ribbon is the region where the dark matter Axion further solves the strong CP problem of the QCD.  The light green area shows the estimated detection range and sensitivity for the DAQ in MnBi$_2$Te$_4$. }
\label{Fig4}
\end{figure*}

$\alpha$ has two contributions, spin and Berry curvature. The spin part comes from the localized magnetic ions (e.g. $3d$ of Mn or Cr); The Berry curvature part comes from the orbital motion of itinerant electrons. Their relative contribution depends on the specific electronic structure of a system. To investigate this in MnBi$_2$Te$_4$, we measure its static $\alpha$ as a function of the charge density $n$. Our data shows that $\alpha$ varies with $n$ (Fig.~\ref{Fig4}\textbf{b}). Importantly, changing $n$ corresponds to changing the occupation of the itinerant electrons, whereas the localized Mn $3d$ orbitals are not affected.  This can be clearly seen in the MnBi$_2$Te$_4$ band structure shown in Extended Data Fig.~\ref{Theory1}, where the Mn $3d$ bands are far (a few eV) away from the Fermi energy.  Therefore, the observed $n$ dependence provides a qualitative experimental evidence that the magnetoelectric coupling $\alpha$ in MnBi$_2$Te$_4$ is dominated by the orbital contribution from the itinerant electrons.


To achieve quantitative studies, we directly compute the different contributions of $\alpha$ based on the first-principles band structure of 6-layer MnBi$_2$Te$_4$ (Fig.~\ref{Fig4}\textbf{c}). Interestingly, the calculated Berry curvature contribution strongly dominates over the spin contribution, providing another supporting evidence. Moreover, the calculated $n$ dependence show reasonable agreement with our experimental data. Therefore, we conclude that $\alpha$ in MnBi$_2$Te$_4$ is dominated by the orbital Berry curvature contribution. Specifically, this contribution equals the Berry curvature real space dipole $\mathcal{D}$, which we provide a band structure understanding as follows. A topological insulator features Dirac fermions on the top and bottom surfaces, which are gapped by the antiferromagnetic order, resulting in opposite Berry curvatures on opposite surfaces ($\Omega_{\textrm{T}}=-\Omega_{\textrm{B}}$). Hence, there is a Berry curvature real space dipole $\mathcal{D}=\frac{e^2}{4\pi h}\int_\mathbf{k}(\Omega_{\textrm{T}}-\Omega_{\textrm{B}})$ (Fig.~\ref{Fig4}\textbf{a}). One can show from Berry phase theory that the orbital contribution of $\alpha$ equals $\mathcal{D}$ (see derivation in Methods.5).

We can now achieve further evidence for $\alpha$ equals $\mathcal{D}$. Even-layer MnBi$_2$Te$_4$ hosts an electric-field-induced anomalous Hall effect (AHE) - the so-called layer Hall effect \cite{gao2021layer}. This electrical transport Hall measurement, which is independent from the optical measurement above, provides another probe of $\mathcal{D}$. Specifically, one can show that the electric-field-induced AHE $\sigma_{xy}$ measures $d\mathcal{D}/dn$ (derivation in SI.III). On the other hand, we obtain $d\alpha/dn$ by taking a derivative of our $\alpha$ data in Fig.~\ref{Fig4}\textbf{b}. Shown in Extended Data Fig.~\ref{LHE}, the agreement between $d\mathcal{D}/dn$ and $d\alpha/dn$ further supports $\alpha=\mathcal{D}$, i.e., the Berry curvature origin for the magnetoelectric coupling in MnBi$_2$Te$_4$.

\vspace{0.5cm}
\textbf{Ultrafast control of Berry curvature by antiferromagnetic magnons}



Our experimental observation in Fig.~\ref{Fig3}\textbf{c} of ultrafast modulation of $\alpha$ naturally establishes an ultrafast modulation of Berry curvature real space dipole because we showed $\alpha=\mathcal{D}$ above. To further confirm this, we directly compute the band structure at different spin angles of the magnon oscillation under the frozen magnon approximation. For the out-of-phase magnon, we see (Fig.~\ref{Fig4}\textbf{e}) that the top and bottom Berry curvature oscillates with opposite phase. As such, their difference (the grey area of Fig.~\ref{Fig4}\textbf{e}), which is the Berry curvature real space dipole $\mathcal{D}$, also oscillates in time.  

We can further try to understand why the in-phase magnon does not induce $\alpha$ oscillation. For the in-phase magnons, our calculation shows that the top and bottom Berry curvature oscillates in time but with the same phase (Extended Data Fig.~\ref{Theory2}). So the Berry curvature real space dipole $\mathcal{D}$ is invariant in time.  Therefore, the coherent oscillation of Berry curvature real space dipole generates the DAQ in even-layer MnBi$_2$Te$_4$.

 We explain why the DAQ is strong in even-layered MnBi$_2$Te$_4$. Beyond Berry curvature, it also requires hybridization between the top and bottom surface states due to finite thickness. The strength of DAQ is measured by $\frac{\delta\theta}{\delta L_z}$ \cite{li2010dynamical}, i.e., the change of $\theta$ per change of the antiferromagnetic order $L_z$. If we reduce the hybridization strength, $\theta$ becomes robust when $L_z$ changes, leading to a small $\frac{\delta\theta}{\delta L_z}$. Only when the hybridization is comparable to the magnetism-inducd Zeeman gap, a large $\frac{\delta\theta}{\delta L_z}$ is achieved. \color{black} Our calculations in Extended Data Fig.~\ref{Theory3} indeed shows that 6SL MnBi$_2$Te$_4$ is a good choice because of its large $\frac{\delta\theta}{\delta L_z}$. The thickness dependent $\frac{\delta\theta}{\delta L_z}$ calculation also provides a guidance of how the strength of the effect may evolve as a function of layer number.  \color{black}.  


\vspace{0.5cm}
\textbf{Sensitivity of dark matter Axion detection in the meV mass regime }

 Searching for dark matter Axion is one of the most challenging yet exciting topics in fundamental physics. Below, we describe the state-of-the-art in this field   in terms of existing detectors and proposed future detectors   and then estimate the axion detection frequency range and sensitivity based on our experimental results. This will justify why our DAQ approach can break new ground and make important contributions. The Axion mass is not known but astrophysics experiments have excluded mass $>1$ eV. Therefore, the mission is to search for the Axion below $1$ eV, which is represented by the entire area below the horizontal dotted line (the astrophysics limit) in Fig.~\ref{Fig4}\textbf{g}. Given the vast frequency range, detectors should have tunable frequency. For example, the $\mu$eV regime, which corresponds to microwave frequencies, has been experimentally explored (vertical streaks in Fig.~\ref{Fig4}\textbf{g}), because microwave photon cavities with tunable frequencies can serve as the Axion detectors in that regime. 

By contrast, the meV mass regime is particular promising according to cosmology and astrophysics \cite{gorghetto2021more, saikawa2024spectrum} (the grey region in Fig.~\ref{Fig4}\textbf{g}), but  \textit{there is no Axion detector for the meV regime}. This regime is more technologically challenging: Cavity methods struggle, since tuning over wide ranges becomes difficult. As such, any detector in the meV regime that goes below the astrophysics limit breaks new ground, and there are only two proposed future detectors known as BRASS and BREAD ~\cite{horns2013searching,liu2022broadband}. Interestingly, theory \cite{marsh2019proposal,  schutte2021axion, chigusa2021axion} predicts that the DAQ can serve as a detector in this regime (Fig.~\ref{Fig4}\textbf{f}). The basic working principle is explained in Methods.6. A unique advantage is that the detection frequency can be continuously tuned over a wide range by $B_\parallel$, because the detection frequency is given by $\sqrt{m_{\textrm{DAQ}}^2+C^2B_{\parallel}^2}$, where $m_{\textrm{DAQ}}=44$ GHz$=0.18$ meV and $C$ is a constant (see Methods.6). However, this proposal has not been prominent in experimental particle physics since the DAQ remained a proposal. Going beyond the theory proposal \cite{marsh2019proposal,  schutte2021axion, chigusa2021axion}, (1) Our work provides the realization of the DAQ, serving as an essential step forward. (2) we made the conceptional advance that the out-of-phase magnon under $B_{\|}$ can effectively replace the antiferromagnetic amplitude mode. (3) We have experimentally determined the DAQ frequency.

Using our experimental data and theoretical calculations, we calculate the DAQ sensitivity $g_{a\gamma}$ in Fig.~\ref{Fig4}\textbf{g} (see Methods for details). The DAQ is able to go below astrophysical limit (the horizontal dotted line) over a wide frequency range, and even reach sensitivity to the QCD Axion (yellow band in Fig.~\ref{Fig4}\textbf{g}) in a certain range.  To build a well-functioning detector, more experimental works and preparations are needed such as addressing the DAQ sample size issue, and experimentally measuring the THz losses. We describe our solutions to these issues in Methods.6 and SI.II.2.  In SI.II.3, we compare our DAQ approach to the other two proposed future detectors BRASS and BREAD~\cite{horns2013searching,liu2022broadband} and show that our approach has unique advantages.  

\vspace{0.5cm}
\textbf{DAQ materials beyond MnBi$_2$Te$_4$ and other condensed matter outlooks}

It is of interest to find other materials with large DAQ. We need large but non-quantized $\theta$ that changes significantly upon a small change of the magnetic order. For example, while $\theta$ in Cr$_2$O$_3$ and CrI$_3$ are small, some multiferroics are reported to have large $\alpha$, which can be potentially promising for large DAQ. The microscopic mechanism will be different, as the Berry curvature will be nearly zero in wide-gap insulators but there can be a large spin contribution. 


In future, theory works \cite{li2010dynamical, wang2019dynamic, zhang2020large,  wang2020dynamical, zhu2021tunable, roising2021axion, liu2022dynamical, lhachemi2024phononic, shiozaki2014dynamical, sekine2016chiral, sekine2017electric, taguchi2018electromagnetic, terccas2018axion, xiao2021nonlinear,   curtis2023finite,marsh2019proposal,  schutte2021axion, chigusa2021axion} have predicted a wide range of novel phenomena, such as the Axion polariton \cite{li2010dynamical}, the dynamical chiral magnetic effect \cite{sekine2016chiral}, finite momentum instability \cite{curtis2023finite} etc. In particular, the DAQ arises from the magnetoelectric coupling and the antiferromagnetic magnons, two phenomena of importance in spintronics. The observed coherent oscillation of $\alpha(t)$ means that a DC $E$ field can control ultrafast spin polarization, $M(t)=\alpha(t)E$, potentially relevant to the recent coherent antiferromagnetic spintronics \cite{han2023coherent}. The coherent control of Berry curvature demonstrated here can be generalized. Berry curvature is only the imaginary part of quantum geometry, whereas the real part is the quantum metric, which has recently been observed in MnBi$_2$Te$_4$ \cite{wang2023quantum,gao2023quantum}. The magnons may be able to coherently modulate its quantum metric, which can be detected by measuring the nonlinear Hall effect with time-resolution. Beyond MnBi$_2$Te$_4$, in magnetic Weyl semimetals such as Co$_3$Sn$_2$S$_2$ or Mn$_3$Sn, the magnetic spin direction dictates the $k$-space locations of the Weyl nodes. The magnons may coherently control the Weyl node locations, leading to time-dependent AHE. Finally, nonlinearly driven magnons \cite{zhang2024terahertz} may induce new non-equilibrium phases in analogy to the light-induced superconductivity driven by phonons \cite{mitrano2016possible}.

\vspace{0.5cm}

\vspace{0.5cm}
\vspace{0.3cm}
\textbf{Methods}
\\
\textbf{1. Bulk crystal growth and 2D sample fabrication}\\Our MnBi$_2$Te$_4$ bulk crystals were grown by the Bi$_2$Te$_3$ flux method \cite{Yan2019}. Elemental Mn, Bi and Te were mixed at a molar ratio of $15:170:270$, loaded in a crucible, and sealed in a quartz tube under one-third atmospheric pressure of Ar. The ampule was first heated to $900^{\circ}$C for $5$ hours. It was then moved to another furnace where it slowly cooled from $597^{\circ}$C to $587^{\circ}$C and stayed at $587^{\circ}$C for one day. The 2D device fabrication processes were completed in an argon environment without exposure to air, chemicals, or heat (O$_2$ and H$_2$O level below $0.01$ ppm and a dew point below $-96^{\circ}$C). The glovebox was attached to an e-beam evaporator, allowing us to make metal deposition without exposure to air. MnBi$_2$Te$_4$ was mechanically exfoliated onto a baked $300$-nm SiO$_2/$Si wafer using Scotch tape. A stencil mask technique \cite{gao2021layer} was used to make Cr/Au contacts on top of MnBi$_2$Te$_4$. BN flakes were directly exfoliated onto a polydimethylsiloxane (PDMS) film, and a 10-30 nm thick BN flake was identified and transferred onto the MnBi$_2$Te$_4$ as the top gate dielectric layer. Next, a metal gate was evaporated onto the BN/MnBi$_2$Te$_4$ heterostructure. \color{black} SI.I.3 presents additional measurements to assess the sample quality following the processing of the electric contacts. \color{black}  
\vspace{3mm}
\\
\textbf{2. How we characterized the magnon frequencies}\\ We followed the experimental methods in \cite{zhang2020gate}, which measured the magnon frequencies in bilayer CrI$_3$, also a fully-compensated layered antiferromagnet. In this method, a monolayer WSe$_2$ is stacked on top of the layered antiferromagnet (Extended Data Fig.~\ref{Magnon}\textbf{e}). This monolayer WSe$_2$ breaks the layer degeneracy, allowing us to selectively probe the top layer information \cite{zhang2020gate,zhong2020}. Pump-probe Kerr rotation under normal incidence was performed on this heterostructure. The pump laser launches the magnons. The probe Kerr rotation measures the out-of-plane magnetization $M_z$ of the top layer preferentially because of the WSe$_2$. The data is shown in Extended Data Figs.~\ref{Magnon}\textbf{f,g}. Coherent oscillations composed of two distinct frequencies are clearly observed, corresponding to the in-phase and the out-of-phase magnons. The monolayer WSe$_2$ allows us to selectively probe the top layer, which is crucial for observing both modes. We emphasize that the monolayer WSe$_2$ is ONLY used in this dataset (Extended Data Fig.~\ref{Magnon}) in order to characterize the magnon frequencies. For all other measurements (everything in the main text), we used even-layer MnBi$_2$Te$_4$ without WSe$_2$. \color{black} Additional pump photon polarization dependence (SI.I.2) shows that the excitation mechanism is laser-heating induced coherent precession. In future, one way to selectively excite the in-phase and out-of-phase magnons may be to use microwave/THz photons that are resonant with the corresponding magnon polariton frequencies.\color{black}


Key experimental parameters: The fs laser pulse was generated by an amplified Yb:KGW laser (Pharos, LightConversion) with pulse duration $168$ fs, wavelength $1030$ nm and repetition rate $100$ kHz. For the pump beam, an optical parametric amplifier (Orpheus, LightConversion) was used to tune the wavelength to $720$ nm (pump fluence $\sim3\mu$J/cm$^{2}$). For the probe beam, a BBO crystal was used to convert the laser wavelength to $515$ nm. The pump beam went through an optical chopper at $1000$ Hz and was combined with the probe beam by a dichroic mirror, both of which were focused on the sample (spot size $\sim 1 \mu$m). The reflected beam was filtered to remove the pump beam, passed through a half-wave plate and a Wollaston prism, and entered the balanced photodiode detector. The balanced photodiode signal was analyzed by a lock-in amplifier at the chopper frequency. 
\vspace{3mm}
\\
\textbf{3. How we converted $\alpha$ to the unit of $e^2/2h$ }\\ The antiferromagnetic ground state of even-layer MnBi$_2$Te$_4$ features the magnetoelectric coupling, which is the electric field induced magnetization $\alpha=\frac{d M}{d E}$. The out-of-plane electric field $E_{z}$ was applied by the top and bottom gates; The net out-of-plane magnetization $M_z$ was probed by the optical Kerr rotation under normal incidence  \cite{huang2017}, $\textrm{Kerr rotation}= M_{z}/\gamma$ ($\gamma$ is a material specific conversion coefficient). Therefore, we have $\alpha =\gamma\frac{d \textrm{Kerr rotation}}{d E_z}$. To converted $\alpha$ from the directly measured unit of $\mu\textrm{rad}\cdot V^{-1}\cdot \textrm{nm}$ to the unit of $\frac{e^2}{2h}$, we need to determine the coefficient $\gamma$.

We (semi-)quantitatively determined $\gamma$ using a method reported in \cite{jiang2018}. This method uses the fact that MnBi$_2$Te$_4$'s magnetic order can be controlled by the out-of-plane magnetic field $B_{\perp}$. With increasing $B_{\perp}$, the magnetic order changes from the layered antiferromagnetic state to a spin-flop state. In the spin-flop state at $B_{\perp}=6$ T, we measured both the Kerr rotation and the $M_{z}$, from which we determined the value of $\gamma$ (Extended Data Fig.~\ref{Conversion}). Using this $\gamma$, we converted the $\alpha$ of the antiferromagnetic state to the unit of $\frac{e^2}{2h}$. In this method, we needed to assume that the spin flop state at $B_{\perp}=6$ T and the antiferromagnetic state at $B_{\perp}=0$ T have the same $\gamma$. This is an approximation. 
\vspace{3mm}
\\
\textbf{4. How we measured the DAQ}\\ The DAQ manifests as a coherent oscillation of the magnetoelectric coupling, which requires us to measure $\alpha(t)$ with fs time-resolution. This was achieved by combining ultrafast pump-probe optics with 2D electronic devices. We built a dual-gated 6L MnBi$_2$Te$_4$ device (no WSe$_2$). The probe beam combined with the gate-applied $E_{z}$ measures $\alpha$, whereas the pump beam excites the magnons. By varying the delay time $t$, we can measure $\alpha(t)$ with fs time-resolution.

Experimentally, this was achieved by connecting two lock-in amplifiers in series as illustrated in Extended Data Fig.~\ref{Ultrafast_ME}\textbf{a}. An optical chopper modulated the pump laser at frequency $\omega_1=1000$ Hz. A functional generator modulated the electric field at frequency $\omega_2=0.7$ Hz \color{black} and with AC modulation amplitude $\Delta\epsilon E=0.84$ V/nm \color{black} The signal collected by the balanced photodiode detector was first fed into a lock-in at the chopper frequency $\omega_1=1000$ Hz and then into the second lock-in at the functional generator ($E_{z}$) frequency $\omega_2=0.7$ Hz. \color{black} We used the fundamental beam of our Yb:KGW laser as the pump (1030 nm and fluence is $\sim$ 160  $\mu$J/$cm^{2}$) and its second-harmonic as the probe (515 nm). SI.I.3 shows photon energy dependence, confirming that the specific choice of photon energy does not affect our conclusion. \color{black}

\vspace{3mm}
\textbf{5. $\alpha$ originating from Berry curvature real space dipole $\mathcal{D}$}\\
In a $\mathcal{PT}$-symmetric material, the magneto-electric coupling $\alpha$ manifests as an electric field $E_z$-induced magnetization. Here, we focus on the orbital contribution of $\alpha$, so $\alpha^{\textrm{orb.}}=\frac{M^{\textrm{orb.}}_z}{E_z}$. According to \cite{xiao2005,thonhauser2005,ceresoli2006}, the orbital magnetization $M^{\textrm{orb.}}_z$ { of a quasi-2D system with thickness $d$} is given by

\begin{align}
\notag
M^{\textrm{orb.}}_z
&=\frac{e}{\hbar d}{\rm Im}\sum_n\int_{\varepsilon_{n\bf k}\le \mu} \frac{dk_xdk_y}{(2\pi)^2}
\langle \partial_{k_x}u_{n\bf k}|H_{\bf k}+\varepsilon_{n\bf k}-2\mu | \partial_{k_y}u_{n\bf k}\rangle,
\end{align}

Therefore, we have~\cite{ceresoli2006,bianco2016}
\begin{align}
\frac{\partial M^z}{\partial \mu}
= \frac{e}{h d}\frac{1}{2\pi}\sum_n\int_{\varepsilon_{n\bf k}\le \mu} dk_xdk_y \Omega_{{ n}},
\end{align}
Here, we have utilized the fact that the Berry curvature $\Omega_{{ n}}=-2{\rm Im}\langle \partial_{k_x}u_{n\bf k} |\partial_{k_y}u_{n\bf k}\rangle$.
{ Since the Berry curvature vanishes inside the bulk, we can focus on the surface degrees of freedom.}

Under a uniform vertical electric field $E_z$, the chemical potentials for the top (T) and bottom (B) surfaces change by $\delta(\mu^\textrm{T}-\mu^\textrm{B})=eE_zd$, such that for the total out-of-plane orbital magnetization $M^{\textrm{orb.}}_z=M^{\textrm{T, orb.}}_z+M^{\textrm{B, orb.}}_z$ changes by
\begin{align}
\notag
\alpha_{zz}
\equiv \frac{\partial M^{\textrm{orb.}}_z}{\partial E_z} &=  \frac{\partial M^{\textrm{T, orb.}}_z}{\partial E_z} +\frac{\partial M^{\textrm{B, orb.}}_z}{\partial E_z} = \frac{\partial M^{\textrm{T, orb.}}_z}{\partial \mu^\textrm{T}} \frac{\partial \mu^\textrm{T}}{\partial E_z} +\frac{\partial M^{\textrm{B, orb.}}_z}{\partial \mu^\textrm{B}} \frac{\partial \mu^\textrm{B}}{\partial E_z} 
\end{align}
We then can rewrite $\mu^\textrm{T}$ and $\mu^\textrm{B}$ as $\frac{\mu^\textrm{T}+\mu^\textrm{B}}{2}+\frac{\mu^\textrm{T}-\mu^\textrm{B}}{2}$ and $\frac{\mu^\textrm{T}+\mu^\textrm{B}}{2}-\frac{\mu^\textrm{T}-\mu^\textrm{B}}{2}$, respectively. Then, we have:
\begin{equation}
\notag
\begin{aligned}
\alpha_{zz} &= \frac{1}{2}(\frac{\partial M^{\textrm{T, orb.}}_z}{\partial \mu^\textrm{T}}  +\frac{\partial M^{\textrm{B, orb.}}_z}{\partial \mu^\textrm{B}} )\frac{\partial (\mu^\textrm{T}+\mu^\textrm{B})}{\partial E_z} +\frac{1}{2}(\frac{\partial M^{\textrm{T, orb.}}_z}{\partial \mu^\textrm{T}}  -\frac{\partial M^{\textrm{B, orb.}}_z}{\partial \mu^\textrm{B}} )\frac{\partial (\mu^\textrm{T}-\mu^\textrm{B})}{\partial E_z} 
\end{aligned}
\end{equation}
Owing to $\mathcal{PT}$ symmetry, we have $\frac{\partial M^{\textrm{T, orb.}}_z}{\partial \mu^\textrm{T}}  +\frac{\partial M^{\textrm{B, orb.}}_z}{\partial \mu^\textrm{B}} =0$. Also, we know that $\frac{\partial (\mu^\textrm{T}-\mu^\textrm{B})}{\partial E_z} = ed$. As such, we can get
\begin{equation}
\alpha_{zz}=\frac{ed}{2}(\frac{\partial M^{\textrm{T, orb.}}_z}{\partial \mu^\textrm{T}}  -\frac{\partial M^{\textrm{B, orb.}}_z}{\partial \mu^\textrm{B}})
\end{equation}

Plugging Eq(1) into Eq. (2), we get

\begin{equation}
\alpha_{zz}=\frac{e^2}{2h}\frac{1}{2\pi}{ \sum_n}\int_{\varepsilon_{n\bf k}\le \mu} dk_xdk_y (\Omega_{{ n}}^{\textrm{T}}-\Omega_{{n}}^{\textrm{B}})=\mathcal{D}
\end{equation}

\vspace{3mm}
\textbf{6. Estimation of the dark matter Axion detection sensitivity}\\
\underline{Detection scheme 1 - photon counting:} In the presence of external magnetic field $B_\parallel$, the dark matter (DM) Axion  of mass $m_\textrm{DM}$ can be converted into a photon at the same energy. In the DAQ material, that THZ photon and the DAQ can resonantly couple to form a DAQ polariton \cite{li2010dynamical}. The resonant interaction leads to an enhanced electromagnetic signals (i.e., photon amplification) \cite{marsh2019proposal,schutte2021axion, chadha2022,svrcek2006}, which are eventually emitted off the DAQ material and detected by a single photon detector (Fig.~\ref{Fig4}\textbf{f}). Moreover, resonant frequency between the DAQ and the photon can be directly controlled by the external $B_\parallel$ field, given by $\omega_{\textrm{resonance}}=\sqrt{m_{\rm DAQ}^2+b^2}$, where $b=\frac{e^2}{2\sqrt{2\epsilon}\pi f_\Theta}B_\parallel$, $m_{\rm DAQ}$ is the DAQ mass, which is $\sim44$ GHz in our materials, $\epsilon_0$ and $\epsilon$ are the permittivity of vacuum and of the material, $f_\Theta=82$ eV for our material. \color{black} The small correction $\delta\omega_j^2$ (see Eq. (4.33) in Ref. \cite{schutte2021axion}) is negligible, as shown in SI.II.1. \color{black}

\underline{Sensitivity estimation:} The detection sensitivity $g_{a\gamma}$ based on photon counting is shown in Fig.~\ref{Fig4}\textbf{g}. $g_{a\gamma}$ is given by ~\cite{schutte2021axion}
$$g_{a\gamma}=\frac{m_{\rm DM}^{3/2}}{B_\parallel\beta \sqrt{S\eta  \rho_{\rm DM}}}\cdot \sqrt{\frac{1}{\tau}+2\sqrt{\frac{\lambda_d}{\tau}}}$$
In the expression above, $\rho_{\rm DM}\sim 0.4$ GeV/cm$^3$ is the local Axion dark matter density. $\tau$ is the measurement time at a specific magnetic field, which was on the order of minutes in order to scan one decade in axion dark matter mass if the total experimental run time is 3 years. $\eta$ is the detection efficiency of THz single photon detector (set to $95\%$). $\lambda_d$ is the dark count rate of the photon detector (set to $10^{-5}$). $S$ is the sample area, which was set to $0.16$ m$^2$. $\beta$ is the boost factor, which was estimated to be 100 in our material using the following expression, $\beta=\frac{4}{\frac{4\pi^2}{m_{\rm DM}^2 d^2}+(\frac{m_{\rm DM}^2\Gamma_m}{b^2}+\Gamma_{\rho})dn^2}$. Here $n$ is the dielectric constant of the material ($\sim$ 6.4), and $d$ is the thickness of the material (we assumed optimal thickness $0.4$ mm). $\Gamma_m$ is the magnetic impurity density, which is estimated to be $0.7\times10^{-3}$. $\Gamma_{\rho}$ is the imaginary part of the dielectric constant at the frequency of $m_{\rm DM}$ ($\sim 0.2\times10^{-3}$). SI.II details how we estimated the values of the parameters above based on our experimental and theoretical results. \color{black} We also note that in Fig.~4\textbf{g}, each vertical streak denotes a particular previous experiments (data extracted from \cite{ohare2020b}). \color{black}

Future experimental preparations that are needed to actually build a functional detector. For example, this photon counting approach requires a large sample with optimal thickness $\sim0.4$ mm. However, the DAQ in 6L MnBi$_2$Te$_4$ relies on the finite thickness hybridization. SI.II describes our proposed future works to address all issues including sample size and THz single photon detectors. For example, we propose to grow thick samples that consist of repeating superlattice between 6L MnBi$_2$Te$_4$ and spacer layer. 

\underline{Detection scheme 2 - Kerr:} We propose a different detection approach that does not need a thick sample. While this approach is less sensitive (see Extended Data Fig.~\ref{DM_Kerr}), it serves as a first step that is readily applicable to our existing 6L MnBi$_2$Te$_4$. Meanwhile, we address the sample thickness issue (SI.II.2). This Kerr method is very similar to the experiments in the main text, but replace the pump laser (visible photons) by the dark matter Axion. In the presence of $B_\parallel$ field, the dark matter Axion can be converted into a meV (sub THz) photon, and that photon and the DAQ can resonantly couple to form an axion polariton, which is essentially a coherent oscillation of $\theta(\omega)$, where $\omega=\sqrt{m_{\rm DAQ}^2+b^2}$. By applying an out-of-plane electric field $E_z$, such a coherent oscillation of $\theta(\omega)$ will lead to an oscillating magnetization $M_z(\omega)=\theta(\omega)E_z$. We propose to use MOKE with start-of-the-art sensitivity to measure this oscillating magnetization. In this way, the sample size is not an issue, because Kerr measures the magnetization which is per volume. 

\underline{Sensitivity estimation:} The sensitivity $g_{a\gamma}$ of this approach is shown in Extended Data Fig.~\ref{DM_Kerr}. $g_{a\gamma}$ can be expressed by 

$$g_{a\gamma}=\frac{2h\gamma S_{\rm Kerr}f_{\Theta} m_{\rm DM} m_{\rm DAQ} \Gamma_m }{e^2E_z B_\parallel \sqrt{2 \tau_{\rm Kerr} \rho_{\rm DM} }}$$. 


Here $E_z$ is the applied electric field, which we take the value of 2 V/nm. $\gamma$ is the conversion factor between Kerr rotation and magnetization, which is estimated as $1.4\times 10^{4}$ $\rm A\cdot m^{-1} \cdot rad^{-1}$  in the best case scenario  \cite{tao2024}. $S_{\rm Kerr}$ is the sensitivity for Kerr rotation measurement, which is estimated to be 0.7 $\rm nrad/\sqrt{\rm Hz}$ under ac modulation \cite{ma2022}. $h$ is the Planck constant and $e$ is the electron charge. $\tau_{\rm Kerr}$ is the measurement time for Kerr rotation, which is estimated to be 3 months. Meanwhile, we note that $\rho_{\rm DM}$, $\Gamma_{\rm m}$, $m_{\rm DAQ}$, $f_{\Theta}$, $m_{\rm DM}$ and $B_\parallel$ has been defined and estimated in the previous section.

\vspace{5mm}
\textbf{7. Electronic structure calculations}\\
\vspace{-2mm}
First-principles calculations were performed using the projector augmented wave (PAW) method as implemented in the VASP. To simulate thin film MnBi$_2$Te$_4$, a $9\times9\times1$ $\Gamma$-centered $k$-grid was used for the Brillouin zone integration, and the kinetic energy cutoff was set to 400 eV. The exchange-correlation was approximated within the GGA framework. The Wannier models of MnBi$_2$Te$_4$ were built using Mn $d$, Bi $p$, and Te $p$ orbitals. Onsite Coulomb potentials of $U=5$ eV for Mn $d$ were applied. We also used the tight-binding model for 6L MnBi$_2$Te$_4$ (see SI.IV for details), which was described in Ref. \cite{Liu2020}. Each layer contained two orbitals and two spins (four bands) and different layers were coupled with symmetry-allowed interlayer hybridization. The antiferromagnetic order was described by a layer-dependent Zeeman energy.

\vspace{3mm}
7.1 The $\theta$ is the trace part of the magnetoelectric coupling $\alpha$. 

\vspace{-2mm}
\begin{align}
\theta&=\pi\frac{2h}{3e^2}\sum_{i=x,y,z}\alpha_{ii}
\end{align}. 

\color{black} We note that the above expression only holds only when the Dirac surface states are gapped by magnetic order. In the presence of perfect time-reversal symmetry both in the bulk and on the surface of a sample, $\alpha$ should be strictly zero irrespective of topological nature. \color{black} We also note that in the main text we assumed that $\alpha$ is isotropic $\alpha_{xx}=\alpha_{yy}=\alpha_{zz}$, so that $\theta=\pi\frac{2h}{e^2}\alpha$. This is a good approximation for MnBi$_2$Te$_4$ but not necessarily true for other materials. $\alpha_{ii}$ was directly computed by the electronic structure by the following expressions,

\begin{align}
\alpha_{xx} &= {\frac{e^2 }{V}\sum_{m,n}\int\frac{f_{n}-f_{m}}{\varepsilon_{m}-\varepsilon_{n}}}\ {\rm Re}[r_{{nm}}^x\braket{m(\textbf{k})|-\frac{1}{2}\left(\hat{v}^y\hat{r}^z+\hat{r}^z\hat{v}^y\right)+{m_e^{-1}}\hat{S}_x|n(\textbf{k})}], \\
\alpha_{yy} &= {\frac{e^2 }{V}\sum_{m,n}\int\frac{f_{n}-f_{m}}{\varepsilon_{m}-\varepsilon_{n}}}\ {\rm Re}[r_{{ nm}}^y\braket{m(\textbf{k})|\frac{1}{2}\left(\hat{v}^x\hat{r}^z+\hat{r}^z\hat{v}^x\right)+{ m_e^{-1}}\hat{S}_y|n(\textbf{k})}],\\
\alpha_{zz} &= {\frac{e^2 }{V}\sum_{m,n}\int\frac{f_{n}-f_{m}}{\varepsilon_{m}-\varepsilon_{n}}\ {\rm Re}[\frac{1}{2}\sum_{p;\varepsilon_p\neq\varepsilon_m}(r_{nm}^zr_{mp}^xv_{pn}^y-r_{np}^zr_{pm}^xv_{mn}^y
+r^z_{nm}r^x_{mn}v^y_{mm}
-(x\leftrightarrow y))}
+r_{mn}^z{m_e^{-1}}(\hat{S}_z)_{mn}],
\end{align}
In the equations above, $m,n$ are the band indices, $\int=\int dk_xdk_y/(2\pi)^2$, $f_n$ is the Fermi Dirac function with energy $\epsilon_n$ of band $n$, $\hat{r}$, $\hat{v}$ and $\hat{S}$ are the position, velocity and spin operators, and $m_e$ is the electron mass. As explained in the main text, $\alpha$ has two microscopic contributions, spin and orbital. The last term of each equation, related to the spin operator $\hat{S}$, accounts for the spin contribution; The other terms account for the orbital contribution. We note that the expressions above are valid when the Fermi level is inside the band gap, which is true for all experiments except the carrier density dependence in Figs.~\ref{Fig4}\textbf{b,c}. When the Fermi level cuts into the band, $\alpha_{zz}$ has an additional Fermi surface contribution term $\frac{e^2 }{V}\sum_{n}\int\frac{\partial f_{n}}{\partial\epsilon_n}\frac{1}{2}(v^y_ng^{zx}_n-v^x_ng^{zy}_n)$, where $g^{ij}_n=\sum_mr^i_{nm}r^j_{mn}$ is the quantum metric of band $n$.

\vspace{3mm}
7.2 The Berry curvature real space dipole $\mathcal{D}$ is 
\begin{align}\mathcal{D}&=\frac{e^2}{4\pi h}{\sum_n}(\int_{\varepsilon_{n\bf k}\le \mu}\Omega_{n}^{\textrm{T}}-\int_{\varepsilon_{n\bf k}\le \mu}\Omega_{n}^{\textrm{B}})\end{align}

$\Omega_{n}^{\textrm{T(B)}}$ is the layer-resolved Berry curvature of the $n^{\textrm{th}}$ band as defined in \cite{Varnava2018} (T/B is the top/bottom surface), which is given by

\begin{align}\Omega_n^{\textrm{T(B)}}=-2\textrm{Im}\sum_{m,n'}\frac{\left \langle n(\textbf{\textit{k}})|v_x|m(\textbf{\textit{k}}) \right \rangle \left \langle m(\textbf{\textit{k}})| v_y|n'(\textbf{\textit{k}}) \right \rangle}{(\varepsilon_{m}-\varepsilon_{n})^2} \rho_{nn'}^{{\textrm{T(B)}}}(\textbf{\textit{k}})\end{align}

where $\rho_{nn'}^{{\textrm{T(B)}}}(\textbf{\textit{k}})=\sum_{j\in {\textrm{T(B)}}}{\left \langle n(\textbf{\textit{k}})|j\rangle\langle j|n'(\textbf{\textit{k}}) \right \rangle}$ describes the projection on to the top or bottom layer.


\vspace{0.5cm}

\textbf{Author contributions:}  SYX conceived the experiments and supervised the project. JXQ fabricated the devices, performed the measurements and analyzed data with help from AG, CT, HCL, YFL, DB, TD, TH, JMB, CF, QM, RM, RJM, TQ and NN grew the bulk MnBi$_2$Te$_4$ single crystals. BG, YTY, MS, TVT, JA, IP, OL, EMB, PN, TRC, AB, HL,PPO, IM and AV made the theoretical studies including first-principles calculations and effective modeling. KW and TT grew the bulk hBN single crystals. JSE, JXQ, KCF, DJEM, and SYX made the calculation of sensitivity for dark matter Axion detection. SYX and JXQ wrote the manuscript with input from all authors. 

\textbf{Acknowledgement:} JXQ, JA, AG, HL, MS, JMB, PPO, QM, RM, RJM, IM, AV and SYX were supported through the Center for the Advancement of Topological Semimetals (CATS), an Energy Frontier Research Center (EFRC) funded by the US Department of Energy (DOE) Office of Science, through the Ames National Laboratory under contract DE-AC0207CH11358.  The work in SYX group was supported by CATS (task that CATS supported) and the Air Force Office of Scientific Research (AFOSR) grant FA9550-23-1-0040 (data analysis and manuscript writing). S.-Y.X. acknowledges the Sloan foundation and Corning Fund for Faculty Development. S.Y.X. and D.B. were supported by the National Science Foundation (NSF; Career Grant No. DMR-2143177). Bulk single crystal growth and characterization at UCLA were supported by the U. S. Department of Energy (DOE), Office of Science, Office of Basic Energy Sciences (BES) under Award Number DE-SC0021117. JSE was supported by the National Science Foundation under cooperative agreement 202027 and by the by Japan Science and Technology Agency (JST) as part of Adopting Sustainable Partnerships for Innovative Research Ecosystem (ASPIRE), Grant Number JPMJAP2318.  KW and TT acknowledge support from the JSPS KAKENHI (Grant Numbers 21H05233 and 23H02052) , the CREST (JPMJCR24A5), JST and World Premier International Research Center Initiative (WPI), MEXT, Japan.  CF was supported by Deutsche Forschungsgemeinschaft (DFG, German Research Foundation) through SFB 1143 (project ID. 24731007) and the W\"urzburg-Dresden Cluster of Excellence on Complexity and Topology in Quantum Matter-ct.qmat (EXC 2147, project No. 390858490). HL acknowledges the support by Academia Sinica in Taiwan under grant number AS-iMATE-113-15. TRC was supported by National Science and Technology Council (NSTC) in Taiwan (Program No. MOST111-2628-M-006-003-MY3 and NSTC113-2124-M-006-009-MY3), National Cheng Kung University (NCKU), Taiwan, and National Center for Theoretical Sciences, Taiwan. This research was supported, in part, by the Higher Education Sprout Project, Ministry of Education to the Headquarters of University Advancement at NCKU. TRC thanks the National Center for High-performance Computing (NCHC) of National Applied Research Laboratories (NARLabs) in Taiwan for providing computational and storage resources. The work at Northeastern University (AB and BG) was supported by the National Science Foundation through the Expand-QISE award NSF-OMA-2329067 and benefited from the resources of Northeastern University’s Advanced Scientific Computation Center, the Discovery Cluster, the Massachusetts Technology Collaborative award MTC-22032, and the Quantum Materials and Sensing Institute. For the computational work at S.N. Bose National Center for Basic Sciences (SNBNCBS), BG acknowledge National Supercomputing Mission (NSM) for providing computing resources of ‘PARAM RUDRA’ at SNBNCBS, Saltlake, Kolkata-700106, India, which is implemented by C-DAC and supported by the Ministry of Electronics and Information Technology (MeitY) and Department of Science and Technology (DST), Government of India. JMB and RM's experimental activities were performed at the National High Magnetic Field Laboratory, which is supported by National Science Foundation Cooperative Agreement No. DMR-2128556 and the State of Florida. DJEM is supported by an Ernest Rutherford Fellowship (ST/T004037/1) and by a Science and Technologies Facilities Council grant (ST/X000753/1). Q.M. also acknowledges support from the National Science Foundation (NSF) CAREER award DMR-2143426 and Sloan Fellowship. O.L., I.P., E.M.B, and P.N. were supported by the Quantum Science Center (QSC), a National Quantum Information Science Research Center of the U.S. Department of Energy (DOE).  O.L., I.P., E.M.B, and P.N. also gratefully acknowledge support from the Gordon and Betty Moore Foundation grants No. 8048 \& No. 12976, and from the John Simon Guggenheim Memorial Foundation (Guggenheim Fellowship).\color{black}

\textbf{Competing financial interests:} The authors declare no competing financial interests.

\clearpage

\clearpage


\clearpage

\setcounter{figure}{0}
\renewcommand{\figurename}{\textbf{Extended Data Fig.}}

\clearpage

\setcounter{figure}{0}
\renewcommand{\figurename}{\textbf{Extended Data Fig.}}

\clearpage


\begin{figure*}[htb]
\includegraphics[width=16cm]{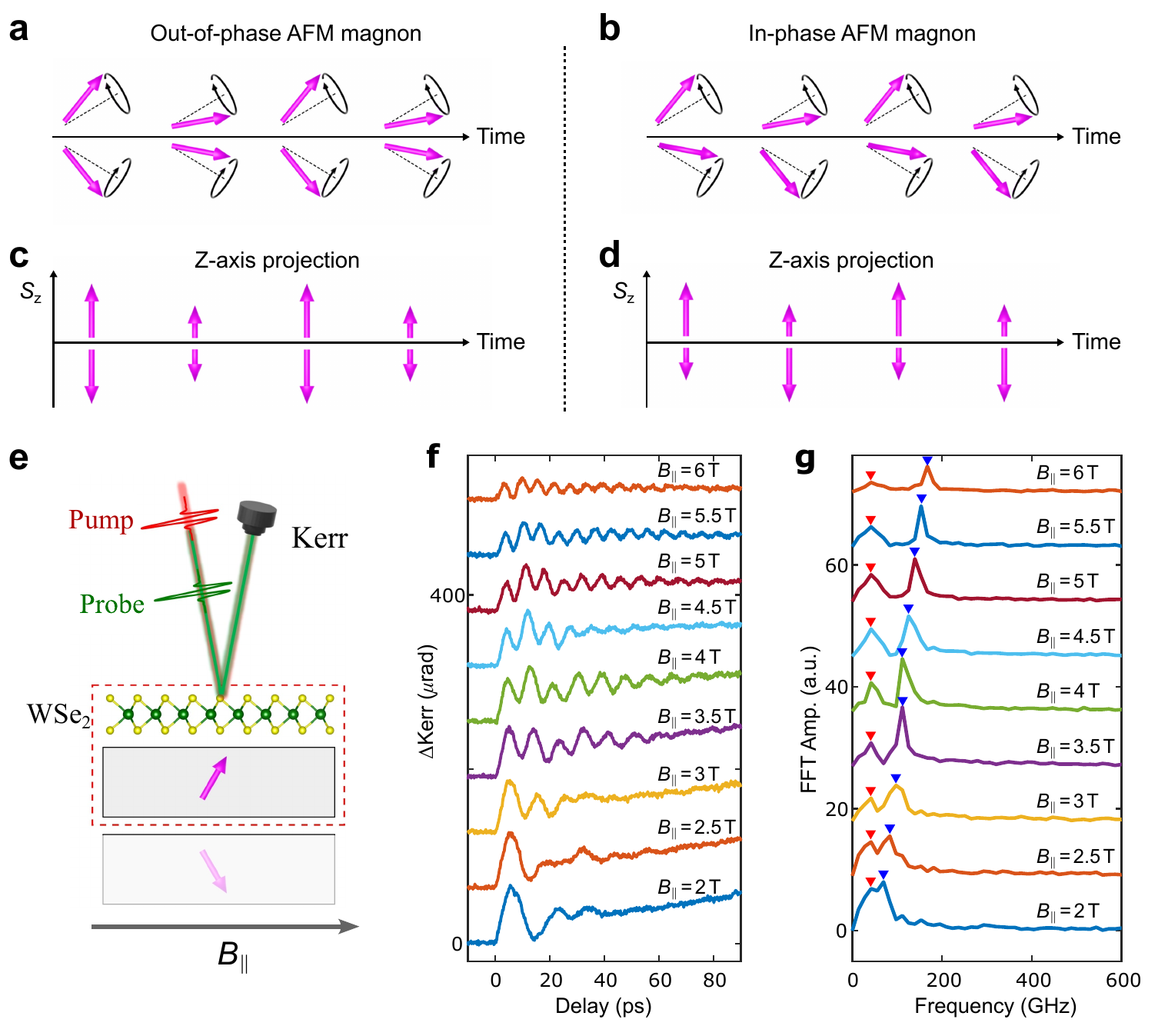}
\vspace{-0.5mm}
\caption{{\bf Characterization of the magnon frequency in 6-layer MnBi$_2$Te$_4$.} \textbf{a,} Time evolution of out-of-phase magnon. \textbf{c,} $\hat{z}$ projection of the out-of-phase magnon, which resembles the antiferromagnetic amplitude mode. \textbf{b,d,} Same as panels (\textbf{a,c}) but for in-phase magnon, which features an oscillation of net magnetization along $\hat{z}$. \textbf{e,} We followed the method established in \cite{zhang2020gate} (See detailed discussion in Methods.2). A monolayer WSe$_2$ was stacked on top of MnBi$_2$Te$_4$, which breaks the layer degeneracy, allowing us to selectively probe the top layer information \cite{zhang2020gate}. Pump-probe Kerr rotation under normal incidence was performed on this heterostructure. The pump laser launches the magnons. The probe Kerr rotation measures the out-of-plane magnetization $M_z$ of the top layer preferentially because of the WSe$_2$. \textbf{f,g,} Pump-induced $\Delta$Kerr data and FFT at different $B_{\parallel}$.}
\label{Magnon}
\end{figure*}

\clearpage
\begin{figure*}[h]
\includegraphics[width=17cm]{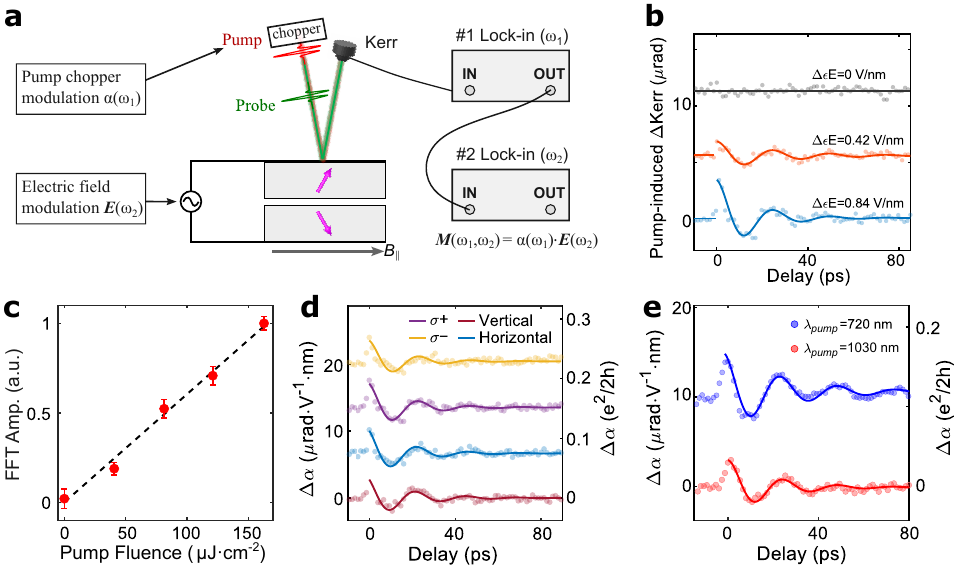}
\vspace{-0.5mm}
\caption{{\bf Experimental setup for measuring the DAQ}. \textbf{a,} The DAQ manifests as a coherent oscillation of the magnetoelectric coupling, which requires us to measure $\alpha(t)$ with fs time-resolution. This was achieved by combining ultrafast pump-probe optics with 2D electronic devices. We built a dual-gated 6L MnBi$_2$Te$_4$ device (no WSe$_2$). The probe beam combined with the gate-applied $E_{z}$ measures $\alpha$, whereas the pump beam excites the magnons. By varying the delay time $t$, we can measure $\alpha(t)$ with fs time-resolution. Experimentally, this was achieved by connecting two lock-in amplifiers. An optical chopper modulated the pump laser at frequency $\omega_1=1000$ Hz. A functional generator modulated the gate $E_{z}$ at frequency $\omega_2=0.7$ Hz. The signal collected by the balanced photodiode detector was first fed into a lock-in at the chopper frequency $\omega_1=1000$ Hz and then into the second lock-in at the $E_{z}$ frequency $\omega_2=0.7$ Hz. The wavelength of the pump beam was set to 1030 nm, and the pump fluence is $\sim$ 160  $\mu$J/$cm^{2}$. \textbf{b}, Pump-induced Kerr rotation at different AC $E$ field modulation amplitudes. \textbf{c,} Pump fluence dependence of the oscillation amplitude of $\Delta\alpha$. \textbf{d,} Coherent oscillation of $\Delta\alpha$ as a function of pump light polarization. The indifference of pump light polarization suggests excitation mechanism is laser heating induced coherent oscillation of spins \cite{kirilyuk2010ultrafast}.  \textbf{e,} Coherent oscillation of $\Delta\alpha$ as a function of pump wavelength. }
\label{Ultrafast_ME}
\end{figure*}

\clearpage
\begin{figure*}[h]
\includegraphics[width=14cm]{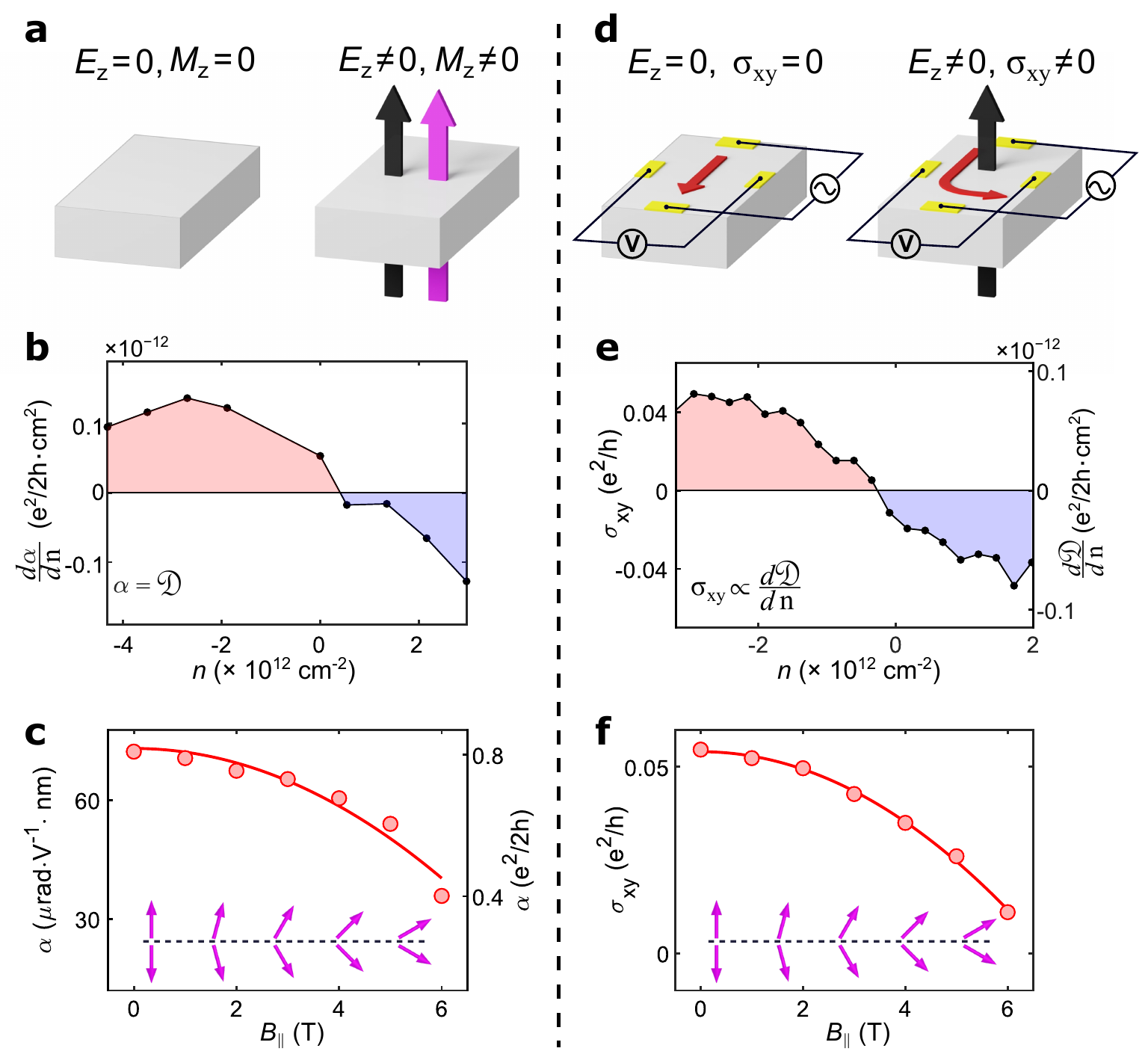}
\vspace{-0.5mm}
\caption{{\bf The magnetoelectric coupling and the layer Hall effect in MnBi$_2$Te$_4$. } \textbf{a,} Schematics for the magnetoelectric coupling ($\alpha = \frac{M_z}{E_z}$). \textbf{b,} Maintext Fig.~\ref{Fig4}\textbf{b} shows the measured $\alpha$ as a function of the charge density $n$. By taking a derivative of this data, we get $\frac{d\alpha}{d n}$ as a function of $n$. \textbf{c,} Measured $\alpha$ as a function of in-plane magnetic field $B_{\parallel}$. \textbf{d,} Schematics for the layer Hall effect. An out-of-plane electric field $E_z$ induces an anomalous Hall effect (finite $\sigma_{xy}$) in 6L MnBi$_2$Te$_4$.  \textbf{e,} $E_z$ induced $\sigma_{xy}$ as a function of carrier density $n$. This $E_z$ induced $\sigma_{xy}$ directly measures $\frac{d \mathcal{D}}{d n}$ ($\mathcal{D}$ is the Berry curvature real space dipole), which is marked in the right axis. \textbf{f,}   $E_z$ induced $\sigma_{xy}$ as a function of $B_{\parallel}$.}
\label{LHE}
\end{figure*}

\clearpage
\begin{figure*}[h]
\includegraphics[width=14cm]{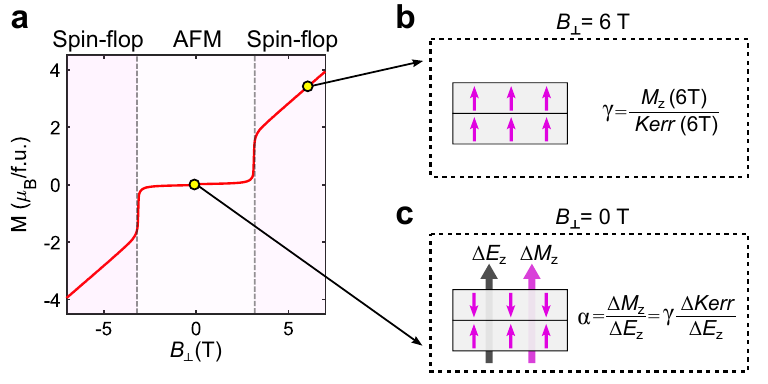}
\vspace{-0.5mm}
\caption{{\bf Determining the conversion factor $\gamma$}. \textbf{a,} Out-of-plane magnetization as function of out-of-plane magnetic field $B_{\perp}$ for bulk MnBi$_2$Te$_4$ measured by SQUID. With increasing $B_{\perp}$, the magnetic order changes from the layered antiferromagnetic state to a spin-flop state. \textbf{b,} In the spin-flop state at $B_{\perp}=6$ T, we measured both the Kerr rotation and the $M_{z}$, from which we determined the value of $\gamma$. \textbf{c,} In the antiferromagnetic ground state at $B_{\perp}=0$ T, 6L MnBi$_2$Te$_4$ features an electric field induced magnetization. The magneto-electric coupling $\alpha$ is given by $\alpha=\gamma \frac{d Kerr}{d E_z}$. Therefore, by using the $\gamma$ determined in the spin-flop state, we converted $\alpha$ of the antiferromagnetic state to the unit of $\frac{e^2}{2h}$. In this method, we needed to assume that the spin flop state at $B_{\perp}=6$ T and the antiferromagnetic state at $B_{\perp}=0$ T have the same $\gamma$. This is an approximation.}
\label{Conversion}
\end{figure*}
\clearpage
\begin{figure*}[h]
\includegraphics[width=16cm]{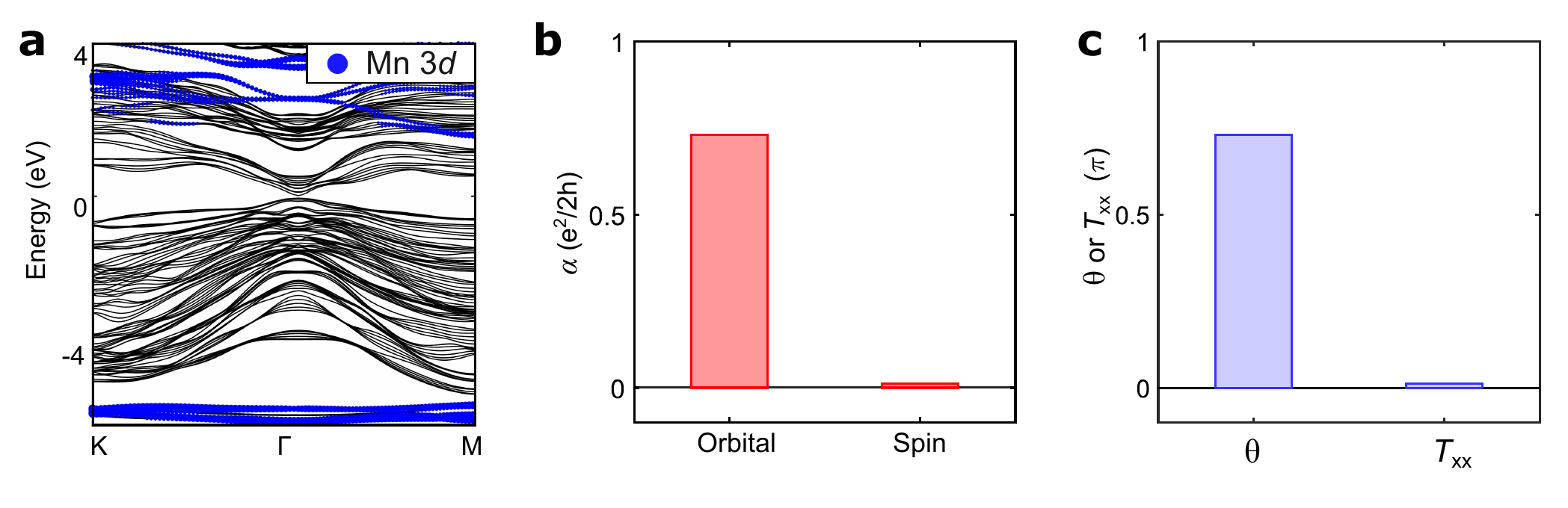}
\vspace{-0.5mm}
\caption{{\bf Microscopic mechanism for the magnetoelectric coupling in 6L MnBi$_2$Te$_4$ } \textbf{a,} First-principles band structures of  6L MnBi$_2$Te$_4$  with the Mn $3d$ orbitals highlighted. \textbf{b,} Calculated $\alpha_{zz}$ from the spin and orbital contributions. The total $\alpha_{zz}$ is the sum of the two contributions. \textbf{c,} Comparison of $\theta$ and $T_{xx}$, which are the trace part and traceless part of $\alpha_{ii}$, respectively (normalized by $e^2/2h$).}
\label{Theory1}
\end{figure*}

\clearpage
\begin{figure*}[h]
\includegraphics[width=14cm]{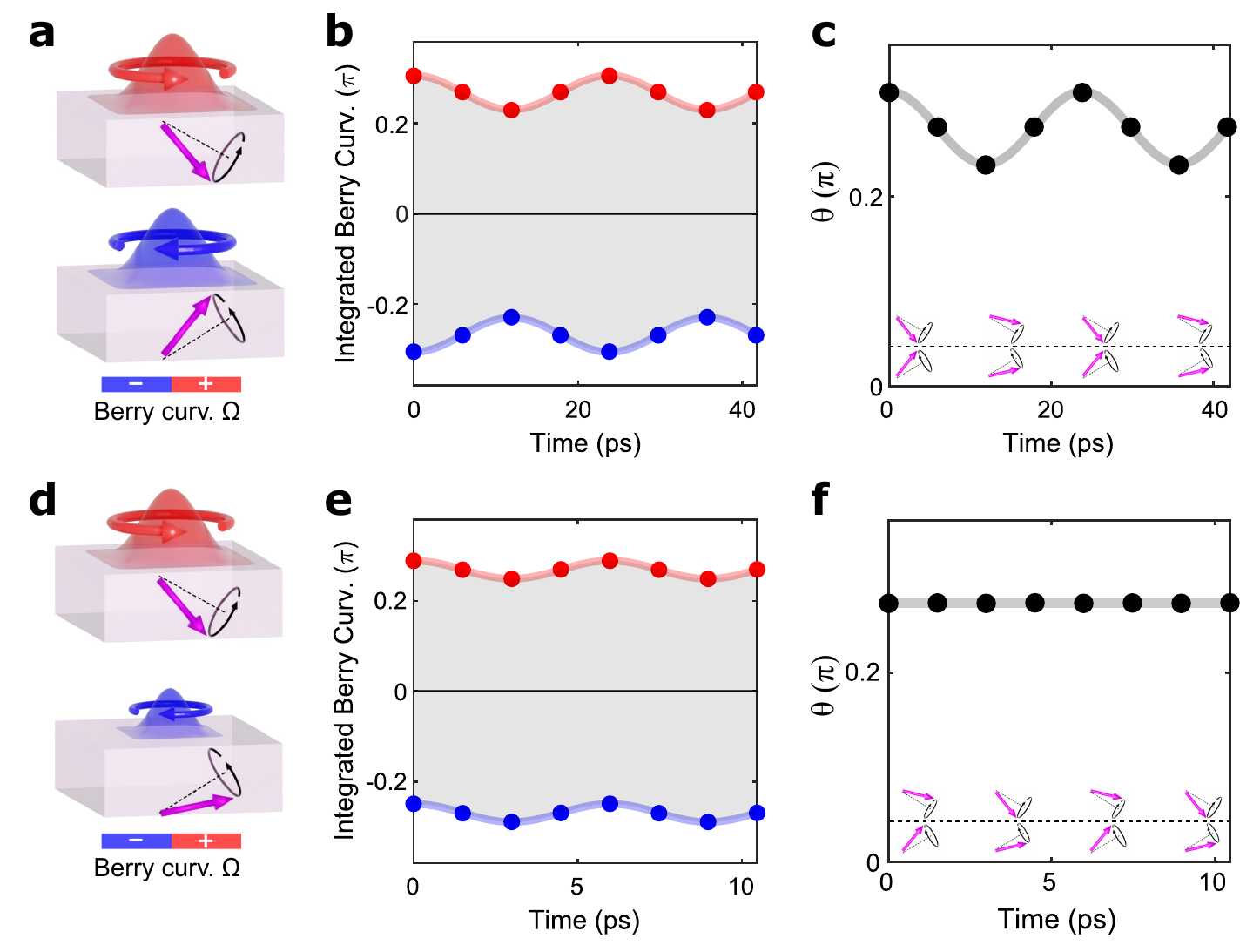}
\vspace{-0.5mm}
\caption{{\bf Ultrafast Berry curvature oscillation by antiferromagnetic magnons.} \textbf{a,d,} In the even-layer MnBi$_2$Te$_4$, the antiferromagnetic order couples to the Dirac surface states, generating large Berry curvature on the top and bottom surfaces. We study how the Berry curvature responds upon exciting the out-of-phase or the in-phase magnon. \textbf{b,e,} Calculated Berry curvature sum of the top and bottom surfaces at different spin angles during the magnon oscillation under the frozen magnon approximation. The grey area (i.e., the difference of Berry curvature from top and bottom surfaces) is the Berry curvature real space dipole $\mathcal{D}$ ($\mathcal{D}=\alpha$). \textbf{c,f,} Calculated $\theta$ at different spin angles during the magnon oscillation.}
\label{Theory2}
\end{figure*}

\clearpage
\begin{figure*}[h]
\includegraphics[width=12cm]{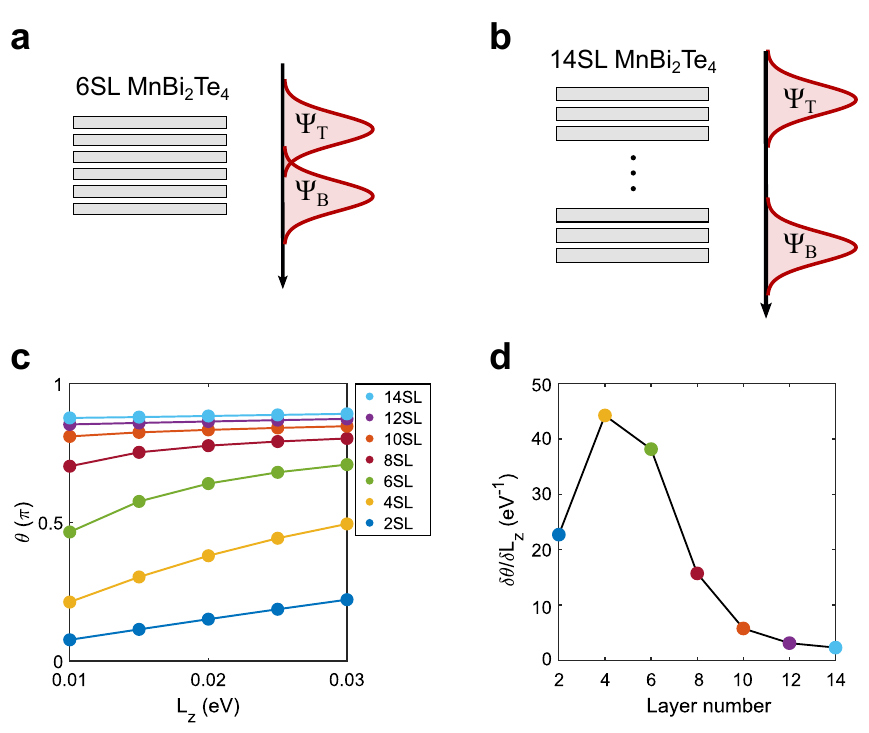}
\vspace{-0.5mm}
\caption{{\bf Calculated DAQ strength of 2D even-layer MnBi$_2$Te$_4$.} The strength of DAQ is measured by the change of $\theta$ per change of the antiferromagnetic order parameter $L$, $\frac{\delta\theta}{\delta L}$ \cite{li2010dynamical}. In 2D even-layer MnBi$_2$Te$_4$, the top and bottom surface state wavefunction can overlap and hybridize. This hybridization gap competes with with magnetism induced Zeeman gap, which leads to a large but non-quantized $\theta$. We theoretically study the $\frac{\delta\theta}{\delta L}$ by calculating $\theta$ as a function of $L$ for different thicknesses. \textbf{a-b}, Wavefunction hybridization for 6SL and 14SL. \textbf{c}, Calculated $\theta$ vs. AFM order $L_z$ for different thicknesses. \textbf{d}, $\delta\theta/\delta L_z$ as a function of thickness. }
\label{Theory3}
\end{figure*}

\clearpage
\begin{figure*}[h]
\includegraphics[width=16cm]{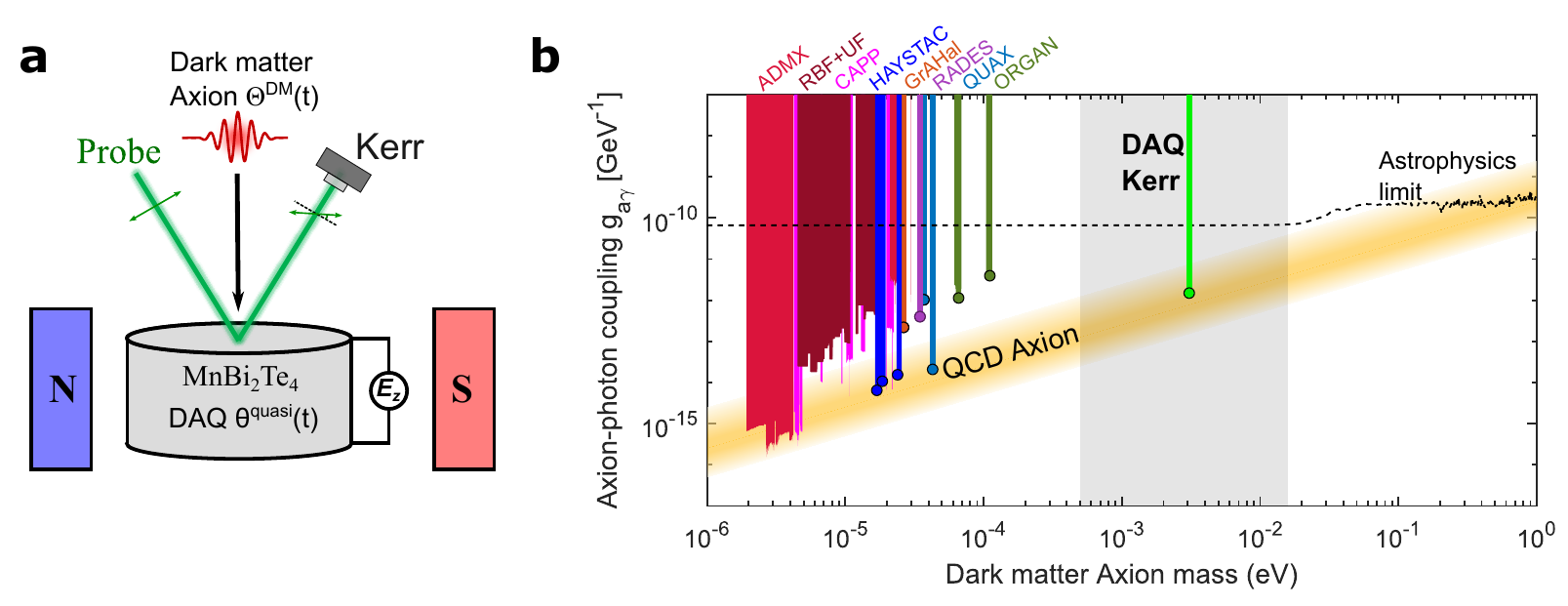}
\vspace{-0.5mm}
\caption{{\bf Kerr effect scheme of dark matter Axion detection using DAQ.} \textbf{a,} The Kerr effect scheme: A dark matter Axion resonantly excite an Axion polariton inside the DAQ material under an external $B_\parallel$ field ($B_\parallel=5$ T). The axion polariton is essentially a coherent oscillation of $\theta(\omega)$, where $\omega=\sqrt{m_{\rm DAQ}^2+b^2}$. By applying an out-of-plane electric field $E_z$, such a coherent oscillation of $\theta(\omega)$ will lead to an oscillating magnetization $M_z(\omega)=\theta(\omega)E_z$. We propose to use MOKE with to measure this oscillating magnetization.  \textbf{b,} Dark matter detection sensitivity ($g_{a\gamma}$) as a function of the Axion mass using the Kerr scheme (see details in Methods.6).}
\label{DM_Kerr}
\end{figure*}


\pagebreak
\newpage

	\vspace{3cm}
	\Large
	\begin{center}
		Supplemental Information for \\ Observation of the Axion quasiparticle in 2D MnBi$_2$Te$_4$
	\end{center}
	\vspace{0.45cm}
	\textbf{
		\begin{center}
			{This file includes:}
		\end{center}
	}
	\vspace{1cm}
	\large
	\begin{tabular}{l l}
		\underline{I.} & Additional data\\
		\enspace I.1. & Magnon data of 5SL MnBi$_2$Te$_4$\\
		\enspace I.2. & Additional magnon data of 6SL MnBi$_2$Te$_4$\\
		\enspace I.3. & Additional static Kerr data\\
		\enspace I.4. & Additional DAQ data of 6SL MnBi$_2$Te$_4$\\
		\enspace I.5. & Phonon data of 6SL MnBi$_2$Te$_4$\\
		\enspace I.6. & Other addition data and sanity check\\
		\underline{II.} & AFM Kerr and DAQ probed by DC $E$ field \\
		\underline{III.} & Dark matter Axion detection \\
		\enspace  III.1. & Detection sensitivity calculation \\
		\enspace  III.2. & Future steps to build a functioning Axion detector \\
		\enspace  III.3 & Comparison between the meV DAQ detector and other proposed detectors\\

		\underline{IV.} & Theoretical derivation for the layer Hall effect\\
		\underline{V.} & Tight binding model for MnBi$_2$Te$_4$ \\
		\underline{VI.} & Additional discussion \\
	\end{tabular}

	\date{\today}

	\normalsize

	
	\clearpage
	\section*{I. Additional data}

	\subsection*{I.1. Magnon data for 5SL MnBi$_2$Te$_4$}
\begin{figure*}[!htb]
		\centering
		\includegraphics[width=11cm]{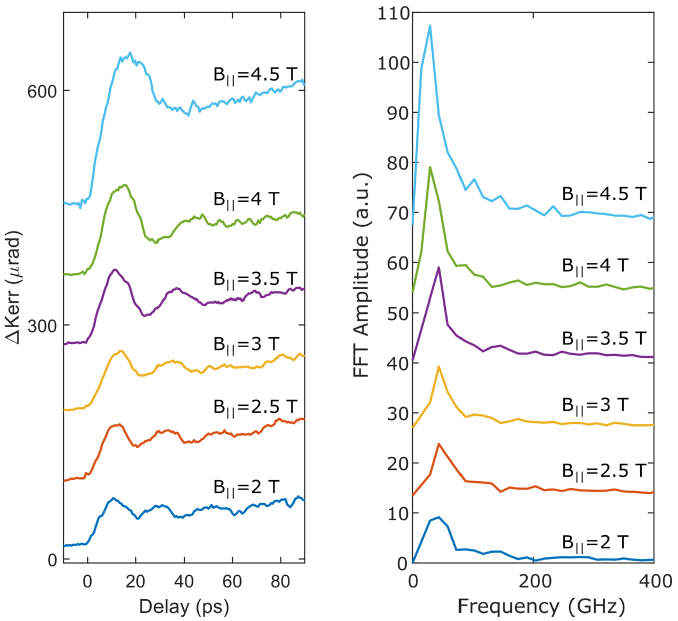}
		\caption{\textbf{a}, TRMOKE data and FFT analysis of 5SL MnBi$_2$Te$_4$ ($\lambda_{\rm pump}$= 1030 nm, $\lambda_{\rm probe}$= 515 nm)}
		\label{SI_5SL_TRMOKE_overview_v3}
	\end{figure*}

\begin{figure*}[!htb]
		\centering
		\includegraphics[width=17cm]{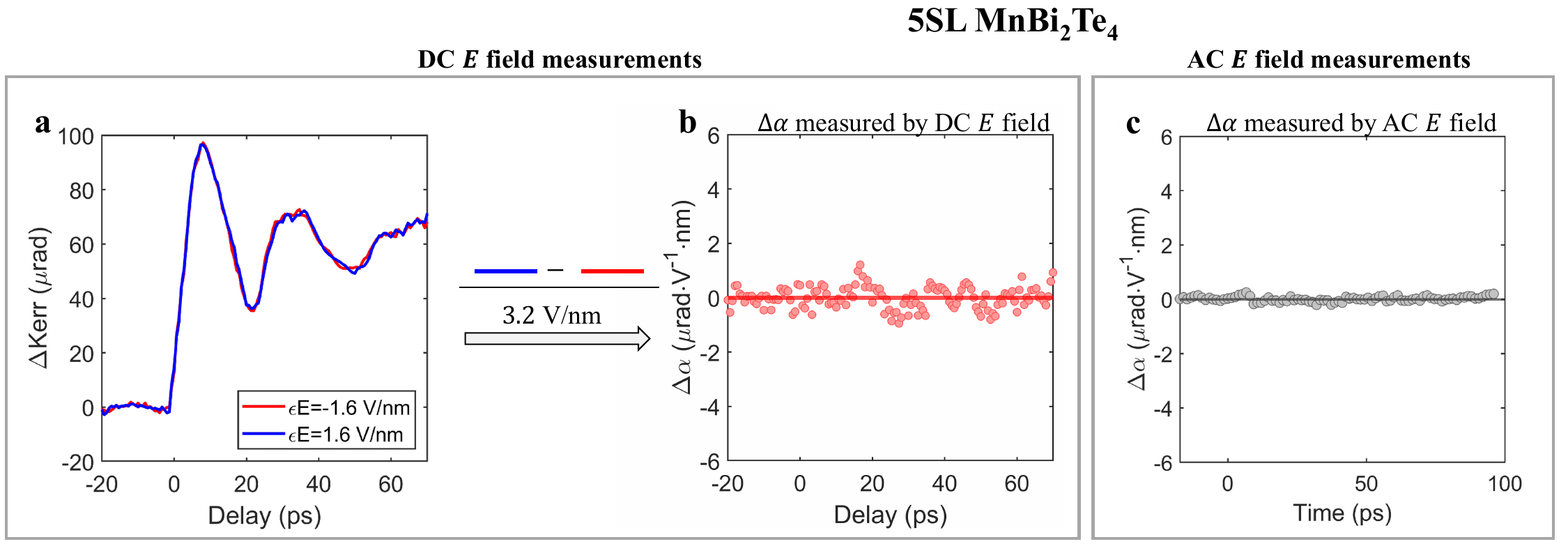}
		\caption{\textbf{$E$-field dependence of TRMOKE data of 5SL MnBi$_2$Te$_4$.} \textbf{a}, TRMOKE data at different DC $E$ field values ($B_{\parallel}$=3.5 T). \textbf{b}, Time-resolved $\Delta\alpha$ is obtained by the difference of the TRMOKE data at $\pm$1.6 V/nm divided by 3.2 V/nm. \textbf{c}, Time-resolved $\Delta\alpha$ measured by AC $E$ field dependent measurements. }
		\label{SI_5SL_AC_DC_v3}
	\end{figure*}

	In the main text, we focused on 6SL MnBi$_2$Te$_4$. For completeness, here we show the magnon data (and the effect of $E$-field) in 5SL MnBi$_2$Te$_4$.  As shown in the TRMOKE data of 5SL MnBi$_2$Te$_4$ (Fig.~\ref{SI_5SL_TRMOKE_overview_v3}), the out-of-phase magnon is the most prominent mode, consistent with previous results\cite{bartram2023real}. We then study the time-dependent magnetoelectric coupling $\alpha$ of 5SL MnBi$_2$Te$_4$. As shown in Fig.~\ref{SI_5SL_AC_DC_v3}, both the DC and AC $E$ field measurements do not show observable $\alpha$ oscillation. 
	
	\pagebreak
	
	\subsection*{I.2. Additional magnon data of 6SL MnBi$_2$Te$_4$}
	
	\subsubsection*{I.2.1 Theoretical calculation of magnon in 6SL MnBi$_2$Te$_4$}
\begin{figure*}[!htb]
		\centering
		\includegraphics[width=17cm]{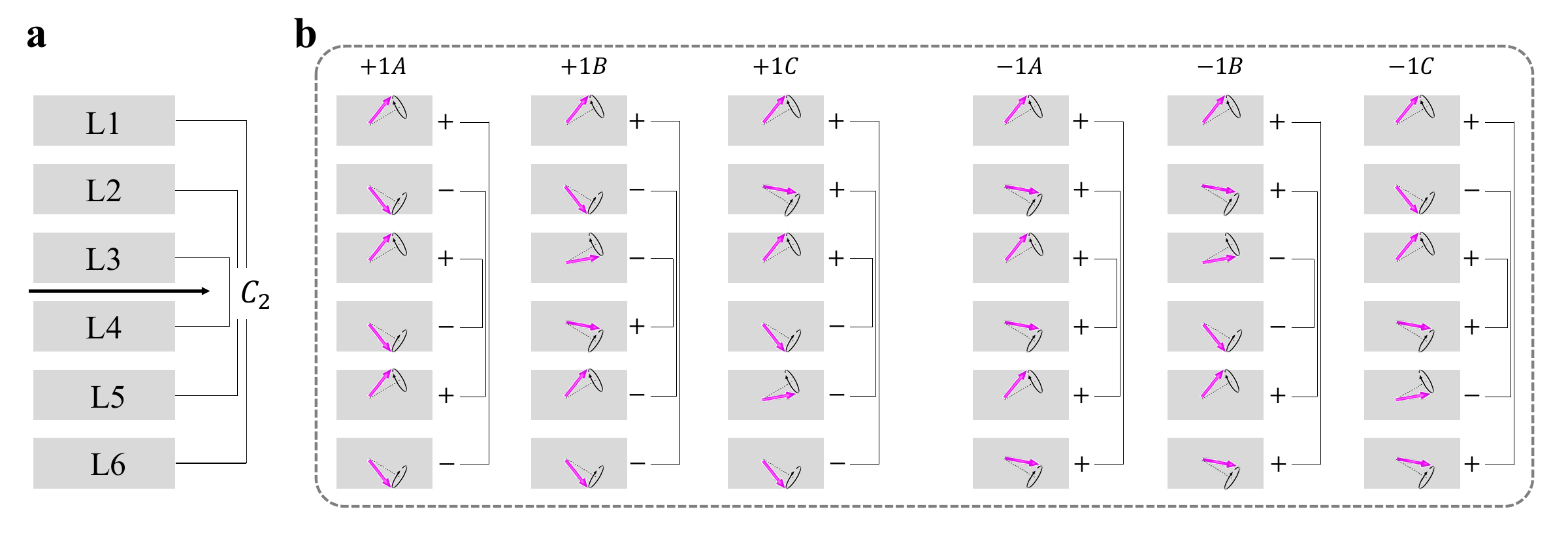}
		\vspace{-1cm}
		\caption{\textbf{a,} C$_2$ pairs in 6SL MnBi$_2$Te$_4$. \textbf{b}, Schematic illustration of the six magnon modes. The initial phase of the spin is denoted by the pink vector as well as the $+,-$ on the right-hand side of the layer. In this figure, the equilibrium angle is assumed to be the same for all layers for simplicity.}
		\label{SI_Magnons_v3}
	\end{figure*}

\begin{figure*}[!htb]
		\centering
		\includegraphics[width=14cm]{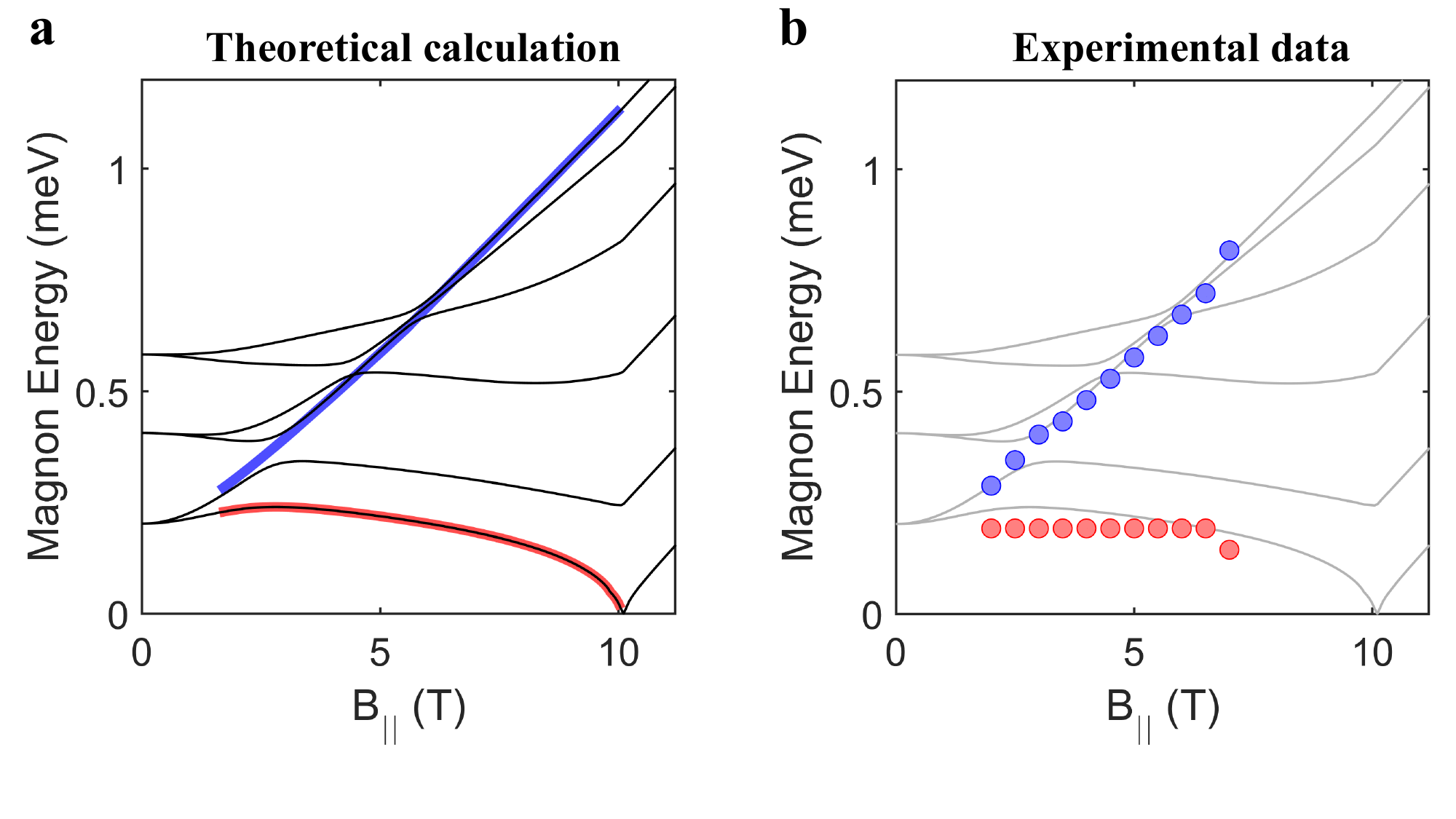}
		\vspace{-1cm}
		\caption{\textbf{a}, Calculated magnon energy as a function of $B_{\parallel}$ for 6SL MnBi$_2$Te$_4$. The red and blue highlights the $+1A$ and $-1A$ modes. \textbf{b}, Red and Blue circles denote our experimentally measured energy of the out-of-phase and in-phase magnons. The black lines are the theoretical calculations same as panel (\textbf{a}). $J=0.2954$ meV and $\kappa= 0.03130$ meV.}
		\label{SI_Magnon_calculation_data_v2}
	\end{figure*}

\begin{figure*}[h]
		\centering
		\includegraphics[width=17cm]{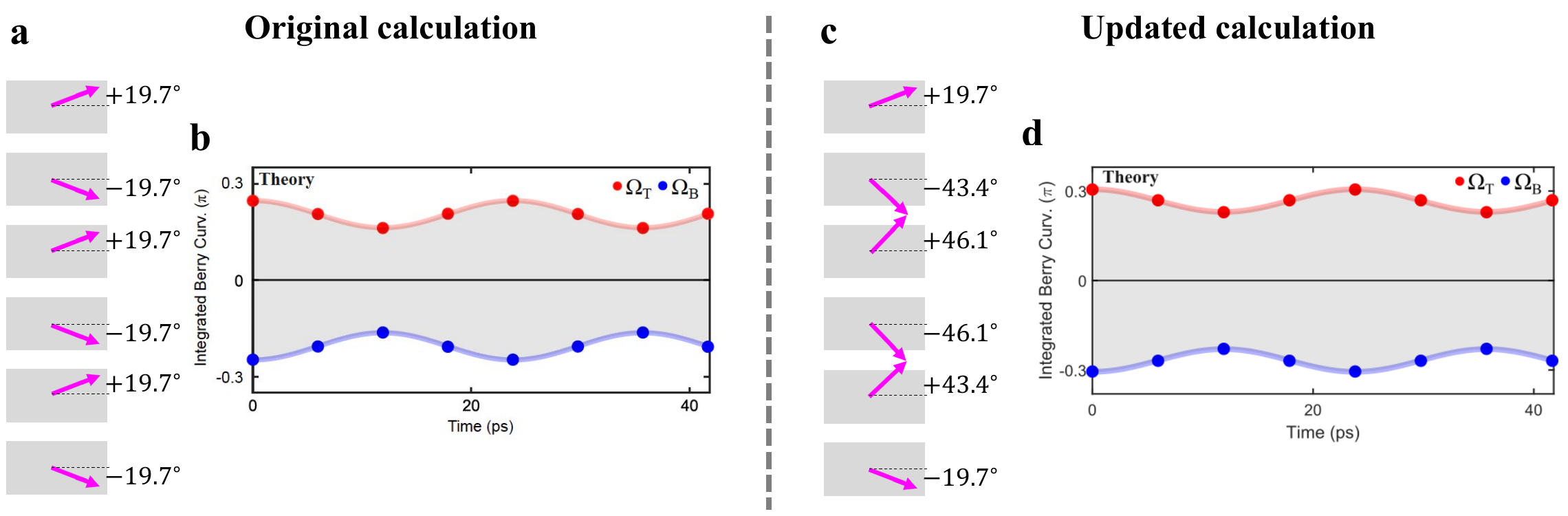}
		\vspace{-0cm}
		\caption{\textbf{a}, Calculation of Berry curvature oscillation of 6SL MnBi$_2$Te$_4$. \textbf{b,} The spin equilibrium angles are assumed to be uniform ($\pm 19.7^{\circ}$) across all layers. The spin precession angle is $5^{\circ}$. \textbf{c,} The spin equilibrium angles are calculated based on the Heisenberg model. We set the angles of the top and bottom layers to be the same ($\pm 19.7^{\circ}$). The angles of inner layers are larger because the inner layers feel stronger exchange coupling. \textbf{d,}  Corresponding Berry curvature oscillation.}
		\label{SI_Spin_angle_BC}
	\end{figure*}

	Using the standard Heisenberg model, we theoretically calculated the magnon modes in 6SL MnBi$_2$Te$_4$. Theoretically, we obtained six magnon modes, which can be classified based on the $C_2$ symmetry (Fig.~\ref{SI_Magnons_v3}). Specifically, we found that two out of the six magnon modes ($+1A$/$-1A$) resemble the in-phase and out-of-phase modes in bilayer or in the bulk limit. By further comparing our calculation with our experimental data (Fig.~\ref{SI_Magnon_calculation_data_v2}), we found that these two modes are likely to be the two experimentally observed magnon modes. Figure~\ref{SI_Spin_angle_BC} studies how the equilibrium spin angles affect the Berry curvature oscillation. In Figs.~\ref{SI_Spin_angle_BC}\textbf{a,b}, the spin equilibrium angle was set to the the same ($\pm 19.7^{\circ}$) across all layers. In Figs.~\ref{SI_Spin_angle_BC}\textbf{c,d} should be different for different layers. The spin equilibrium angles are calculated based on the Heisenberg model. We set the angles of the top and bottom layers to be the same ($\pm 19.7^{\circ}$). In particular, the angles of inner layers are larger because the inner layers feel stronger exchange coupling. The corresponding Berry curvature oscillation are similar to the previous, except for a small global shift (Fig.~\ref{SI_Spin_angle_BC}).
	
	\vspace{1cm}
	\subsubsection*{I.2.2 Additional magnon data of WSe$_2$/6SL MnBi$_2$Te$_4$}
\begin{figure*}[!htb]
		\hspace{-1cm}
		\includegraphics[width=18cm]{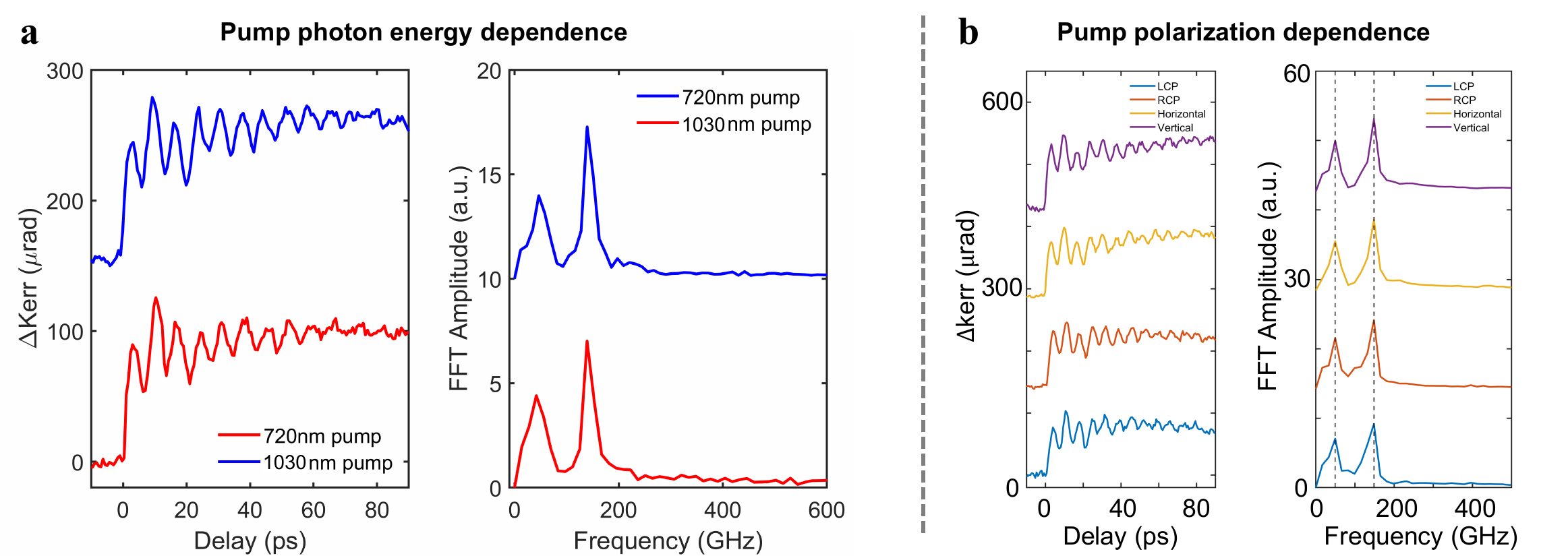}
		\caption{\textbf{a}, TRMOKE of WSe$_2$/6SL MnBi$_2$Te$_4$ using pump photon energies of 720 nm and 1000 nm. \textbf{b}, TRMOKE of WSe$_2$/6SL MnBi$_2$Te$_4$ using different pump polarizations.}
		\label{SI_Wavelength_WSe2MBT_v2}
	\end{figure*}

	Fig.~\ref{SI_Wavelength_WSe2MBT_v2} presents the pump photon energy and pump polarization dependence of magnon data in WSe$_2$/6SL MnBi$_2$Te$_4$. The magnon data does not depend on the pump polarization or pump photon energy, which suggests that the excitation mechanism is laser-heating induced coherent precession\cite{van2002all}.

	\pagebreak

	\subsection*{I.3. Additional static Kerr data}
	
\begin{figure*}[!htb]
		\centering
		\includegraphics[width=17cm]{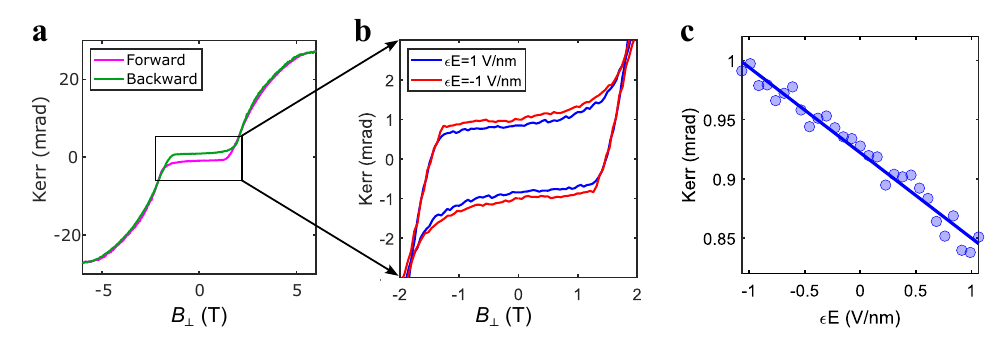}
		\caption{\textbf{a}, Static MOKE data of 6SL MnBi$_2$Te$_4$ as a function of out-of-plane $B_{\perp}$ field. \textbf{b}, Zoomed-in view of the antiferromagnetic state at two different electric field $E$ values. The amplitude of the MOKE changes by $E$ field due to the magnetoelectric coupling. \textbf{c}, MOKE as a function of DC $E$ field at $B_{\perp}=0$. The linear $E$ dependence of the MOKE amplitude shows the magnetoelectric coupling. $\lambda_{\rm{probe}}$ = 515 nm. }
		\label{SI_Layer_hall3}
	\end{figure*}

	Firstly, Fig.~\ref{SI_Layer_hall3} shows the static Kerr vs $B_{\perp}$ data in a 6SL MnBi$_2$Te$_4$ dual-gated device, which is consistent with \cite{Qiu2023axion}, therefore confirming that the sample remains even-layered structure after microfabrications. Additional electric field dependence further demonstrates the magnetoelectric effect. 
	
	\begin{figure*}[h]
		\centering
		\includegraphics[width=14cm]{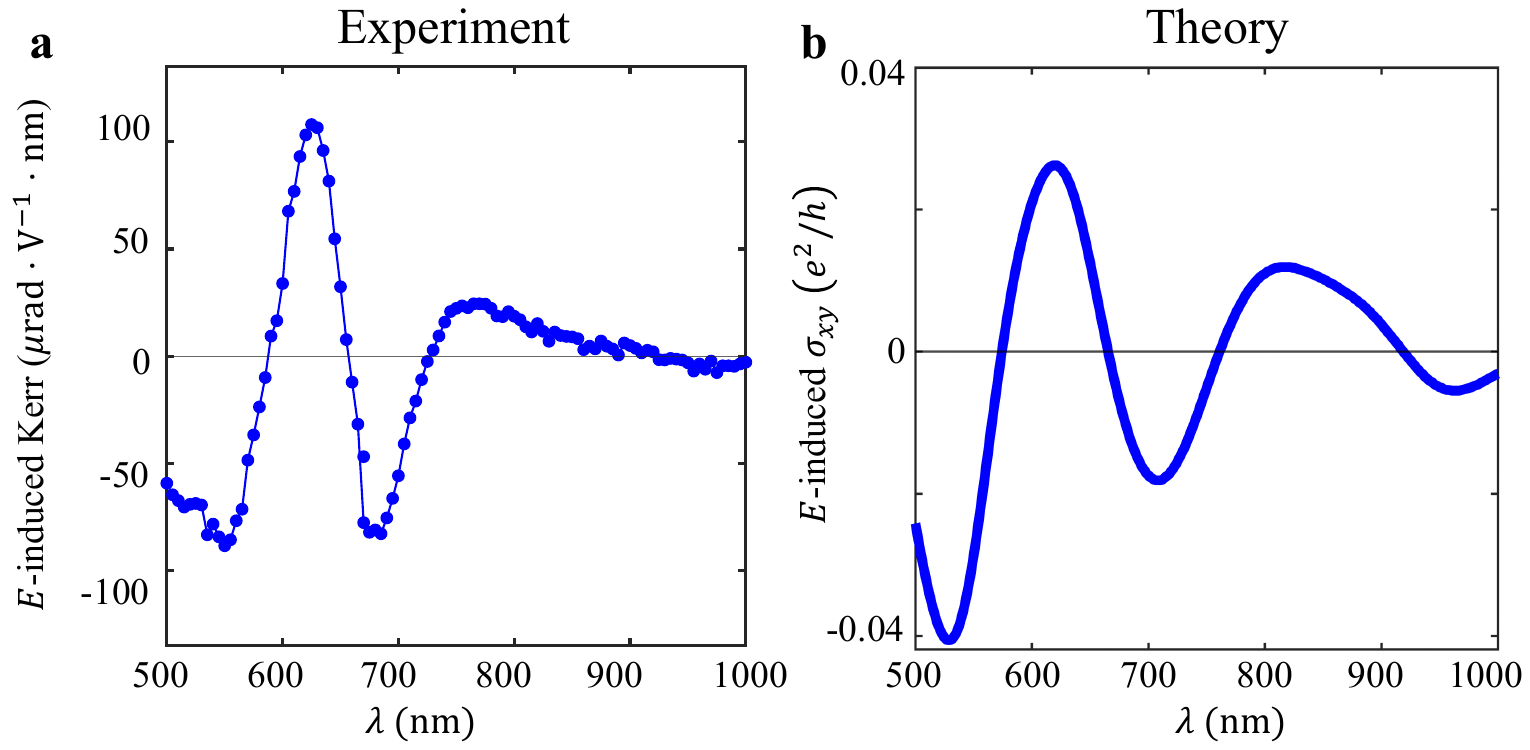}
		\caption{\textbf{a}, Experimentally measured $E$-field induced Kerr at different wavelengths. \textbf{b}, Theoretically calculated $E$-field induced $\sigma_{\rm{xy}}$ at different wavelengths.}
		\label{SI_E_sigma_xy_DATA_THY_v4}
	\end{figure*}
	
\begin{figure*}[!htb]
		\centering
		\includegraphics[width=11cm]{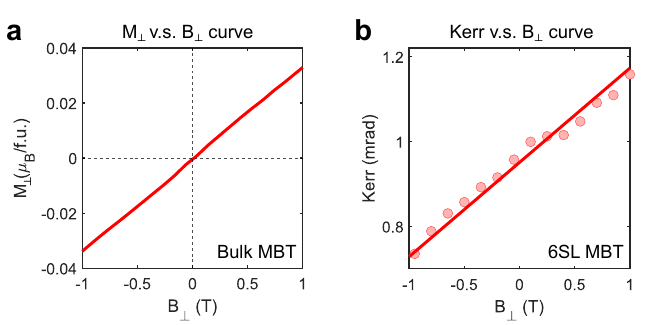}
		\vspace{0cm}
		\caption{\textbf{a}, $M_z$-$B_z$ of bulk MnBi$_2$Te$_4$  measured by SQUID. \textbf{b}, Kerr-$B_z$ of 6SL MnBi$_2$Te$_4$.}
		\label{SI_M_H_Kerr_analysis}
	\end{figure*}

\begin{figure*}[!htb]
		\centering
		\includegraphics[width=11cm]{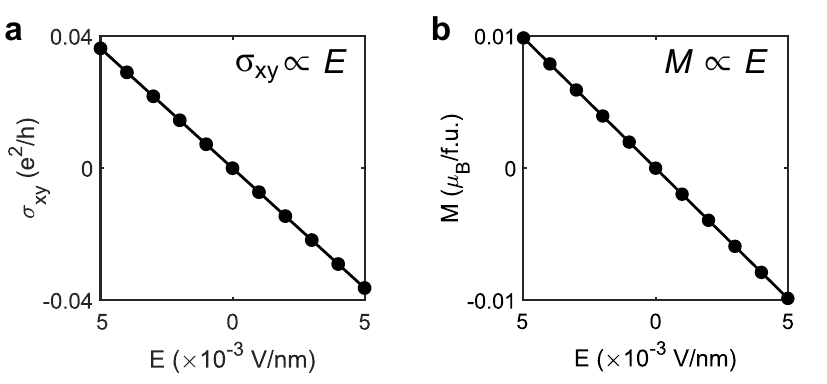}
		\vspace{0cm}
		\caption{\textbf{a}, Theoretical calculation of $\sigma_{ xy}$-$E$ of 6SL MnBi$_2$Te$_4$. The wavelength of $\sigma_{xy}$ is 515 nm. \textbf{b}, Theoretical calculation of $M_z$-$E_z$ of 6SL MnBi$_2$Te$_4$.}
		\label{SI_Linear_calculation_M_Sigma_E}
	\end{figure*}

\begin{figure*}[!htb]
		\centering
		\includegraphics[width=6cm]{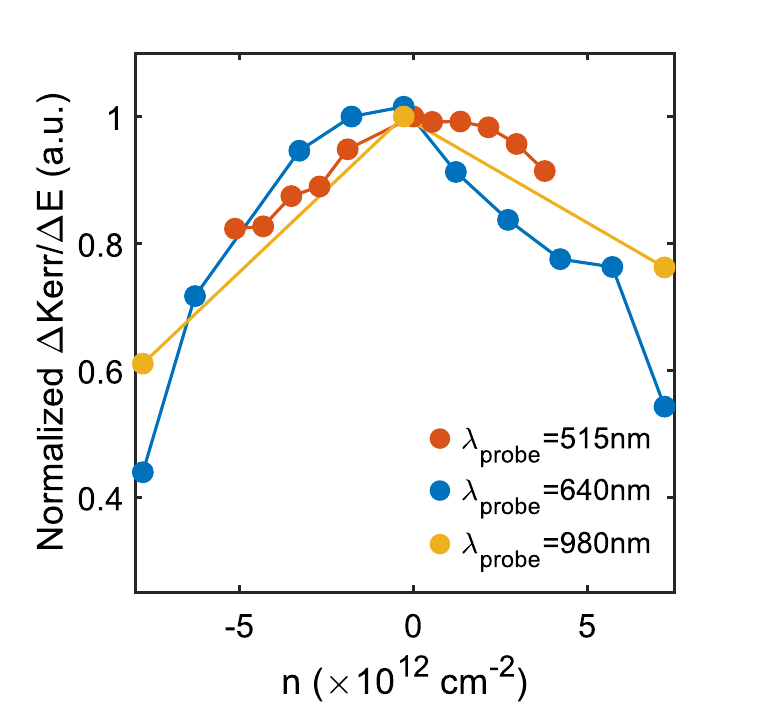}
		\caption{$\frac{\Delta \rm Kerr}{\Delta E}$ vs. $n$ measurement using three wavelengths (normalized at $n=0$).}
		\label{SI_ME_doping_wavelength_data}
	\end{figure*}

\begin{figure*}[!htb]
		\centering
		\includegraphics[width=12cm]{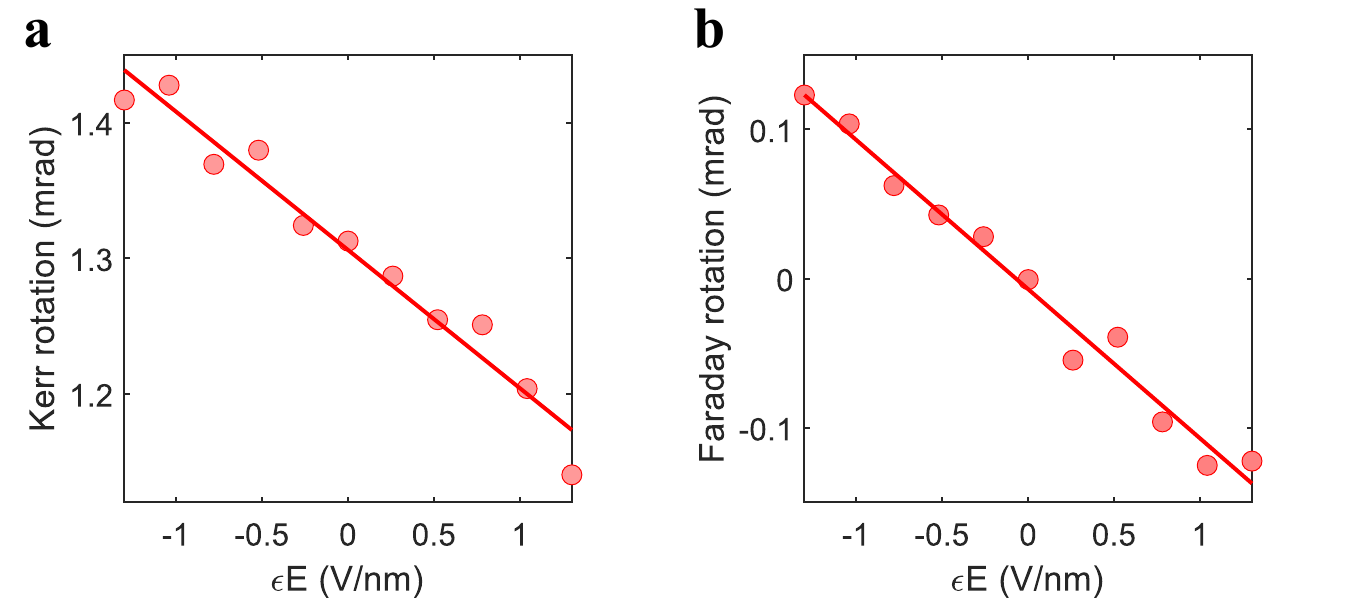}
		\caption{Simultaneous Kerr (\textbf{a}) and Faraday rotation (\textbf{b}) measurements as a function of $E$ field for 6SL MnBi$_2$Te$_4$.}
		\label{SI_Kerr_Faraday}
	\end{figure*}

	Secondly, In addition the optical detection of magneto-electric coupling is based on the assumption that Kerr rotation measures magnetization. 
	\begin{enumerate}
		\item As shown in Fig.~\ref{SI_M_H_Kerr_analysis}, we can measure the magnetization $M_z$  of bulk MnBi$_2$Te$_4$ as a function of $B_z$ using SQUID and compare with the Kerr vs. $B_z$ of 6SL MnBi$_2$Te$_4$. Both the Kerr and $M_z$ show a linear dependence on $B_z$, therefore suggesting that Kerr is proportional to magnetization in our system (Kerr $\propto B \propto  M_z$).
		
		\item Fig.~\ref{SI_E_sigma_xy_DATA_THY_v4}\textbf{a} shows the probe wavelength dependence of the $E$-field induced Kerr rotation. Furthermore, we theoretically calculated the $E$-field induced $\sigma_{\rm xy}$, which shows a good agreement between data and calculation, based on $\rm{Kerr} \propto \sigma_{\rm xy}(\lambda) \propto \gamma(\lambda)M_z $. This result therefore suggests our conclusion that the measured $E$-field induced Kerr rotation in 6SL MnBi$_2$Te$_4$ indeed reflects the $E$-field induced magnetization.
		
		\item To further strengthen point 2 above, in our DFT calculation, we could calculate the $E$-induced $\sigma_{\rm xy}$ and $E$-induced $M_z$.  As shown in Fig.~\ref{SI_Linear_calculation_M_Sigma_E}, both $\sigma_{xy}$ and $M_z$ are linearly dependent on $E$. Hence, we could obtain Kerr $\propto \sigma_{xy} \propto E \propto M$, which demonstrates the proportionality between Kerr rotation and $M_z$.
		
		\item We further measured the doping $n$ dependence of the $\frac{\Delta \rm Kerr}{\Delta E}$ (similar to the Fig. 4b of main text) using three wavelengths. If $\frac{\Delta \rm Kerr}{\Delta E}$  vs. $n$ is a good measure of $\alpha(n)$, then different wavelengths should give consistent results. Indeed, our data (Fig.~\ref{SI_ME_doping_wavelength_data}) show similar behavior. We further elaborate on the logic: Because $\rm{Kerr} \propto \gamma M_z $, its variation as a function of $n$, $\Delta \textrm{Kerr}/\Delta n$, may come from $\Delta M_z/\Delta n$ or $\Delta \gamma/\Delta n$. If it comes from $M_z$, the $n$ dependence should be the same function at various wavelengths upon normalization. By contrast, if it comes from $\gamma$, then $n$ dependence of $\gamma$ should be different functions at various wavelengths. Furthermore, our measured $\frac{\Delta \rm Kerr}{\Delta E}$ vs. $n$ shows good agreement with the theoretically calculated $\alpha(n)$. Therefore, collectively, these evidences suggest that the measured n dependence mostly arises from changes in $M_z$. 
		
		\item Furthermore, we can achieve a symmetry-based, direct proof that our $E$-induced Kerr is measuring $E$-induced magnetization. As shown in Ref. \cite{Qiu2023axion}, magnetization $M_z$ induces both Kerr (reflection) and Faraday (transmission) rotation. By contrast, the AFM order $L_z$ only shows up in Kerr but is absent in Faraday. Therefore, if the $E$-field induced signal shows up in both reflection (Kerr) and in transmission (Faraday), then it measures $M_z$. Figure~\ref{SI_Kerr_Faraday} shows the simulatneous Kerr and Faraday measurements at different DC $E$ field. we see that the linear-to-$E$ effect exists in both reflection and transmission. But in reflection, there is a nonzero offset at $E=0$ due to the AFM Kerr; In transmission, the Faraday rotation is zero at $E=0$. Therefore, our simultaneous reflection and transmission measurements directly prove that the observed linear-to-$E$ Kerr rotation is due to magnetization $M_z$.
	\end{enumerate}

\begin{figure*}[!htb]
		\centering
		\includegraphics[width=17cm]{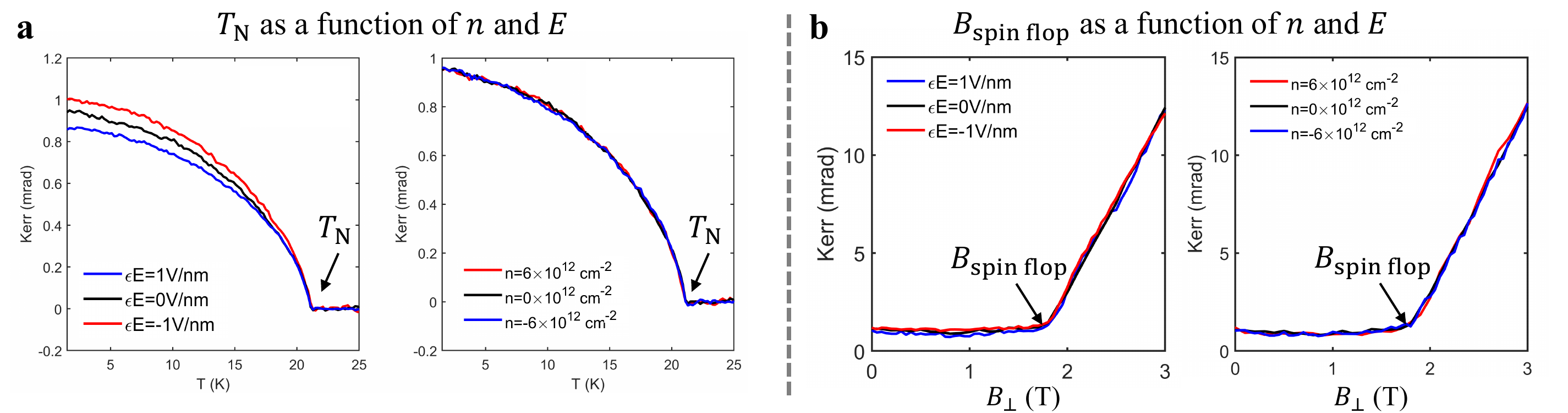}
		\caption{Static Kerr measurements of $T_N$ (\textbf{a}) and $B_{\rm spin flop}$ (\textbf{b}) as a function of $n$ and $E$ ($\lambda_{\rm probe}$=515 nm).}
		\label{SI_Static_TN_Bsf_n_E_v2}
	\end{figure*}

	\vspace{0.5cm}
	Thirdly, static magnetic properties such as $T_N$ and the spin flop field $B_{\rm spin flop}$ can also serve as an experimental check if $J$ and $K$ were strongly modified by gating ($J$: exchange coupling, $K$:anisotropy). As shown in Fig.~\ref{SI_Static_TN_Bsf_n_E_v2}, $T_N$ and $B_{\rm spin flop}$ do not show observable dependence on $n$ and $E$, suggesting that $J$ and $K$ are roughly invariant within the experimentally studied $n$ and $E$ range.
	
	\clearpage
	\subsection*{I.4. Additional DAQ data of 6SL MnBi$_2$Te$_4$}
	\subsubsection*{I.4.1. Addressing competing mechanisms}
	
	We present additional data, to address competing mechanisms. In particular, if the electrostatic gating strongly modifies the exchange coupling $J$ and magnetic anisotropy $K$, it might also change the TRMOKE signal. Thus, we investigated the $n$ and $E$ dependence of the TRMOKE signals. In particular, in even-layer MnBi$_2$Te$_4$, there are in total eight distinct dependences. I.e., how does the \textit{\textbf{frequency}} (or \textit{\textbf{TRMOKE amplitude}}) of the \textit{\textbf{in-phase}} (or \textit{\textbf{out-of-phase}}) magnon depend on $n$ (or $E$). We found that the only observed change is that the TRMOKE amplitude shows a linear $E$ dependence at the out-of-phase magnon frequency. If the exchange coupling $J$ and the anisotropy $K$ were strongly modified by gating, then we would expect both the frequency and TRMOKE amplitude of both magnon modes to show gate dependence (both $n$ and $E$). Therefore, our results strongly suggest that the alternative effect that $J$ and $K$ are strongly modified is unlikely to be origin. Rather, the DAQ is a more plausible and consistent interpretation. Below, we present the corresponding data.
	
	\begin{figure*}[h]
		\centering
		\includegraphics[width=9cm]{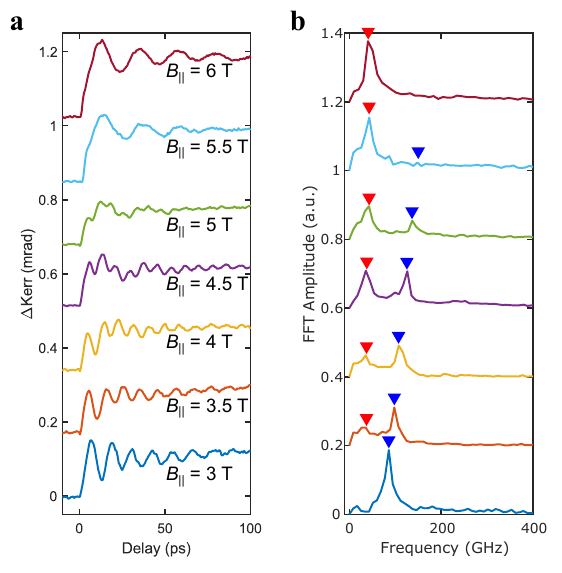}
		\caption{TRMOKE data of 6SL MnBi$_2$Te$_4$ at different $B_{\parallel}$ field (red and blue triangles represent out-of-phase and in-phase magnons).}
		\label{SI_TRMOKE_overview}
	\end{figure*}
	
	Figure~\ref{SI_TRMOKE_overview} shows the TRMOKE data of 6SL MnBi$_2$Te$_4$. The signal consists of both the AFM Kerr and the regular Kerr, therefore can detect both the in-phase and out-of-phase magnons. As shown in Fig.~\ref{SI_TRMOKE_overview}, \textbf{the in-phase magnon is prominent at small $B_{\parallel}$, whereas the out-of-phase magnon becomes strong at large $B_{\parallel}$. }

\begin{figure*}[!htb]
		\centering
		\includegraphics[width=16cm]{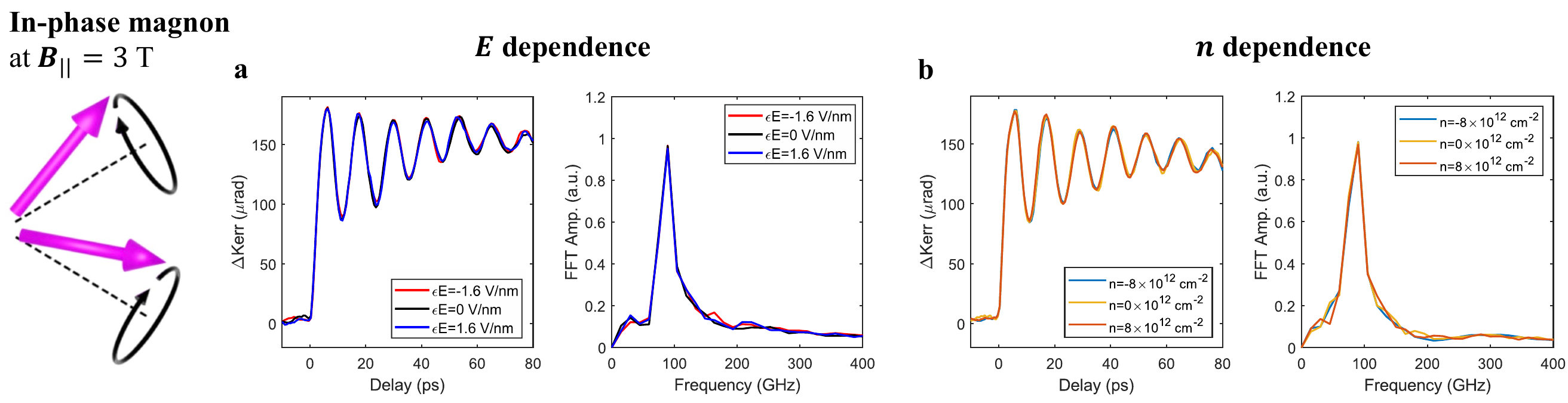}
		\caption{Gate dependent TRMOKE data at $B_{\parallel}=3$ T. The in-phase magnon shows no observable $n$ and $E$ dependence in terms of both the frequency and TRMOKE amplitude. $\lambda_{\rm pump}=1030$ nm, $\lambda_{\rm probe}=515$ nm.}
		\label{SI_In_phase_magnon_n_E_2_v3}
	\end{figure*}

	We set $B_{\parallel}=3$ T, where the in-phase magnon is prominent. As shown in Fig.~\ref{SI_In_phase_magnon_n_E_2_v3}, the in-phase magnon shows no observable $n$ and $E$ dependence in terms of both the frequency and TRMOKE amplitude.

\begin{figure*}[!htb]
		\centering
		\includegraphics[width=16cm]{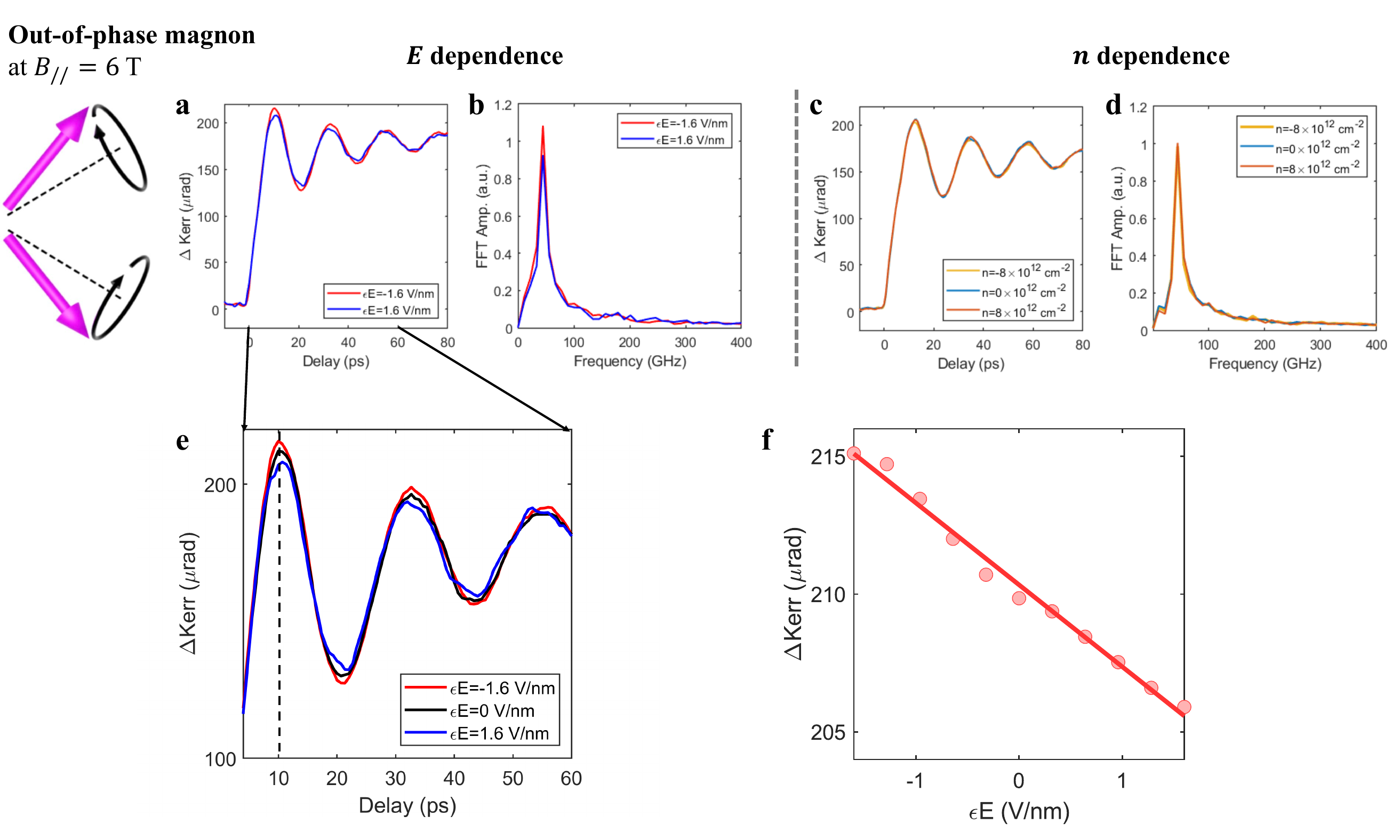}
		\caption{Gate dependent TRMOKE data at $B_{\parallel}=6$ T. Top row: The out-of-phase magnon shows no observable $n$ and $E$ dependence in terms of both the frequency and TRMOKE amplitude. Bottom row: A zoomed-in view of the TRMOKE with three different $E$ values. $E$ dependence TRMOKE at delay time 10 ps (dotted line on the left).  $\lambda_{\rm pump}=1030$ nm, $\lambda_{\rm probe}=515$ nm.}
		\label{SI_Out_of_phase_magnon_n_E_v2}
	\end{figure*}

	We set $B_{\parallel}=6$ T, where the out-of-phase magnon is prominent. As shown in Fig.~\ref{SI_Out_of_phase_magnon_n_E_v2}, the out-of-phase magnon shows no observable $n$ dependence in terms of both the frequency and TRMOKE amplitude as well as no observable $E$ dependence in terms of the frequency. The only observed change is that the TRMOKE amplitude shows a linear $E$ dependence at the out-of-phase magnon frequency. 
	
	\pagebreak
	
	\subsubsection*{I.4.2. $E$ and $n$ dependence of TRMOKE measured by DC and AC methods}
	
\begin{figure*}[!htb]
		\centering
		\includegraphics[width=17cm]{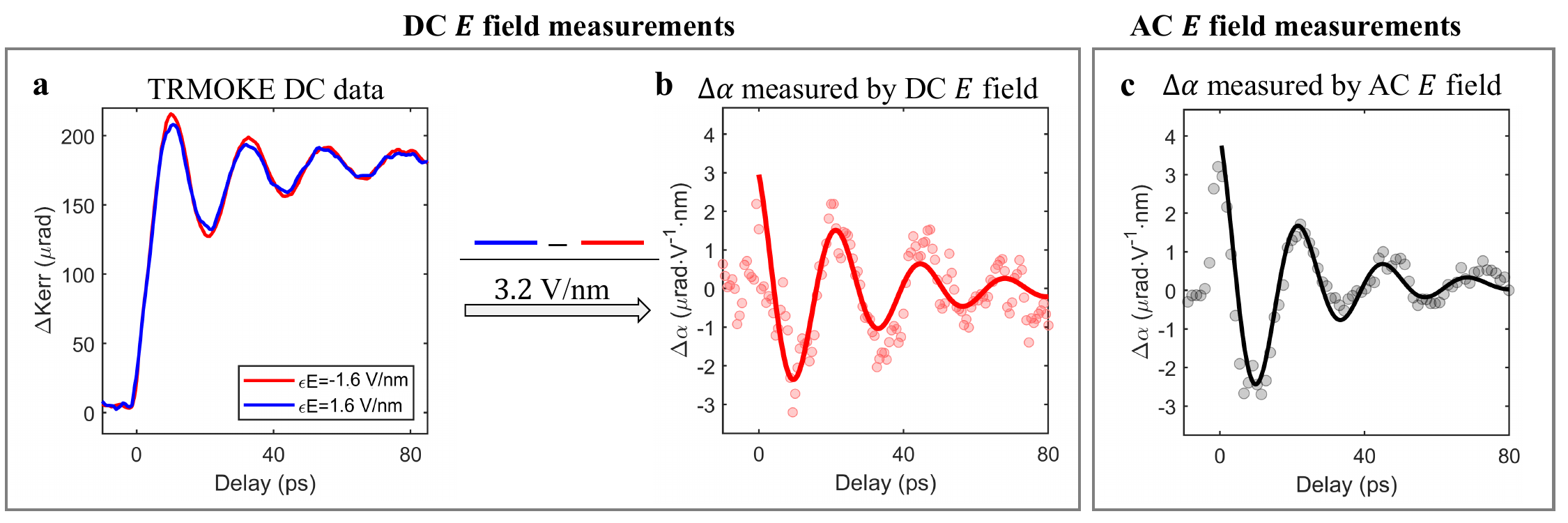}
		\caption{Comparison between DC and AC $E$ field dependent measurements at 6 T. a, TRMOKE data at different DC $E$ field values (same data as Fig.~\ref{SI_Out_of_phase_magnon_n_E_v2}). b, Time-resolved $\Delta \alpha$ is obtained by the difference of the TRMOKE data at $\pm$1.6 V/nm divided by 3.2 V/nm. c, Time-resolved $\Delta \alpha$ measured by AC $E$ field-dependent measurements.}
		\label{SI_AC_DC_v4}
	\end{figure*}

	We now try to obtain the time-resolved $\alpha$ by DC $E$ field method, and show that the result is fully consistent with the AC $E$ field modulation method adapted in the main text. Specifically, we first park the magnetic field at $B_{\parallel}=6$ T, where the out-of-phase magnon is prominent. Figure~\ref{SI_AC_DC_v4}\textbf{a} shows the TRMOKE at different DC $E$ values (same data as Fig.~\ref{SI_Out_of_phase_magnon_n_E_v2}). In order to obtain the time-resolved $\Delta \alpha$, we subtract the TRMOKE data at $E=\pm 1.6$ V/nm and divide it by $3.2$ V/nm. The result, shown in Fig. Fig.~\ref{SI_AC_DC_v4}\textbf{b}, shows an oscillation of $\alpha$. Therefore, we can detect the $\alpha$ oscillation (the DAQ) by DC $E$ field method; The result (Fig.~\ref{SI_AC_DC_v4}\textbf{b}) is consistent with that of AC $E$ field lock-in method shown in Fig.~\ref{SI_AC_DC_v4}\textbf{c} (the AC data has better signal to noise ratio).
	
\begin{figure*}[!htb]
		\centering
		\includegraphics[width=17cm]{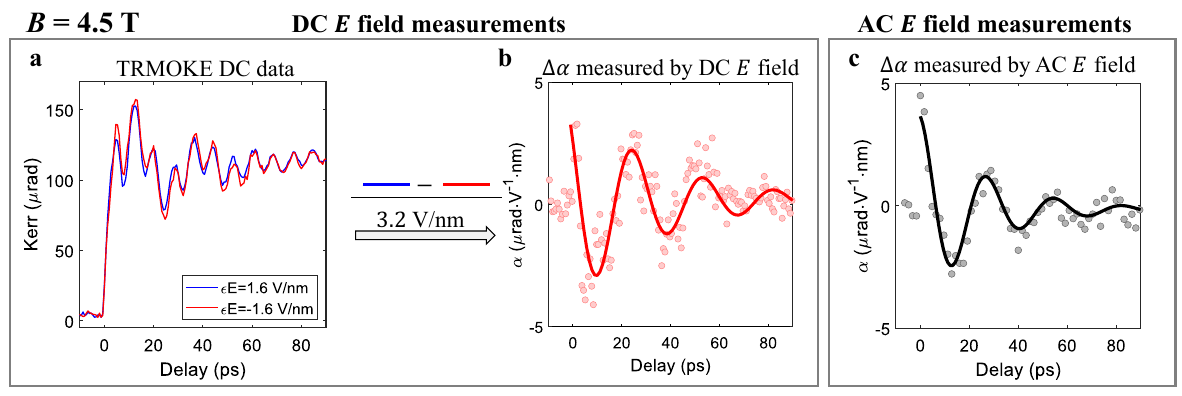}
		\caption{Comparison between DC and AC $E$ field dependent measurements at 4.5 T. a, TRMOKE data at different DC $E$ field values at 4.5 T, both in-phase and out-of-phase magnons are clearly observed. b, Time-resolved $\Delta \alpha$ is obtained by the difference of the TRMOKE data at $\pm1.6$ V/nm divided by $3.2$ V/nm. c, Time-resolved $\Delta \alpha$ measured by AC $E$ field-dependent measurements. \color{black}}
		\label{SI_MBT4p5T_data_AC_DC_field}
	\end{figure*}

	Moreover, we can also the magnetic field at at $B_{\parallel}=4.5$ T, where both the in-phase and out-of-phase magnons are excited. When we perform $E$-field dependence (both DC and AC), we can also see the $\alpha$ oscillation at the out-of-phase magnon frequency with fully consistent results (Fig.~\ref{SI_MBT4p5T_data_AC_DC_field}).
	
\begin{figure*}[!htb]
		\centering
		\includegraphics[width=17cm]{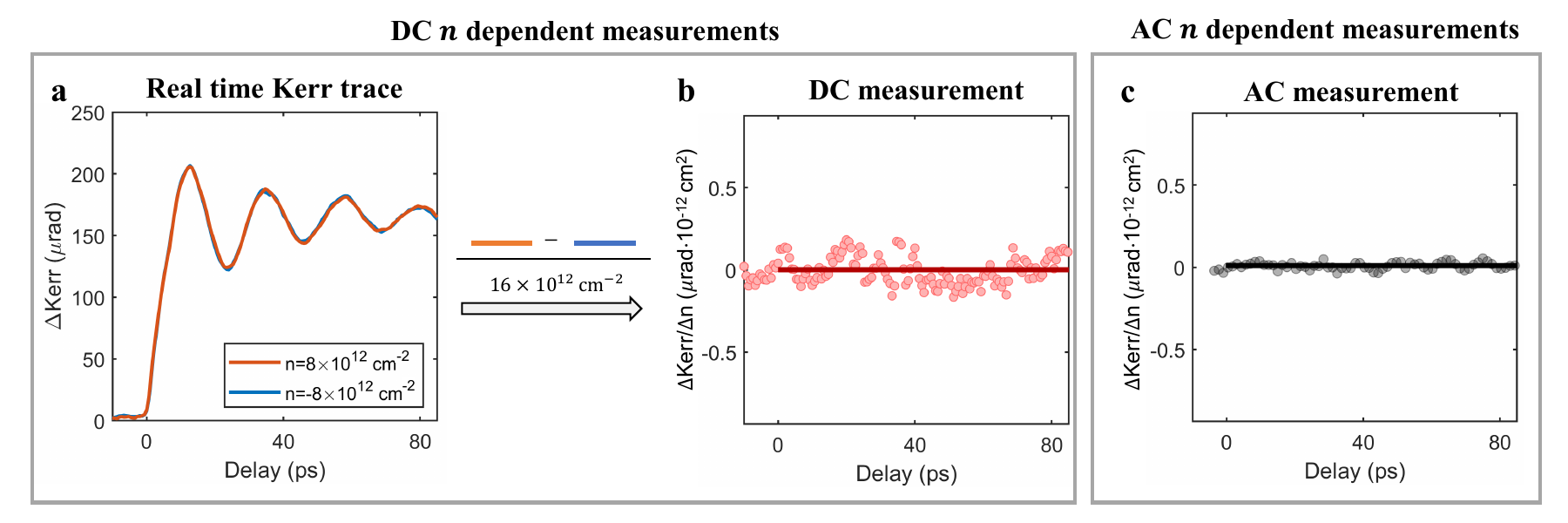}
		\caption{Comparison between DC and AC doping $n$ dependent measurements. $\textbf{a}$, TRMOKE data at different DC $n$ values. $\textbf{b}$, The difference of the TRMOKE data at $\pm 8\times 10^{12}$ cm$^{-2}\cdot$V/nm divided by $16\times 10^{12}$ cm$^{-2}\cdot$V/nm. $\textbf{c}$, AC doping $n$ dependent measurements of the TRMOKE.}
		\label{SI_AC_DC_n_v3}
	\end{figure*}

	Furthermore, we performed similar comparison between DC and AC doping $n$ dependence. Figure~\ref{SI_AC_DC_n_v3} shows the TRMOKE data at $\pm n$ values (same data as Fig.~\ref{SI_Out_of_phase_magnon_n_E_v2}). The TRMOKE does not show significant change with $n$. As such, their difference (Fig.~\ref{SI_AC_DC_n_v3}\textbf{b}) is nearly zero. Consistent result is obtained if one AC modulates the doping $n$ (Fig.~\ref{SI_AC_DC_n_v3}\textbf{c}).
	
	\pagebreak
	
	\subsubsection*{I.4.3. Doping dependence of DAQ data}
\begin{figure*}[h]
		\centering
		\vspace{0cm}
		\includegraphics[width=11cm]{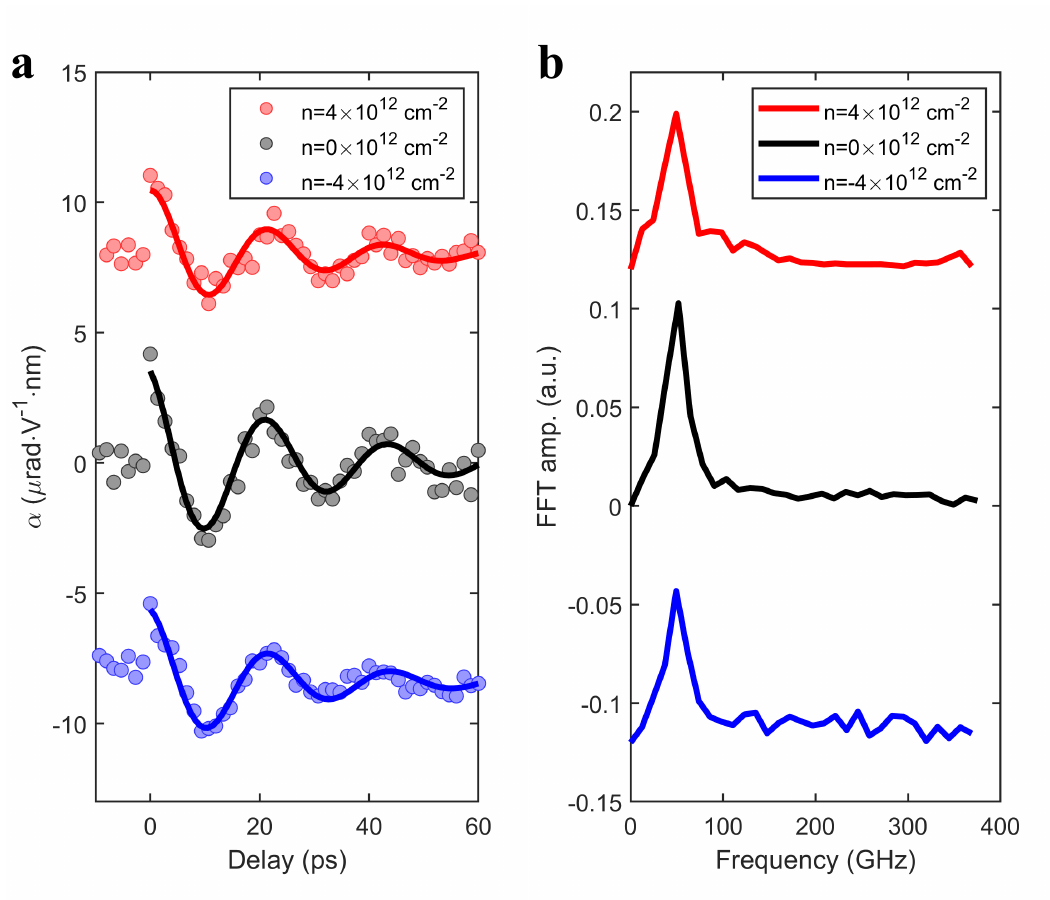}
		\vspace{-0.5cm}
		\caption{Time-resolved $\alpha$ data at different $n$ values ($B_{\parallel}$=6 T). }
		\label{SI_TR_alpha_n_dependence_v2}
	\end{figure*}

	In the main text, all $\Delta\alpha$ measurements were performed with the Fermi level at charge neutrality ($n=0$). Here, we present the $n$ dependence of $\Delta\alpha$ measurements for completeness. As shown in Fig.~\ref{SI_TR_alpha_n_dependence_v2}, the $\Delta\alpha$ oscillation frequency remains the same at different $n$ values, but the oscillation amplitude decreases as one changes from charge neutrality ($n=0$) to both electron-doped and hole-doped regimes. This observation is consistent with our static $\alpha$ versus $n$ measurements shown in Fig.~4\textbf{b} of the main text.

	\clearpage

	\subsubsection*{I.4.4. Reproducibility of DAQ}
\begin{figure*}[h]
		\centering
		\includegraphics[width=14cm]{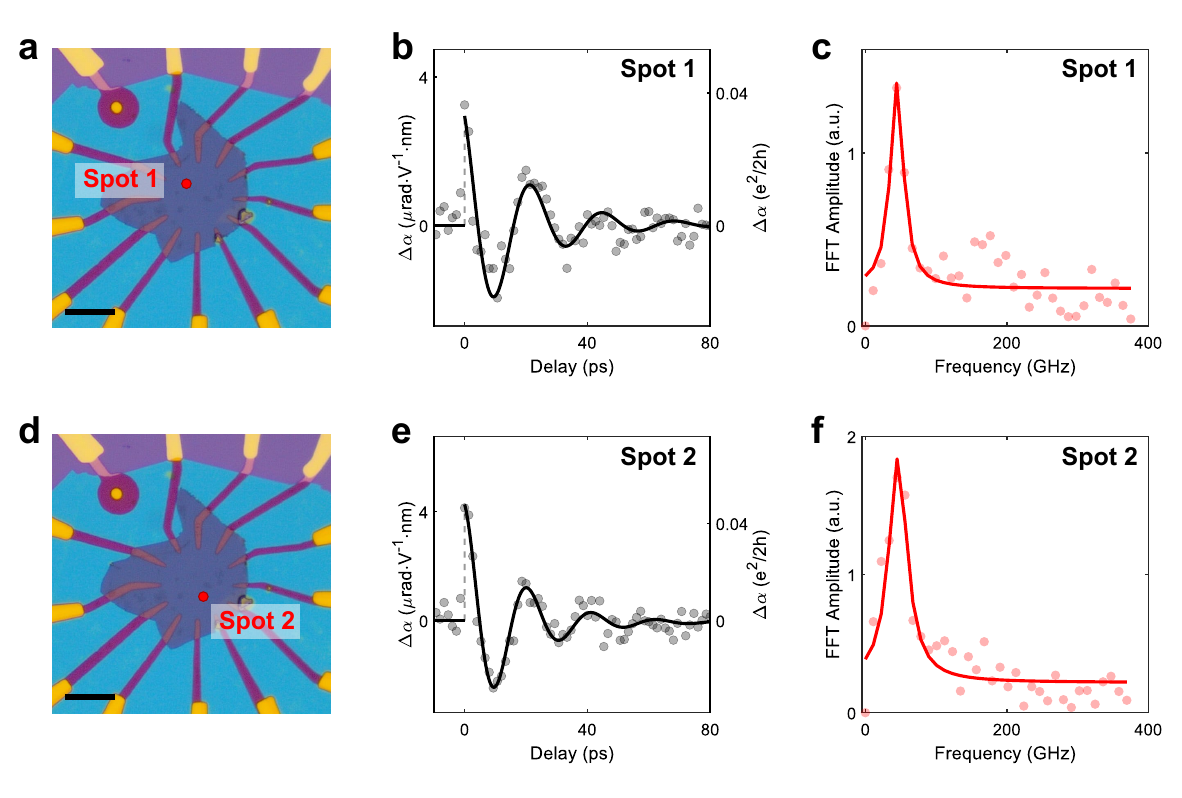}
		\vspace{-0.5cm}
		\caption{\textbf{Reproducibility of dynamical Axion quasiparticle at different locations. }\textbf{a-c} At laser spot-1 location, time-dependent $\alpha$ measurements and FFT.  \textbf{d-f,} Same as \textbf{a-c} for a different location on the sample.  Scale bar: 20 $\rm \mu m$}
		\label{SI_DAQ_location_reproducibility}
	\end{figure*}

	We present the reproducibility of DAQ at different locations on the sample (Fig.~\ref{SI_DAQ_location_reproducibility}).

	\clearpage
	
	\subsubsection*{I.4.5. Direct comparison of DAQ and magnon}
\begin{figure*}[!htb]
		\centering
		\vspace{-0.5cm}
		\includegraphics[width=11cm]{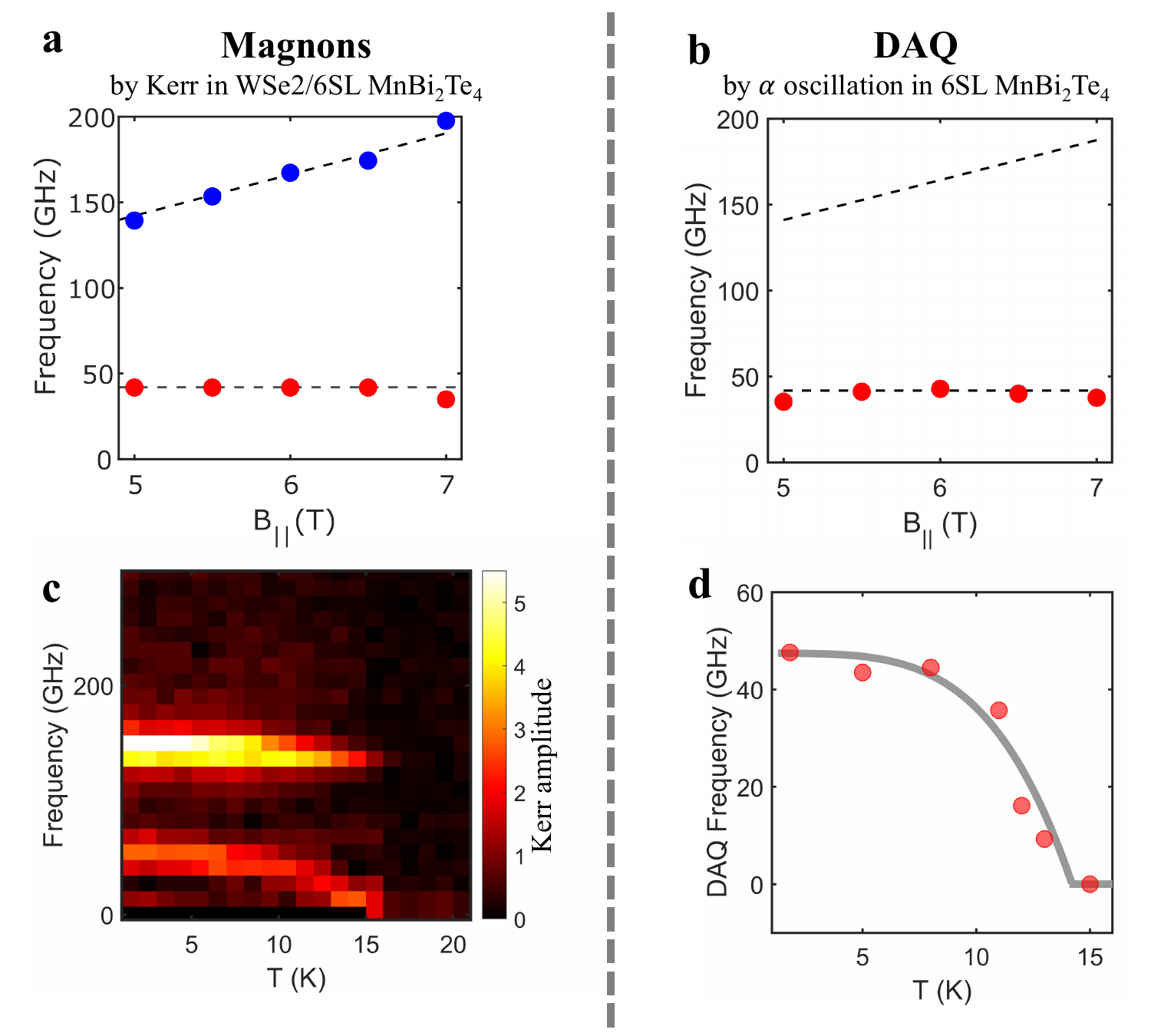}
		\vspace{-0.5cm}
		\caption{\textbf{Comparison between magnon and DAQ} \textbf{a}, Magnon frequency probed by the TRMOKE in WSe$_2$/6SL MnBi$_2$Te$_4$ heterostructure. \textbf{b}, The DAQ by the $\alpha$ oscillation in 6SL MnBi$_2$Te$_4$. \textbf{c-d} Temperature dependence of magnon and DAQ, respectively. }
		\label{SI_DAQ_Magnon_v3}
	\end{figure*}

	Here, we show the direction comparison of DAQ (measured by $\Delta\alpha$) and magnons (measured by TRMOKE) in Fig.~\ref{SI_DAQ_Magnon_v3}. Regarding both the $B_{\parallel}$ and $T$ dependence, the DAQ agrees very well with the out-of-phase magnon. 
	
	\subsection*{I.5. Phonon data of 6SL MnBi$_2$Te$_4$}
	
\begin{figure*}[!htb]
		\centering
		\includegraphics[width=16cm]{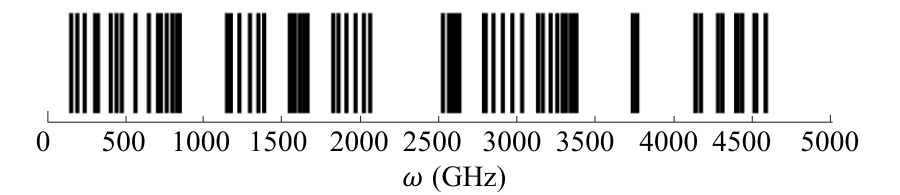}
		\caption{\textbf{Theoretically calculated phonon frequency}}
		\label{SI_Phonon_frequency_v1}
	\end{figure*}

\begin{figure*}[!htb]
		\centering
		\includegraphics[width=12cm]{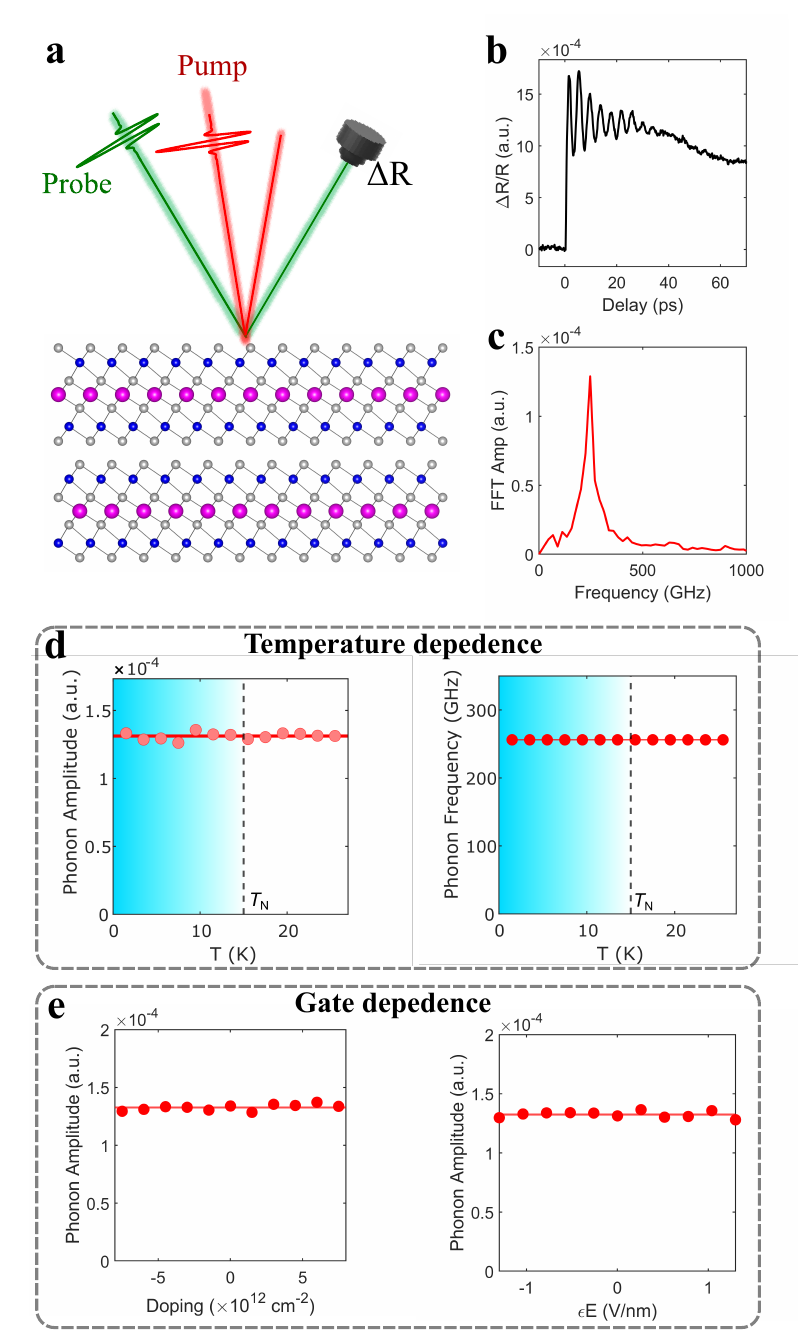}
		\caption{\textbf{a-c}, Time-resolved reflectivity data showing the breathing phonon oscillation. \textbf{d-f}, Temperature and gate dependence of the phonon amplitude (measured by the Fourier peak value). $\lambda_{\rm pump} =1030$ nm, $\lambda_{\rm probe} =515$ nm,  }
		\label{SI_Phonon_data_v3}
	\end{figure*}

	In the main text, we focused on the magnon-induced effect. We discuss phonon induced effect in this section. Theoretically, we use DFT to compute the phonon frequencies at $\Gamma$ point of 6SL MnBi$_2$Te$_4$. As shown in Fig.~\ref{SI_Phonon_frequency_v1}, the lowest phonon frequency is $\sim$160 GHz, which is much higher than the experimentally observed DAQ frequency. Experimentally, we can observe the breathing phonon at $\sim$200GHz by time-resolved reflectivity ($\Delta R/R$) measurements, which is also consistent with \cite{bartram2022ultrafast}. In addition, we also investigated the gate dependence of this phonon mode. As shown in Fig.~\ref{SI_Phonon_data_v3}, our data show no observable gate dependence and no observable temperature dependence across $T_N$.
	
	\newpage
	\pagebreak
	
	\pagebreak

	\newpage
	\pagebreak

	\section*{I.6. Other addition data and sanity check}
\begin{figure*}[!htb]
		\centering
		\includegraphics[width=17cm]{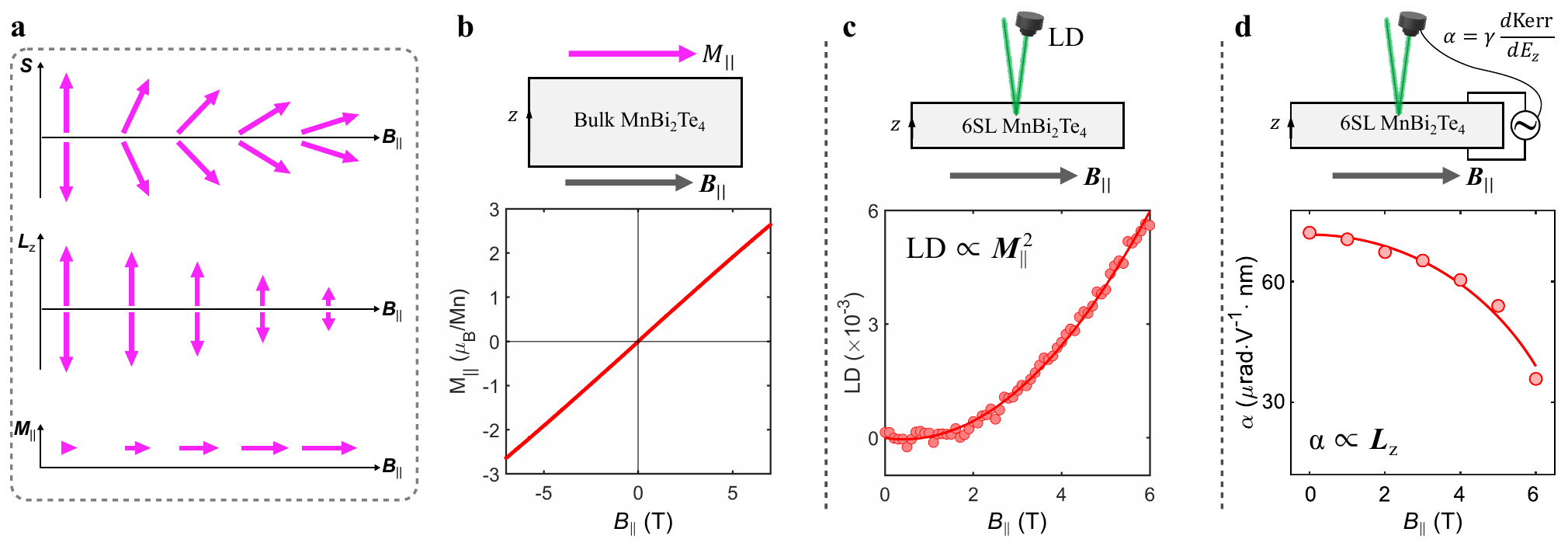}
		\caption{\textbf{a}, Schematic illustration of how the Mn spins cant with increasing $B_{\parallel}$ and the evolution of $L_z$ and $M_{\parallel}$. \textbf{b}, $M_{\parallel}-B_{\parallel}$ data in bulk MnBi$_2$Te$_4$. \textbf{c}, Optical linear dichroism (LD) as a function of $B_{\parallel}$ in 6SL MnBi$_2$Te$_4$. \textbf{d}, $\alpha$ (measured by $E$-field induced Kerr rotation) as a function of $B_{\parallel}$ in 6SL MnBi$_2$Te$_4$.}
		\label{SI_In_plane_spin_test_v2}
	\end{figure*}

\begin{figure*}[!htb]
		\centering
		\includegraphics[width=16cm]{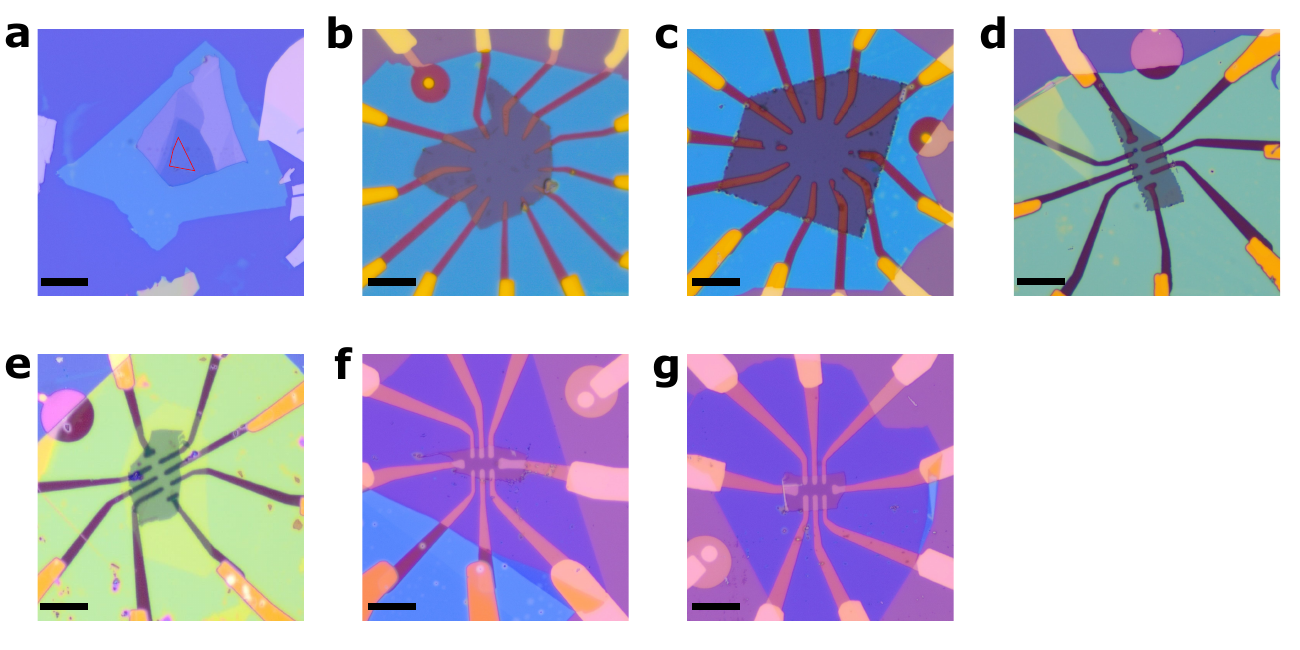}
		\caption{\textbf{Optical microscope device images.} \textbf{a} Dev1 is an heterostructure of  6-SL MnBi$_2$Te$_4$ and monolayer WSe$_2$ (red triangle).  \textbf{b-g} Device images of 6-SL MnBi$_2$Te$_4$  for Dev2-Dev7.  Scale bar: 20 $\rm \mu m$}
		\label{SI_figure_OM_device}
	\end{figure*}

\begin{figure*}[!htb]
		\centering
		\includegraphics[width=16cm]{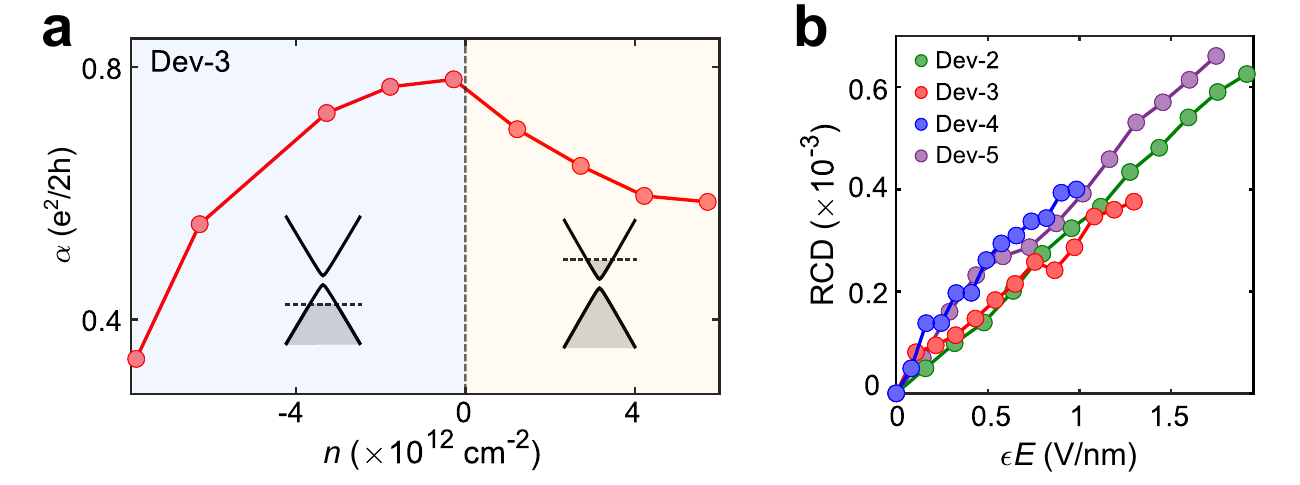}
		\caption{\textbf{Reproducibility data for static magnetoelectric coupling $\alpha$ measurements. } Probe wavelength is at $600$ nm. We note that Kerr and RCD gives similar results.}
		\label{SI_figure_ME_reprop}
	\end{figure*}

\begin{figure*}[!htb]
		\centering
		\includegraphics[width=14cm]{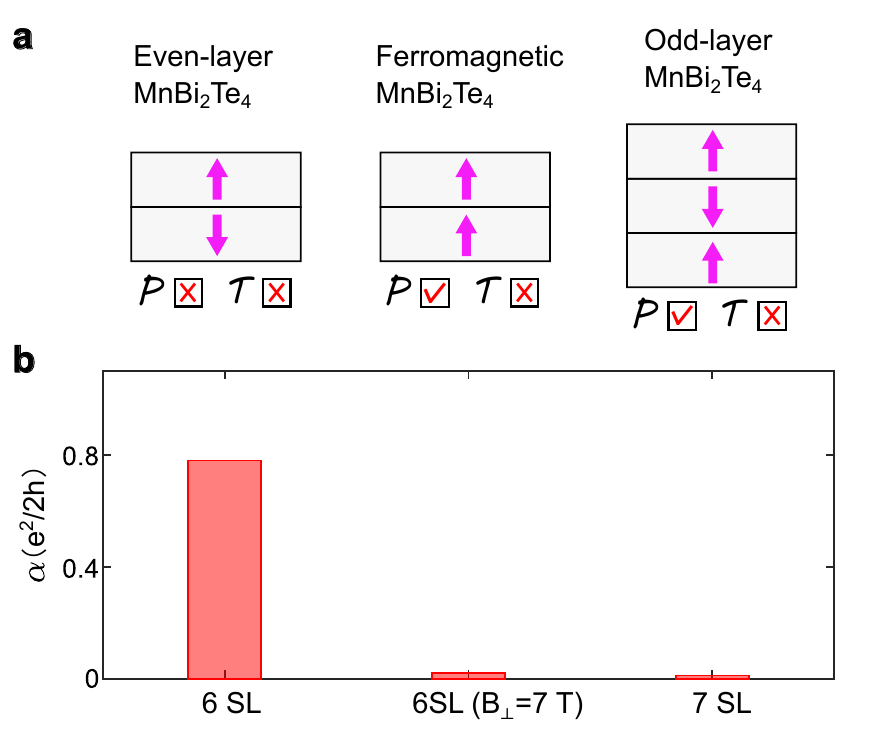}
		\caption{\textbf{Symmetry analysis for static magnetoelectric coupling $\alpha$ measurements. } \textbf{a,} The magnetoelectric coupling $\alpha$ requires the breaking of both time-reversal $\mathcal{T}$ and space-inversion $\mathcal{P}$ symmetries. Even-layer antiferromagnetic MnBi$_2$Te$_4$ breaks both $\mathcal{P}$ and $\mathcal{T}$, therefore is expected to show a nonzero $\alpha$. By contrast, ferromagnetic MnBi$_2$Te$_4$ and odd-layer antiferromagnetic MnBi$_2$Te$_4$ both respect $\mathcal{P}$, therefore are expected to show zero $\alpha$. \textbf{b,} Measured magnetoelectric coupling for 6SL antiferromagnetic MnBi$_2$Te$_4$, ferromagnetic 6SL MnBi$_2$Te$_4$ ($\textit{B}_{\perp}=7$ T) and 7SL antiferromagnetic MnBi$_2$Te$_4$. }
		\label{SI_figure_ME_symm}
	\end{figure*}

	In this section, we present other additional data to support our conclusion in the main text. As shown in Fig.~\ref{SI_In_plane_spin_test_v2}, we have performed additional experiments including the $M_{\parallel}-H_{\parallel}$ measurements on bulk MnBi$_2$Te$_4$, the optical linear dichroism ($\propto M_{\parallel}^2$) in 6SL MnBi$_2$Te$_4$ and the magnetoelectric coupling ($\alpha\propto L_z$) in 6SL MnBi$_2$Te$_4$. These experiments confirmed the expected magnetic ground state under a finite in-plane magnetic field $B_{\parallel}$. In addition, we present the optical images of all devices used in our experiments in Fig.~\ref{SI_figure_OM_device}. Furthermore, Fig.~\ref{SI_figure_ME_reprop} and  Fig.~\ref{SI_figure_ME_symm} showed the reproducibility and symmetry analysis of the magneto-electric effect.
	
	\clearpage
	\subsection*{II The AFM Kerr effect and DAQ probed with DC $E$-field}
	
	\subsubsection*{II.1. The AFM Kerr effect}
	
\begin{figure*}[!htb]
		\centering
		\includegraphics[width=8.5cm]{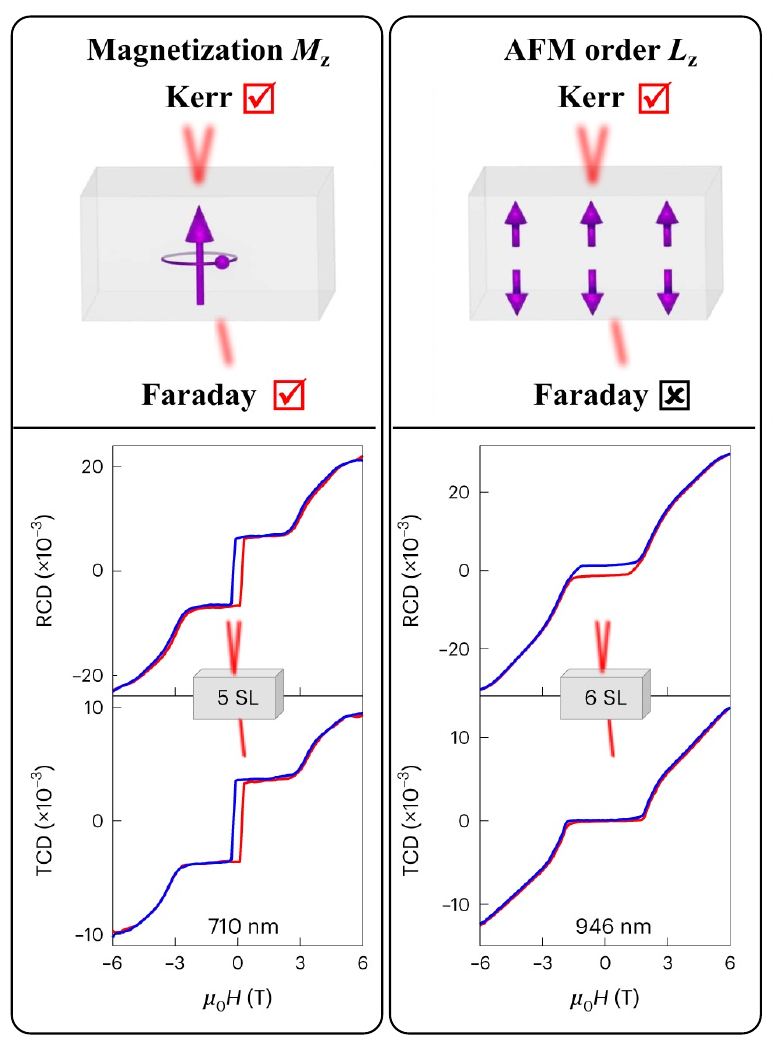}
		\caption{The net magnetization $M_z$ can be probed by both Kerr and Faraday. The AFM order $L_z$ can only be probed by Kerr. These conclusions have been experimentally demonstrated by Ref. \cite{Qiu2023axion}. Here we reproduce the Fig. 3 of Ref. \cite{Qiu2023axion}. We note that Kerr/Faraday have the same symmetry as RCD/TCD (reflection circular dichroism and transmission circular dichroism).}
		\label{AFM_Kerr_v1}
	\end{figure*}

	
	The regular Kerr effect is proportional to magnetization $M_z$. However, in a $PT$-symmetric antiferromagnet with zero magnetization (breaking $P$ and $T$ but preserves $PT$) such as Cr$_2$O$_3$ \cite{Krichevtsov1993spontaneous} or even-layer MnBi$_2$Te$_4$ \cite{Qiu2023axion}, there is an AFM Kerr effect, which is proportional to the AFM order $L_z$. Moreover, the $PT$-symmetric AFM order $L_z$ leads to the AFM Kerr effect but no Faraday effect. By contrast, $M_z$ leads to both nonzero Kerr and Faraday effect. These conclusions have been experimentally demonstrated in MnBi$_2$Te$_4$ by Ref. \cite{Qiu2023axion}. 
	\begin{itemize}
		\item 6SL MnBi$_2$Te$_4$ is a $PT$-symmetric AFM with zero net magnetization. It shows the AFM Kerr effect. Moreover, it shows zero Faraday effect (Fig.~\ref{AFM_Kerr_v1}). 
		\item 5SL MnBi$_2$Te$_4$ has an uncompensated magnetization ($M_z$), which is like a ferromagnet. So it supports both nonzero Kerr effect and Faraday effect, both proportional to $M_z$ (Fig.~\ref{AFM_Kerr_v1}). 
	\end{itemize}
	
	\subsubsection*{II.2. The AFM Kerr rotation shows up as a DC offset at $E$=0}

\begin{figure*}[!htb]
		\centering
		\includegraphics[width=12cm]{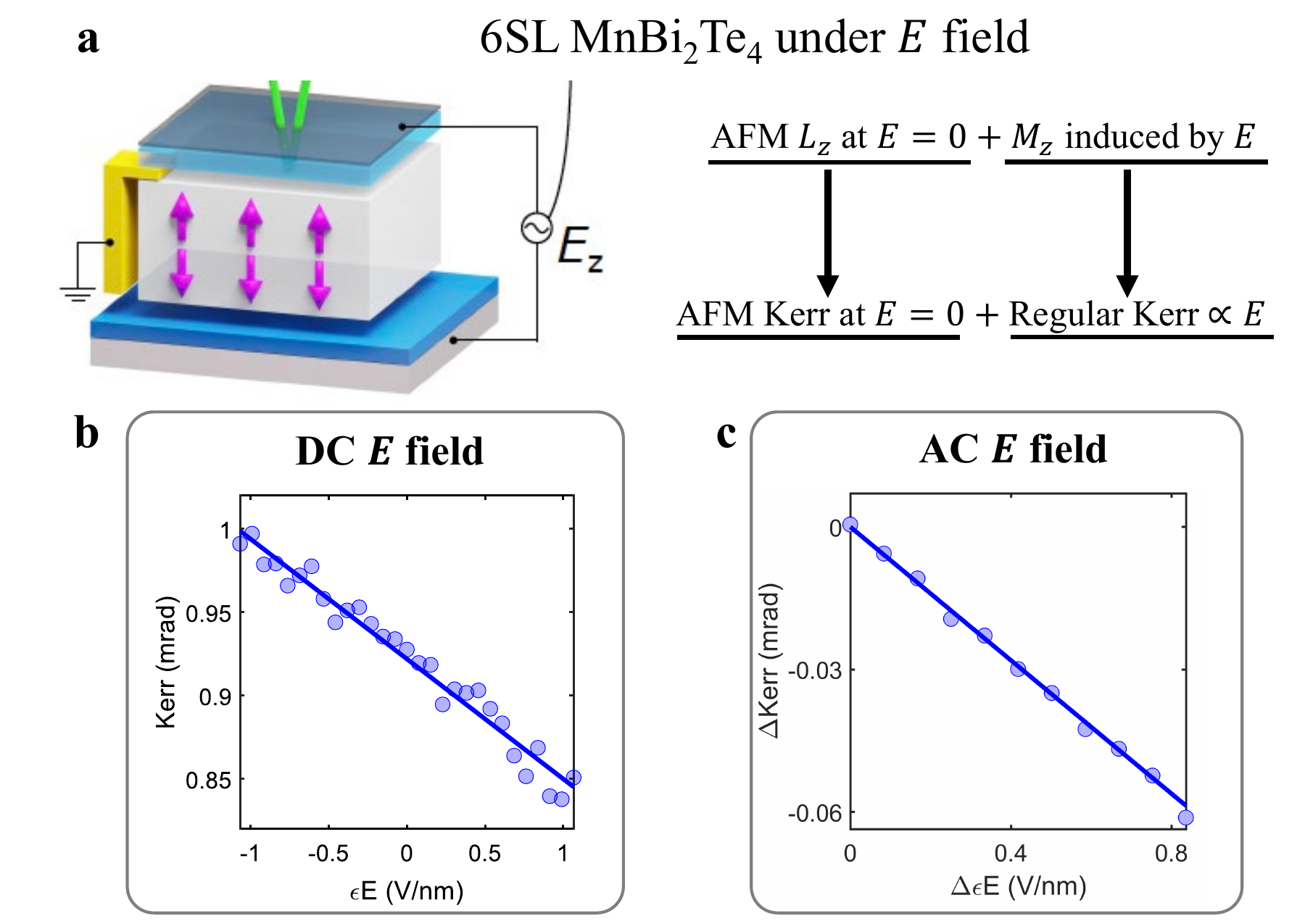}
		\caption{\textbf{a}, Analysis of the different order parameters ($L_z$ and $M_z$) of 6SL MnBi$_2$Te$_4$ under $E$ field. \textbf{b}, Kerr data as a function of DC $E$ field. \textbf{c}, Kerr data as a function of AC $E$ field modulation. }
		\label{SI_AFM_Kerr_2_v5}
	\end{figure*}

	Therefore, for 6SL MnBi$_2$Te$_4$, if we apply a DC $E$ field, we expect that the AFM Kerr ($\propto L_z$) already shows up at $E$=0 (i.e., a nonzero offset); on top of that, there will also be a regular Kerr ($\propto M_z$) that is linearly proportional to $E$ field due to the magnetoelectric coupling $\alpha$. Indeed, this can be seen from our data under DC $E$ field (Fig.~\ref{SI_AFM_Kerr_2_v5}). By contrast, if we apply AC $E$ field modulation, then the lock-in signal would only detect the Kerr rotation that is linear to $E$ (Fig.~\ref{SI_AFM_Kerr_2_v5}), therefore allowing us to isolate the magnetoelectric coupling signal.
	
	\subsubsection*{II.3. The AFM Kerr does not have any linear-in-$E$ contribution}
	
	One important question is whether the AFM Kerr can also support a linear-in-$E$ component. Here, we show that, based on the symmetry analysis, the AFM Kerr can NOT have any linear-in-$E$ contribution. The symmetry analysis is as follows: The AFM Kerr has distinct symmetry properties; it breaks $P$ and $T$ but respects $PT$. As a result, let us suppose the AFM Kerr has a linear-in-$E$ component: AFM Kerr $= \chi E$.  We can perform $PT$ operation to both sides of the equation. Under $PT$, AFM Kerr is invariant, but $E$ becomes $-E$. Therefore $\chi=0$. In other words, the AFM Kerr effect cannot have any linear-in-$E$ component (it can only have even powers of $E$ including $E^0$). In the next subsection, we further prove that the observed linear-in-$E$ Kerr purely comes from the $E$-induced magnetization, i.e., the magnetoelectric coupling.
	
	\subsubsection*{II.4. The observed linear-in-$E$ MOKE arises from $E$-induced magnetization}

	\begin{figure*}[h]
		\centering
		\includegraphics[width=11cm]{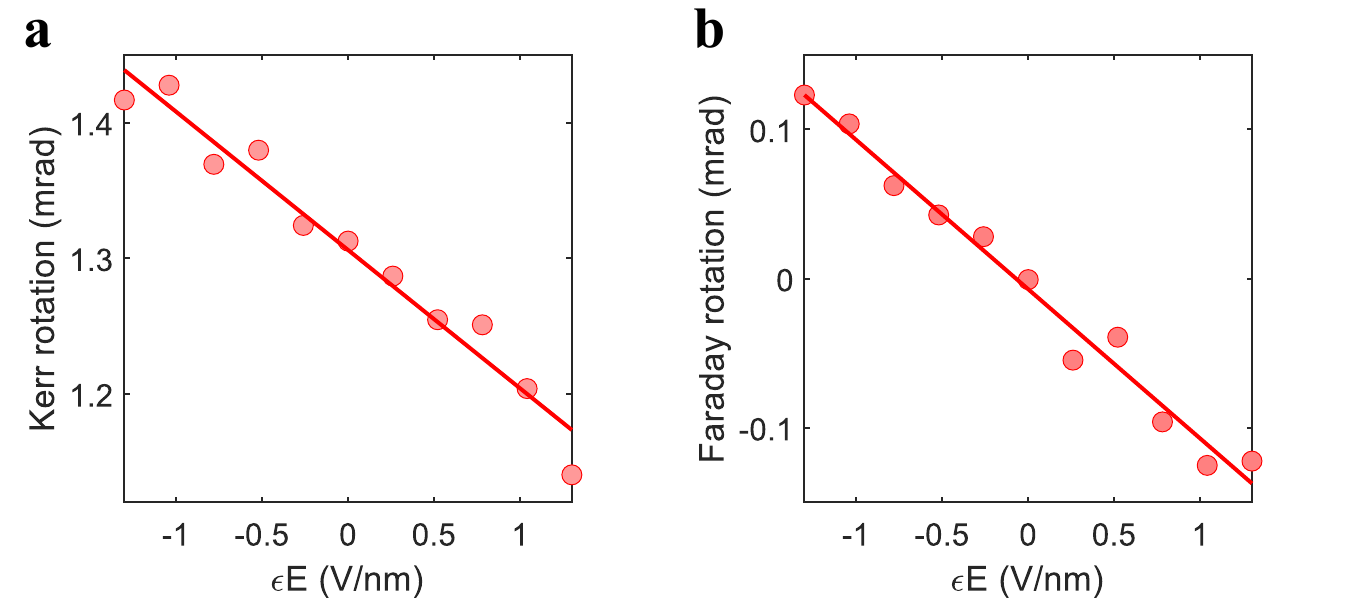}
		\caption{Simultaneous Kerr (\textbf{a}) and Faraday rotation (\textbf{b}) measurements as a function of DC $E$ field for 6SL MnBi$_2$Te$_4$. Both Kerr and Faraday show an linear-in-$E$ effect due to the $E$-induced $M_z$. The Kerr signal has an additional offset at $E=0$ due to the AFM Kerr effect.}
		\label{SI_Kerr_Faraday_v1_2}
	\end{figure*}
	
	We now aim to prove that the observed linear-in-$E$ Kerr comes from the $E$-induced magnetization. As we discussed above, $M_z$ will induced both Faraday and Kerr rotation, but $L_z$ will only induce Kerr rotation. Therefore, by testing if there is also an $E$-induced signal in Faraday effect, we can know if the $E$-induced Kerr signal is AFM Kerr ($\propto L_z$) or regular Kerr ($\propto M_z$). We have performed new measurements to study Kerr and Faraday simultaneously using DC $E$ field method. As shown in Fig.~\ref{SI_Kerr_Faraday_v1_2}, our data clearly shows that the linear-in-$E$ signal exists in both Kerr and Faraday. Therefore, our simultaneous reflection and transmission measurements directly prove that the observed linear-in-$E$ Kerr rotation is the regular Kerr $\propto M_z$. 
	

	\begin{figure*}[h]
		\centering
		\includegraphics[width=14cm]{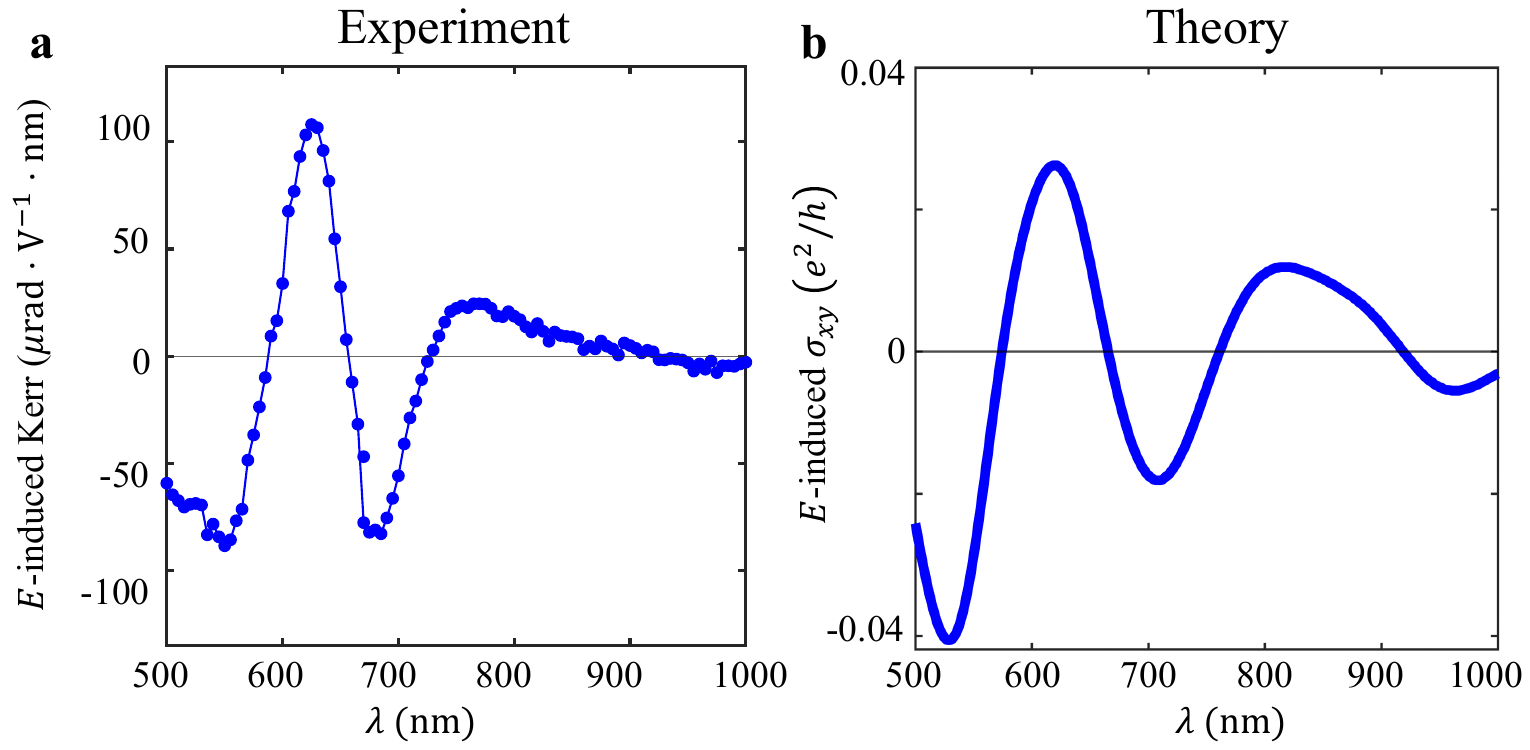}
		\caption{\textbf{a}, Experimentally measured $E$-field induced Kerr at different wavelengths. \textbf{b}, Theoretically calculated $E$-field induced $\sigma_{{xy}}$ at different wavelengths.}
		\label{SI_E_sigma_xy_DATA_THY_v5}
	\end{figure*}
	
	To further prove that the observed $E$-induced Kerr comes from magnetization. We note that formally, the regular Kerr rotation measures the off-diagonal (Hall) part of the optical conductivity $\sigma_{xy}$, which has identical symmetry properties as the out-of-plane magnetization $M_z$. Therefore, we can write $\textrm{Kerr}(\lambda)\propto \sigma_{xy}(\lambda) \propto \gamma(\lambda)M_z$. In order to verify this, we have (1) experimentally measured the $E$-field induced Kerr rotation as a function of the wavelength $\lambda$ and (2) theoretically calculated the wavelength dependence of the $E$-field $\sigma_{xy}(\lambda)$. As shown in Fig.~\ref{SI_E_sigma_xy_DATA_THY_v5}, we see a good agreement between data and calculation, which therefore strengthens our conclusion that the measured the $E$-field induced Kerr rotation in 6SL MnBi$_2$Te$_4$ indeed reflects the $E$-field induced magnetization (i.e., the magnetoelectric coupling $\alpha$).

	
	

	\begin{figure*}[!htb]
		\centering
		\includegraphics[width=11cm]{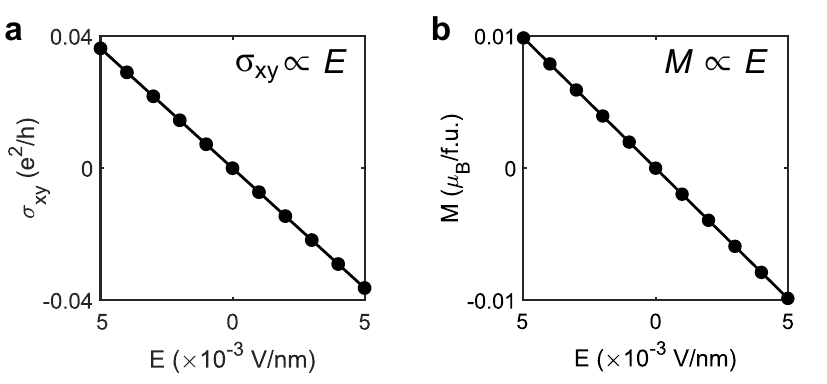}
		\vspace{0cm}
		\caption{\textbf{a}, Theoretical calculation of the optical $\sigma_{xy}$-$E$ of 6SL MnBi$_2$Te$_4$. The wavelength of $\sigma_{xy}$ is 515 nm. \textbf{b}, Theoretical calculation of $M_z$-$E_z$ of 6SL MnBi$_2$Te$_4$. \color{black}}
		\label{SI_Linear_calculation_M_Sigma_E_2}
	\end{figure*}
	
	We can further strengthen the logic of the above proof as follows: First, the good agreement of the wavelength dependence demonstrates that our measured $E$-induced Kerr comes from $E$-field $\sigma_{xy}(\lambda)$. This proof is quite solid because it is not based on a single number (e.g. at one wavelength); Rather it is based on the agreement of an entire curve (e.g. many wavelengths). Second, in our DFT calculation, we obtained the $E$-induced $\sigma_{xy}(\lambda)$ by calculating $\sigma_{xy}(\lambda)$ at different $E$ values. We can also calculate the net magnetization $M_z$ at those $E$ values. Fundamentally, we first compute a band structure of 6SL MnBi$_2$Te$_4$ for every $E$ value, then the corresponding $\sigma_{xy}(\lambda)$ and $M_z$ can be calculated based on the calculated band structure, see expressions below. As shown in Fig.~\ref{SI_Linear_calculation_M_Sigma_E_2}, both $\sigma_{xy}$ and $M_z$ are linearly dependent on $E$. Hence, we showed $\textrm{Kerr} \propto \sigma_{xy} \propto E \propto M_z$. 
	
	\begin{align}
		\notag
		M^{\textrm{orb.}}_z
		&=\frac{e}{\hbar d}{\rm Im}\sum_n\int_{\varepsilon_{n\bf k}\le \mu} \frac{dk_xdk_y}{(2\pi)^2}
		\langle \partial_{k_x}u_{n\bf k}|H_{\bf k}+\varepsilon_{n\bf k}-2\mu | \partial_{k_y}u_{n\bf k}\rangle, \\
		\sigma_{xy}(\omega)&=\frac{e^2}{\hbar d} \int{\frac{dk_xdk_y}{(2\pi)^2}} \sum_{n,m} f_{mn{\bf k}} \frac{i(\varepsilon_{m{\bf k}}-\varepsilon_{n{\bf k}})\braket{u_{n{\bf k}}|i \partial_{k_x}|u_{m{\bf k}}} \braket{u_{m{\bf k}}|i \partial_{k_y}|u_{n{\bf k}}}  }{(\varepsilon_{m{\bf k}}-\varepsilon_{n{\bf k}})-\hbar \omega-i\eta}.
	\end{align}
	
	\subsubsection*{II.5. TRMOKE of 6SL MnBi$_2$Te$_4$} 
	
	\begin{figure*}[h]
		\centering
		\includegraphics[width=9cm]{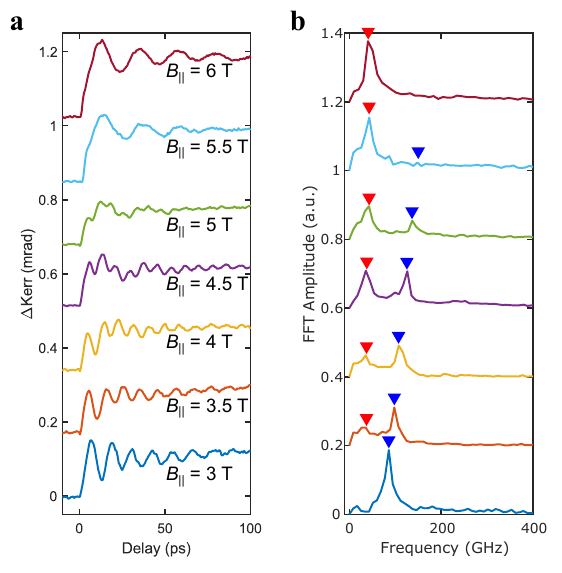}
		\caption{TRMOKE data of 6SL MnBi$_2$Te$_4$ (no WSe$_2$) at different $B_{\parallel}$ field (red and blue triangles represent out-of-phase and in-phase magnons).}
		\label{SI_TRMOKE_overview_v3}
	\end{figure*}
	
	
	We now proceed to the time-resolved MOKE data. We have shown that the MOKE signal consists of two parts: the regular Kerr and the AFM Kerr.
	\begin{itemize}
		\item  The regular Kerr is proportional to the net magnetization $M_z$. Hence, it can detect the in-phase magnon, because the in-phase magnon features an oscillating $M_z$. 
		\item The AFM Kerr is proportional to the AFM order parameter $L_z$ reported in\cite{Qiu2023axion}. Hence, it can detect the out-of-phase magnon, because the out-of-phase magnon features an oscillating $L_z$.  
	\end{itemize}
	As shown in Fig.~\ref{SI_TRMOKE_overview_v3}, both magnon modes are observed. \textbf{The in-phase magnon is prominently excited at small $B_{\parallel}$, whereas the out-of-phase magnon is prominently excited at large $B_{\parallel}$. }
	
	\subsubsection*{II.6. TRMOKE at different DC $E$ and $n$ values} 
	
\begin{figure*}[!htb]
		\centering
		\includegraphics[width=16cm]{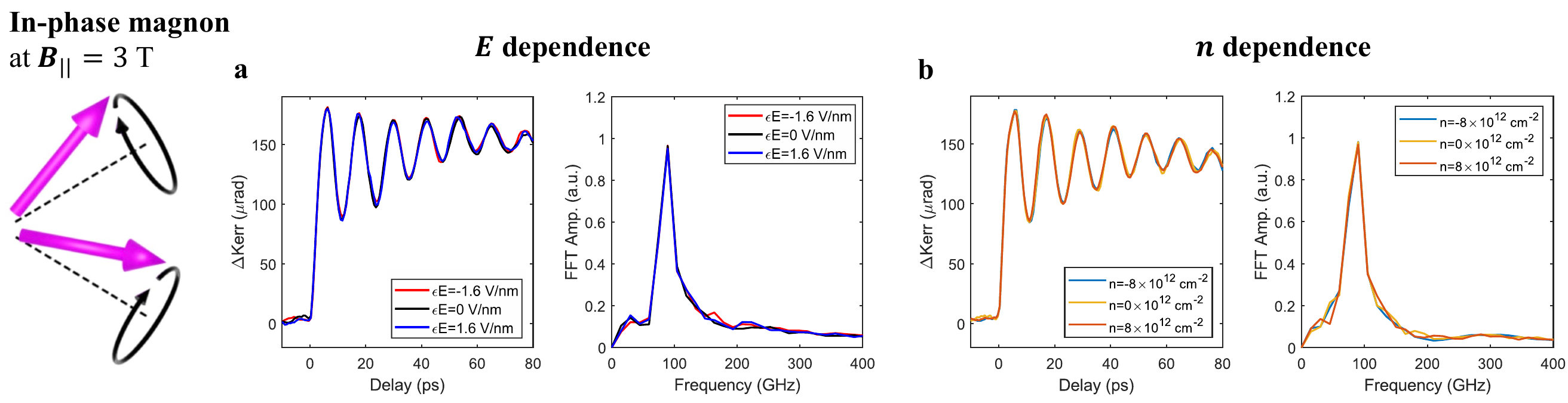}
		\caption{Gate dependent TRMOKE data at $B_{\parallel}=3$ T. The in-phase magnon shows no observable $n$ and $E$ dependence in terms of both the frequency and TRMOKE amplitude. $\lambda_{\rm pump}=1030$ nm, $\lambda_{\rm probe}=515$ nm.}
		\label{SI_In_phase_magnon_n_E_2_v4}
	\end{figure*}

	We set $B_{\parallel}=3$ T, where the in-phase magnon is prominent. As shown in Fig.~\ref{SI_In_phase_magnon_n_E_2_v4}, the in-phase magnon shows no observable $n$ and $E$ dependence in terms of both the frequency and TRMOKE amplitude.

\begin{figure*}[!htb]
		\centering
		\includegraphics[width=16cm]{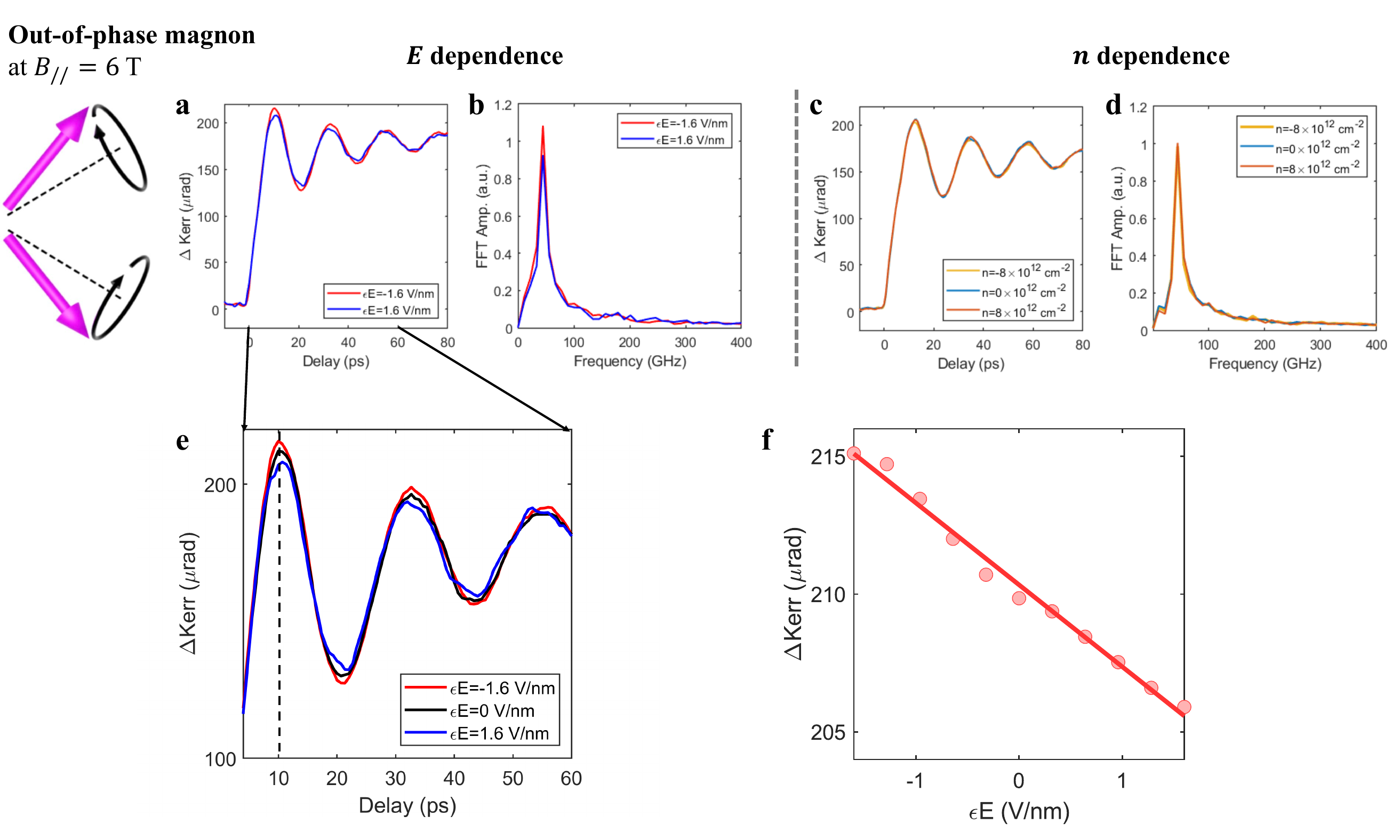}
		\caption{Gate dependent TRMOKE data at $B_{\parallel}=6$ T. Top row: The out-of-phase magnon shows no observable $n$ and $E$ dependence in terms of both the frequency and TRMOKE amplitude. Bottom row: A zoomed-in view of the TRMOKE with three different $E$ values. $E$ dependence TRMOKE at delay time 10 ps (dotted line on the left).  $\lambda_{\rm pump}=1030$ nm, $\lambda_{\rm probe}=515$ nm.}
		\label{SI_Out_of_phase_magnon_n_E_v3}
	\end{figure*}

	We set $B_{\parallel}=6$ T, where the out-of-phase magnon is prominent. As shown in Fig.~\ref{SI_Out_of_phase_magnon_n_E_v3}, the out-of-phase magnon shows no observable $n$ dependence in terms of both the frequency and TRMOKE amplitude as well as no observable $E$ dependence in terms of the frequency. The only observable effect is that the TRMOKE amplitude shows a linear in $E$ dependence at the out-of-phase magnon frequency, which is exactly the $\alpha$ oscillation. These systematic data also provide an important cross-check in terms of what is observable/not observable: 7 out of 8 dependences show no effect NOT due to insufficient resolution; Under the same conditions, we clearly observe linear E dependence of TRMOKE amplitude at the out-of-phase frequency.
	
	
	
	
\begin{figure*}[!htb]
		\centering
		\includegraphics[width=16cm]{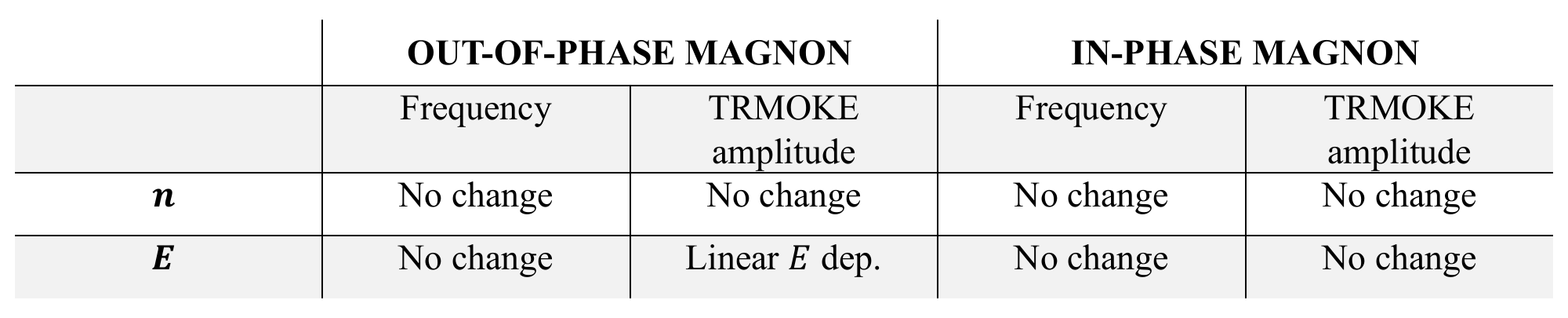}
		\caption{A summary of the gate dependence of the magnon frequency and TRMOKE amplitude in 6SL MnBi$_2$Te$_4$.}
		\label{SI_Summary}
	\end{figure*}

	Fig.~\ref{SI_Summary} presents a summary of the above results. In 6SL MnBi$_2$Te$_4$, we have systematically studied in total eight distinct dependences. I.e., how does the frequency (or TRMOKE amplitude) of the in-phase (or out-of-phase) magnon depend on n (or E). Seven out of eight dependences show no observable effect. The only exception is that the TRMOKE amplitude shows a linear-in$E$ dependence at the out-of-phase magnon frequency, which is the $\alpha$ oscillation. Moreover, we further show in SI.I (see Figs.~\ref{SI_5SL_TRMOKE_overview_v3} and \ref{SI_5SL_AC_DC_v3}) that there is no gate dependence for the TRMOKE of 5SL layer MnBi$_2$Te$_4$. If the exchange coupling $J$ and the anisotropy $\kappa$ were strongly modified by gating, then we would expect both the frequency and TRMOKE amplitude of both magnon modes to show gate dependence (the $n$ dependence is typically even stronger than $E$ dependence, according to previous studies in Cr$_2$Ge$_2$Te$_6$ \cite{hendriks2024electric} and CrI$_3$ \cite{zhang2020gate}). Also, we would expect similar effect in both even and odd layers. By contrast, if $J$ and $\kappa$ are roughly unchanged but the results were due to the DAQ (the $\alpha$ oscillation), then we would only expect a linear-in-$E$ TRMOKE signal at the out-of-phase magnon frequency in even layer. Therefore, our results strongly suggest that the alternative effect that $J$ and $\kappa$ are strongly modified is unlikely to be origin. Rather, the DAQ is a more plausible and consistent interpretation. 
	
	\vspace{0.5cm}
	\subsubsection*{II.7. DAQ probed DC $E$ field} 
	
\begin{figure*}[!htb]
		\centering
		\includegraphics[width=17cm]{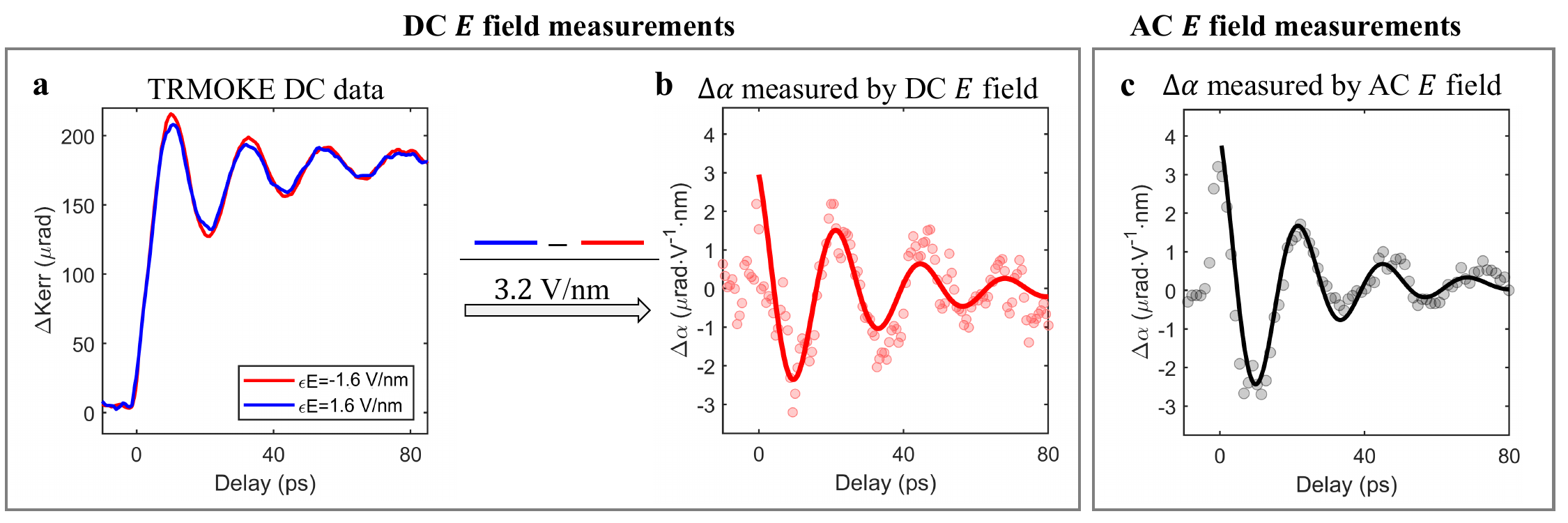}
		\caption{Comparison between DC and AC $E$ field dependent measurements at 6 T\color{black}. a, TRMOKE data at different DC $E$ field values (same data as Fig.~\ref{SI_Out_of_phase_magnon_n_E_v3}). b, Time-resolved $\Delta \alpha$ is obtained by the difference of the TRMOKE data at $\pm$1.6 V/nm divided by $3.2$ V/nm. c, Time-resolved $\Delta \alpha$ measured by AC $E$ field-dependent measurements.}
		\label{SI_AC_DC_v5}
	\end{figure*}

	We now try to obtain the time-resolved $\alpha$ by DC $E$ field method, and show that the result is fully consistent with the AC $E$ field modulation method adapted in the main text. Specifically, we first park the magnetic field at $B_{\parallel}=6$ T, where the out-of-phase magnon is prominent. Figure~\ref{SI_AC_DC_v5}\textbf{a} shows the TRMOKE at different DC $E$ values (same data as Fig.~\ref{SI_Out_of_phase_magnon_n_E_v3}). In order to obtain the time-resolved $\Delta \alpha$, we subtract the TRMOKE data at $E=\pm 1.6$ V/nm and divide it by $3.2$ V/nm. The result, shown in Fig. Fig.~\ref{SI_AC_DC_v5}\textbf{b}, shows an oscillation of $\alpha$. Therefore, we can detect the $\alpha$ oscillation (the DAQ) by DC $E$ field method; The result (Fig.~\ref{SI_AC_DC_v5}\textbf{b}) is consistent with that of AC $E$ field lock-in method shown in Fig.~\ref{SI_AC_DC_v5}\textbf{c} (the AC data has better signal to noise ratio).
	
\begin{figure*}[t]
		\centering
		\includegraphics[width=17cm]{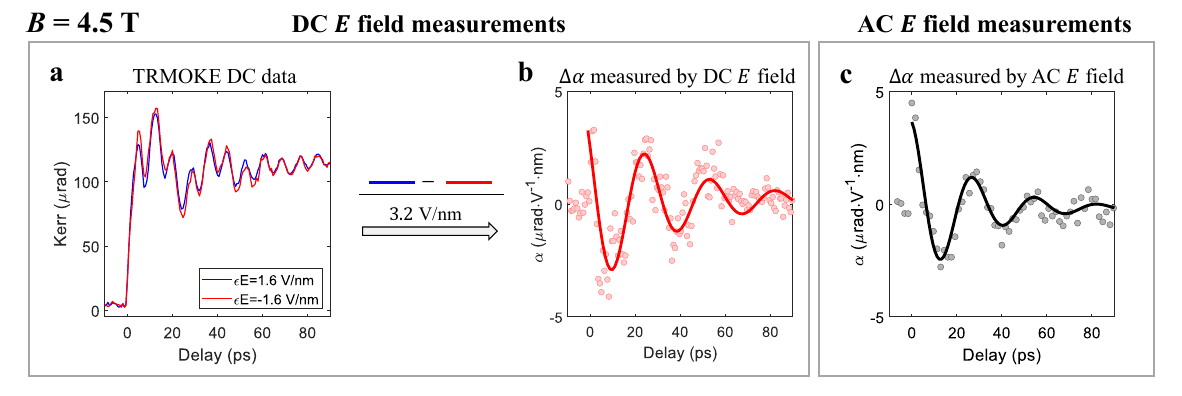}
		\caption{Comparison between DC and AC $E$ field dependent measurements at $4.5$ T. a, TRMOKE data at different DC $E$ field values at $4.5$ T, both in-phase and out-of-phase magnons are clearly observed. b, Time-resolved $\Delta \alpha$ is obtained by the difference of the TRMOKE data at $\pm$1.6 V/nm divided by $3.2$ V/nm. c, Time-resolved $\Delta \alpha$ measured by AC $E$ field-dependent measurements. \color{black}}
		\label{SI_MBT4p5T_data_AC_DC_field_2}
	\end{figure*}

	Moreover, we can also set the magnetic field at at $B_{\parallel}=4.5$ T, where both the in-phase and out-of-phase magnons are excited. When we perform $E$-field dependence (both DC and AC), we can also see the $\alpha$ oscillation at the out-of-phase magnon frequency with fully consistent results (Fig.~\ref{SI_MBT4p5T_data_AC_DC_field_2}).

	\clearpage
	
	\section*{III. Dark Matter Axion Detection}
	
	\subsection*{III.1 Detection sensitivity calculation}
	Table \ref{estimation_Axion_table} describes how we estimated the values of the parameters needed for the sensitivity calculation of the dark matter Axion detection using DAQ~\cite{schutte2021axion}. 
	\begin{table}[h]
		\centering 
		\begin{tabular}{c lll} 
			\hline
			Parameter & Description & Reference Values & Comments \\ [0.5ex]
			\hline 
			$\rho_{\rm DM}$ & Local Dark matter Axion density  & 0.4 GeV/cm$^{3}$ & Estimation from astrophysics~\cite{tanabashi2018review} \\
			$f_\Theta$ & DAQ decay constant & 82 eV & DFT calculations, see below\\ 
			$m_{\rm DAQ}$ & DAQ mass & 0.18 meV & Experimental data in this work \\
			n & Dielectric constant & 6.4 & DFT calculation\\
			$\Gamma_m$ &  Magnetic impurity density & 0.7$\times 10^{-3}$ & Value estimated from ~\cite{bayrakci2013lifetimes}  \\ 
			$\Gamma_\rho$ &  THz conductance & 0.2$\times10^{-3}$ & DFT calculation \\ 
			A & Detector Area & 0.16 m$^2$ & Large sample grown by MBE \cite{zhao2021even} \\
			d & Sample thickness & 0.4 mm & Optimal thickness based on ~\cite{schutte2021axion} \\
			B & External magnetic field & 1-10 T & Typical cryostat conditions\\
			$\eta$ & Single photon detector effiency  & 0.95 & Estimated from ~\cite{lee2020graphene} \\
			$\lambda_d$ & Dark count rate &  $10^{-5}$ Hz & Estimated based on ~\cite{lee2020graphene} \\
			\hline 
		\end{tabular}
		\caption{Parameter estimation values for dark matter Axion detector sensitivity calculation} 
		\label{estimation_Axion_table}
	\end{table}
	
	$f_{\Theta}$ is given by $f_{\Theta}=g\sqrt{J}$ \cite{schutte2021axion}, Where $J$ is the spin wave stiffness, and $g=\delta L / \delta \theta$ describes the strength of the DAQ ($L$ is the antiferromagnetic order parameter). Extended Data Fig. 7  shows $\theta$ vs. $L$ for 6SL MnBi$_2$Te$_4$, from which we get $g=0.042$ eV. $J$ is defined as follows based on \cite{li2010dynamical}. We could write down the Hamitonian for bulk MnBi$_2$Te$_4$:
	$$\mathcal{H}=\sum_{i=1}^{5} d_i \Gamma_{i}$$ 
	\vspace{3mm}
	$$\Gamma_{1,2,..,5}=(\tau_1\sigma_3,-\tau_1\sigma_2,\tau_1\sigma_1,\tau_3\sigma_0,\tau_0\sigma_3)$$
	\vspace{3mm}
	$$d_{1,2,..,5}=(\frac{A_1}{c} \sin(k_z c), \frac{A_2}{a} \sin (k_x a), -\frac{A_2}{a}\cos(k_y a), M_0+2\frac{M_1}{c^2}(1-\cos(k_zc))+2\frac{M_2}{a^2}[2-\cos(k_z a )+\cos(k_y a)],(-1)^lm_5)$$
	, where $l$ denotes the bulk MnBi$_2$Te$_4$ layer number, and $a$, $c$ are lattice constants. The tight-binding model parameters are given by: $A_1=2.7023\ \rm eV\cdot \r{A}$, $A_2=3.1964\ \rm eV\cdot \r{A}$, $M_1=11.9048\ \rm eV\cdot \r{A}^2$, $M_2=9.4048\ \rm eV\cdot \r{A}^2$, $M_0=-0.04\ \rm eV$, $m_5=0.03\ \rm eV$,$a=4.334\ \rm \r{A}$, $c=40.91\ \rm \r{A}$. With the tight-binding model above, the $J$ of bulk MnBi$_2$Te$_4$ can be calculated as: $J= \int \frac{d^3 k}{(2\pi)^3} \frac{\sum_{i=1}^4 d_i^2}{16 \sum_{i=1}^5 d_i^2}$. As a result, we found $J=3.612\times 10^{-4} \rm eV^{-3}\r{A}^{-3}$. By taking the astrophysics convention of $\hbar=c=1$, we could obtain $f_\Theta=g\sqrt{J}=82\rm \ eV$.
	
	\vspace{6mm}
	
	The dielectric constant $\epsilon=\chi+1$, where $\chi$ is the electric susceptibility. $\chi$ was calculated using, 
	
	\begin{equation}
		\chi_{ij}(\omega)=\frac{e^2}{\hbar V} \sum_{n \in occ,m \in unocc,{\bf k}} \frac{2 \omega_{mn}}{\omega_{mn}^2-\omega^2} {\rm Re} [r^i_{nm}r^j_{mn}]
	\end{equation}
	
	Here, V is the system volume, $r^i_{mn}=\braket{\psi_m|\hat{r}^i|\psi_n}$ is the position matrix element, and $\hbar\omega_{mn}=\epsilon_m-\epsilon_n$ is the difference in energy eigenvalues for the Bloch state $\ket{\psi_m}$ and $\ket{\psi_n}$. 
	
	\vspace{2pt}
	\textbf{Resonant frequency:} As shown in Ref. \cite{schutte2021axion}, the resonant frequency for Axion detection is given by
	$$\omega_{\rm r}=\omega_{j}=\sqrt{\omega_{\textrm{DAQ polariton}}^2+\delta \omega_j}=\sqrt{\omega_{\rm DAQ}^2+C^2B^2+\delta \omega_j^2}$$, where $\delta \omega_j^2=\frac{\Delta_j^2b^2}{n^2d^2\omega_{\textrm{DAQ polariton}}^2}$ 
	
\begin{figure*}[!htb]
		\centering
		\includegraphics[width=6cm]{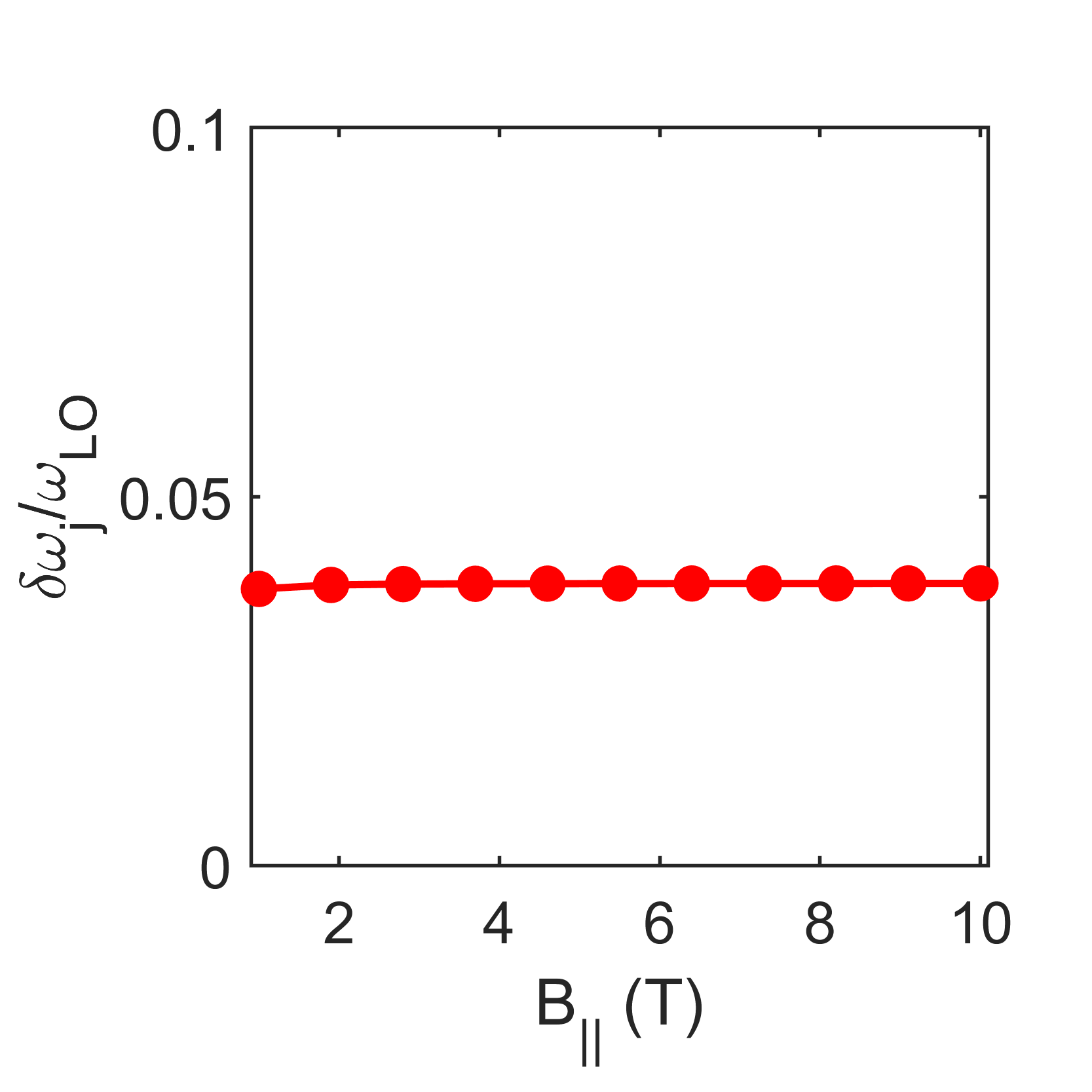}
		\caption{$\delta\omega_j/\omega_LO$ as a function of $B_{\parallel}$}
		\label{SI_omega_j_omega_LO_ratio}
	\end{figure*}

	\vspace{4pt}
	Here we show that $\delta \omega_j$ is indeed neglectably small, as shown in Fig.~\ref{SI_omega_j_omega_LO_ratio}. As such, the Axion-polariton frequency can be simplified as $\sqrt{\omega_{\rm DAQ}^2+C^2B^2}$. 
	
	\vspace{-0.5cm}

	
	\vspace{3mm}
	
	\subsection*{III.2 Future steps to build a functioning Axion detector}
	
	We describe the future experimental preparations that are needed in order to build a functional Axion detector based on our DAQ material.
	
	
	\setlength{\leftskip}{0.5cm}
	\noindent \textbf{1. Sample size:} For the photon counting approach, we need the sample area needs to be $0.16$ m$^2$ and the thickness of $0.4$ mm. MnBi$_2$Te$_4$ has been grown by MBE \cite{bai2024chern} which solves the sample area issue. Large thickness is more challenging. As shown in the main text, the DAQ in 6L MnBi$_2$Te$_4$ relies on the finite thickness hybridization gap. We propose to grow mm-thick samples that consist of repeating superlattice between 6L MnBi$_2$Te$_4$ and spacer layer such as Al$_2$O$_3$. In this way, the 6L MnBi$_2$Te$_4$ films are isolated from each other but the total thickness can be large. We can first test the superlattice with smaller total thickness (e.g. 100nm) to optimize its quality and growth speed. Then we can grow a 0.4 mm thickness superlattice, which we estimate to take multiple weeks \cite{novikov2016growth}. Apart from MnBi$_2$Te$_4$, Mn$_2$Bi$_2$Te$_5$ is a new bulk crystal that can be viewed as a natural superlattice between bilayer MnBi$_2$Te$_4$ and Bi$_2$Te$_3$, where theory has predicted large DAQ \cite{zhang2020large}. Large, mm size single crystals have been synthesized \cite{cao2021growth}. Beyond the Mn-Bi-Te based materials, we have discussed the possibility of searching for large DAQ in multiferroic insulators, which can be a future direction. 
	
	
	\vspace{2mm}
	
	\begin{wrapfigure}{r}{0.3\textwidth}
		\vspace{-3ex}
		\includegraphics[width=0.3\textwidth]{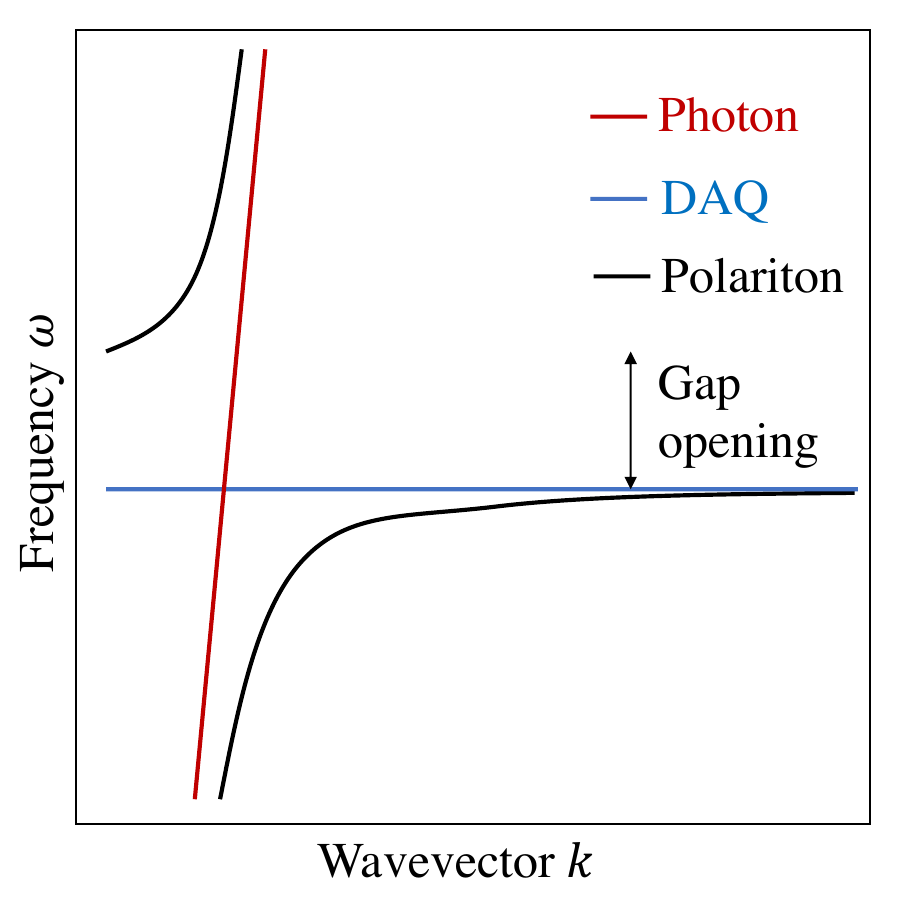}
		\vspace{-6ex}
		\caption{\small Polariton.}
		\label{SI_polariton}
		\vspace{-4ex}
	\end{wrapfigure}
	\noindent \textbf{2. Directly measure Axion polariton:} In our current experiments, we used a pump laser of visble photons ($\hbar\omega\simeq2.5$ eV), which is off resonance and relies on the laser-heating induced coherent spin wave excitation. Hence our pump laser only excited the DAQ (i.e., the out-of-plane magnon) at frequency/mass $m_{\rm DM}=44$ GHz$=0.18$ meV. If we use low energy photons that are close to the resonance, then those photons and the DAQ (i.e., the out-of-plane magnon) can hybridize to form Axion polaritons (magnon polaritons), which can be understood as the gap opening and level-repel picture shown in Fig.~\ref{SI_polariton}. The top of the gap is at energy $\sqrt{m_{\rm DAQ}^2+C^2B_{\parallel}^2}$, which grows with magnetic field $B_\parallel$. Such an Axion polariton can be directly measured on our 6L MnBi$_2$Te$_4$ by the scanning near-field optical microscope (SNORM) \cite{basov2016polaritons}.
	
	\vspace{2mm}
	\noindent \textbf{3. Losses:} Material imperfection (residue conductivity and magnetic impurities) can lead to damping of the electromagnetic waves inside the material. Right now we have $\Gamma_\rho=0.2 \times 10^{-3}$ and $\Gamma_m=0.7 \times 10^{-3}$. If we can improve the sample purity, especially suppressing the residual conductivity and reduce the Mn antisite defects, then one can reach the higher sensitivity than the results in Fig. 4g of main text.
	
	\vspace{2mm}
	\noindent \textbf{4. Single photon detector efficiency:} The single photon detector in the THz regime is rapidly developing. We estimate our sensitivity based on a THz single-photon detector with efficiency $\sim95\%$ and dark-count rate of $10^{-5}$ Hz. This is an reasonable projection because the superconducting nanowire single-photon detector has achieved simultaneously such a low dark count and high efficiency in near infrared regime \cite{marsili2013detecting,gambetta2016investigating}. The same technology is progressing steadily to the longer wavelength \cite{verma2021single} where our DAQ detector shall be able to search for the dark matter axions.

	\vspace{3mm}
	\subsection*{III.3 Comparison between the meV DAQ detector and other proposed detectors}
	
	There is no existing detector in the meV regime. The only two proposed future detectors (called BRASS ~\cite{horns2013searching} and BREAD \cite{liu2022broadband}) are based on the same detection mechanism: under a $B$ field, use a giant focusing mirror to focus all dark matter-induced photons (independent of their frequencies) and detect them with single-photon detectors. The earlier BRASS proposal considered a parabolic focusing mirror, which is difficult to integrate into a dilution refrigerator. The more recent BREAD proposal considered a conical focusing mirror that can be compatible with a dilution refrigerator. 
	
	We list the pros and cons for our DAQ detector when comparing these two proposed detectors. 
	
	\vspace{3mm}
	\textbf{Advantages:}
	\begin{itemize}
		\item \textbf{Frequency resolution and tunablility:} At a given magnetic field, the DAQ detector is only sensitive to a particular frequency given by $\sqrt{m_{\textrm{DAQ}}^2+C^2B^2}$. By varying the magnetic field, we can scan different frequencies. By contrast, the BRASS and BREAD focus all photons independent of their frequencies and detect them at once (if using the single photon detectors with better signal to noise). Therefore, their approach loses all information on the photon frequency, and thus on the axion mass, whereas the DAQ approach would immediately give the axion mass. 
		\item \textbf{Economy:} The BRASS and BREAD essentially collect all photons without resonant enhancement. Therefore, to get a large signal requires a large focusing mirror. To achieve sufficient sensitivity, the area of the mirror was estimated to be $A=10\text{ m}^2$. For BRASS, the parabolic mirror is incompatible with dil fridge. For BREAD, a $A=10\text{ m}^2$ solenoidal focusing mirror requires a very large dil fridge with magnet to house it. This is estimated to cost about tens of millions USD. By contrast, the DAQ sample can easily fit in any commercially available dil fridge, which costs about 600K USD.
	\end{itemize}
	
	\vspace{3mm}
	\textbf{Disadvantages:}
	
	\begin{itemize}
		\item \textbf{Scanning time and sensitivity:} The key disadvantage of the DAQ approach is the need to scan many frequencies. In our estimations in the main text, the measurement time on a single frequency is $\delta \tau = 50$ minutes, while the total measurement time is $\tau = 3$ years. By contrast, the BRASS and BREAD can detect a wide range of frequencies at the same time simultaneously, but with no frequency resolution and no boost factor. Our estimation found that the broadband approach is more sensitive than our proposed DAQ detector by a factor of around 5 (see detailed calculation below). Improving the loss factor of the DAQ material in future (increase $\beta$), and using larger/multiple samples to increase the area then the DAQ sensitivity can be comparable with the broadband approach.
	\end{itemize}
	
	\noindent\makebox[\linewidth]{\rule{\paperwidth}{0.4pt}}

	For the broadband approach, to cover a wide range requires multiple receivers. BRASS estimated $N=9$ receivers in their proposed range~\cite{Bajjali:2023uis}. For the same photon detector, the sensitivity in the broadband approach compared to the DAQ approach is:
	\begin{equation}
		\frac{g^{\rm res.}}{g^{\rm b.b.}} =\frac{1}{\beta} \sqrt{\frac{A^{\rm b.b.}\tau}{A^{\rm res.}\delta \tau}}\, .
	\end{equation}
	For $\beta \approx 100$, $\delta\tau = 50 \text{ min}$ and $\tau = 3 \text{ years}/N$ and equal areas the approaches are 
	comparably sensitive, with $\beta$ compensating the differing measurement time. Using our area of $A^{\rm res}=0.16\text{ m}^2$ compared to BREAD~\cite{liu2022broadband} $A^{\rm b.b.}=10\text{ m}^2$ the broadband approach is more sensitive than our proposed DAQ detector by a factor of around 5. Note that this improved sensitivity is only achieved once the total campaign has been completed. The resonant DAQ detector achieves design sensitivity on each frequency in a shorter time.

	\clearpage
	\section*{IV. Theoretical derivation for the layer Hall effect}
	As a $\mathcal{PT}$-symmetric material, even-layer MnBi$_2$Te$_4$ hosts the so-called layer Hall effect \cite{gao2021layer}, which manifests as an out-of-plane $E_z$-induced AHE. Here we show that the electric-field-induced AHE $\sigma_{xy}$ measures $d\mathcal{D}/dn$. The anomalous Hall conductivity is given by
	
	\begin{align}
		\sigma_{xy}&=\frac{e^2}{2\pi h}\int_{\varepsilon_{n\bf k}\le \mu}\Omega(\textbf{k}) d\textbf{k}\\
		&=\frac{e^2}{2\pi h}\int_{\varepsilon_{n\bf k}\le \mu}[\Omega_\textrm{T}(\textbf{k}) +\Omega_\textrm{B}(\textbf{k}) ]d\textbf{k}
	\end{align}
	
	In the second line, we have used the fact that in MnBi$_2$Te$_4$, the Berry curvature is dominated by the low-energy surface bands localized on the top and bottom surfaces. Because we have $\Omega_\textrm{T}=-\Omega_\textrm{B}$, the AHE conductivity vanishes in even-layer MnBi$_2$Te$_4$. Upon applying the out-of-plane electric field $E_z$, $E_z$ creates an imbalance of chemical potential (charge density, $\Delta n$) between the top and bottom layers. Therefore, the AHE conductivity under finite $E_z$ is given by
	
	\begin{align}
		\sigma_{xy} & = \frac{e^2}{2\pi h}\int_{\varepsilon_{n\bf k}\le \mu}[\Omega_\textrm{T}(\textbf{k},n+\Delta n) +\Omega_\textrm{B}(\textbf{k},n-\Delta n) ]d\textbf{k}\nonumber\\
		& =  \frac{e^2}{2\pi h}\int_{\varepsilon_{n\bf k}\le \mu}[\frac{\partial \ \Omega_\textrm{T}(\textbf{k},n)}{\partial n} \Delta n\ + \frac{\partial \ \Omega_\textrm{B}(\textbf{k},n)}{\partial n} (-\Delta n) ]d\textbf{k}\nonumber\\
		& =  \frac{1}{2}  \frac{\partial \mathcal{D}}{\partial n} (\Delta n) \nonumber
	\end{align}

	The charge density imbalance $\Delta n$ is induced by the out-of-plane electric field $E_z$, which is further related by the top and bottom gate voltages: $\Delta n=\frac{\epsilon E_z}{e}=\frac{\epsilon_0}{2e} (\frac{\epsilon_{\rm SiO_2}V_{\rm BG}}{h_{\rm SiO_2}}-\frac{\epsilon_{\rm hBN}V_{\rm TG}}{h_{\rm hBN}})$. Here $\epsilon_{\rm SiO_2}$ and $\epsilon_{hBN}$ are the dielectric constant of SiO$_2$ and hBN. $\rm h_{\rm SiO_2}$ and $\rm h_{\rm hBN}$ are the thickness of SiO$_2$ and hBN.  Therefore, 
	\begin{align}
		\sigma_{xy} &  =  \frac{1}{2}  \frac{\partial \mathcal{D}}{\partial n} (\frac{\epsilon_0}{2e} (\frac{\epsilon_{\rm SiO_2}V_{\rm BG}}{h_{\rm SiO_2}}-\frac{\epsilon_{\rm hBN}V_{\rm TG}}{h_{\rm hBN}})) \nonumber
	\end{align}
	
	In other words, by measuring the $E_z$-induced $ \sigma_{xy}$ using transport and knowing the value of the applied gate voltages $V_\textrm{TG}$ and $V_\textrm{BG}$, we can obtain $\frac{\partial \mathcal{D}}{\partial n}$ as shown in Extended Data Fig.4.
	
	\clearpage
	\section*{V. Tight binding model for MnBi$_2$Te$_4$ }
	The tight binding model Hamiltonian used in order to describe the few layer MnBi$_2$Te$_4$ was inspired by the earlier works on Bi$_2$Te$_3$, where the low energy bands are represented by the bonding ($P1^+_z$) and the anti-bonding ($P2^-_z$) states formed by Bi $p_z$ and Te $p_z$ orbitals~\cite{zhang2009topological}. The model is designed to  describe four low energy bands (including the spin degrees of freedom) per quintuple layer near the $\Gamma (0,0,0)$, point. The basis set is given by,
	
	\begin{equation}
		\ket{P1^+_z,\uparrow},\ket{P2^-_z,\uparrow},\ket{P1^+_z,\downarrow},\ket{P2^-_z,\downarrow}.
	\end{equation}
	The $\pm$ sign denote the parity of the states, while the $\uparrow, \downarrow$ indicate the spins in the $z$  direction. Various symmetry aspects of this model is explained in detail in Ref~\cite{ahn2022theory} as well as in Ref~\cite{Zhang2019topological}. The bulk Hamiltonian without any magnetization is expressed as,
	
	\begin{equation}
		\begin{split}
			h_{TB}=[e_0-2t_0(\cos k_1a+\cos k_2a+\cos k_3a)-2t_0^z\cos k_4a_z]\Gamma_0 \\
			- t_1(2\sin k_1a- \sin k_2a -\sin k_3a)\Gamma_1 
			-\sqrt{3}t_1(\sin k_2a - \sin k_3a)\Gamma_2 \\
			- 2t_3^z\sin k_4 a_z \Gamma_3 - 2t_4(\sin k_1a+ \sin k_2a+\sin k_3a)\Gamma_4 \\
			+ [e_5-2t_5(\cos k_1a+\cos k_2a+\cos k_3a) - 2t_5^z \cos k_4a_z] \Gamma_5
		\end{split}
	\end{equation}
	The $\Gamma$ matrices are given by, $\Gamma_0=I_4$, $\Gamma_1=s_x\tau_x$, $\Gamma_2=s_y\tau_x$,
	$\Gamma_3=s_z\tau_x$,$\Gamma_4=s_0\tau_y$,$\Gamma_5=s_0\tau_z$. Here the $s$ and $\tau$ represents the spin and the orbital degrees of freedom, respectively. The crystal momentum, $k_1=k_x$, $k_2=\frac{1}{2}(-k_x+\sqrt{3}k_y)$, $k_3=\frac{1}{2}(-k_x-\sqrt{3}k_y)$, $k_4=k_z$.
	
	Here,
	\begin{equation}
		\begin{split}
			e_0=C_0+2C_1/a_z^2+4C_2/a^2,\\
			e_5=M_0+2M_1/a_z^2+4M_2/a^2,\\
			t_0=\frac{2C_2}{3a^2},\\
			t_0^z=\frac{C_1}{a_z^2},\\
			t_1=-\frac{A_2}{3a},\\
			t_3^z=-\frac{A_1}{2a_z},\\
			t_5=\frac{2M_2}{3a^2},\\
			t_5^z=\frac{M_1}{a_z^2}.
		\end{split}
	\end{equation}
	
	The parameters used are listed below:
	
	\begin{equation}
		\begin{split}
			C_0=-0.0048 ~{\rm eV},\\
			C_1=2.7232 ~{\rm eV \AA^2}, \\
			C_2=0 ~{\rm eV \AA^2}, \\
			M_0=-0.1165 ~{\rm eV},\\
			M_1=11.9048 ~{\rm eV \AA^2},\\
			M_2=9.4048 ~{\rm eV \AA^2},\\
			A_1=4.0535 ~{\rm eV \AA},\\
			A_2=3.1964 ~{\rm eV \AA},\\
			a=4.334 ~{\rm \AA},\\
			a_z=\frac{1}{3}c=\frac{40.91}{3}=13.64 ~{\rm \AA}.
		\end{split}
	\end{equation}
	
	The few-layer thick MnBi$_2$Te$_4$ is modeled by discretizing the bulk Hamiltonian along the $z$ direction and coupling individual MnBi$_2$Te$_4$ layers with symmetry allowed interlayer hoppings. The AFM order is included by adding a layer-dependent exchange coupling to the non-magnetic Hamiltonian~\cite{Zhang2020}. In the layer basis, the magnetic part of the Hamiltonian is expressed as,
	
	\begin{equation}
		h_{AFM}=\begin{bmatrix}
			\mathbf{{m_A}.{s}} \otimes \tau_0 & 0 \\
			0 &  \mathbf{{m_B}.{s}} \otimes \tau_0
		\end{bmatrix}
	\end{equation}
	Here, A and B represent the two layers of the AFM unit cell. Different spin configurations representative of the magnon modes are described by controlling the magnetization parameter of each layer, ${\bf m_i}=m_i\hat{m_i}$, where, the $\hat{m_i}=(\cos \phi_i \sin \theta_i, \sin \phi_i \sin \theta_i, \cos \theta_i)$ is the unit vector along the direction of the magnetic moment. The $\theta$ and $\phi$ represent the angles of the spherical polar coordinates. In order to model the pristine MnBi$_2$Te$_4$ thinfilm with the out-of-plane AFM order, we have used $m_A=m_B=30$ meV, and $\theta_A=\pi$, $\theta_B=0$, $\phi_A=\phi_B=0$.   
	
	\clearpage
	\section*{VI. Additional discussion}
	
	\begin{enumerate}
		\item The polar Kerr rotation can arise from the normal Kerr proportional to $M_z$ as well as the AFM Kerr proportional to $L_z$. By the symmetry analysis, one can show that the AFM Kerr can NOT have any linear-to-$E$ contribution, so the linear-to-E Kerr rotation can only arise from the normal Kerr proportional to $M_z$. The symmetry analysis is as follows: The AFM Kerr has distinct symmetry properties; it breaks $P$ and $T$ but respects $PT$. As a result, suppose the AFM Kerr has a linear-$E$ component: $\textrm{AFM Kerr} =\chi E$.  We can perform $PT$ operation to both sides of the equation. Under $PT$, AFM Kerr is invariant, but $E$ becomes -$E$. Therefore $\chi=0$. In other words, the AFM Kerr effect cannot have any linear-$E$ component (it can only have even powers of $E$ including $E^0$). Therefore, the linear-$E$ dependent Kerr purely comes from the $E$-induced magnetization, i.e., the magnetoelectric coupling.
		
		\item Our experiments focused on a relatively small doping range $\pm8\times10^{12}$ cm$^{-2}$ near charge neutrality; Within this small range near charge neutrality, our data demonstrate that the magnetic properties of MnBi$_2$Te$_4$ are roughly unchanged. On the other hand, if one accesses much larger doping level, it is entirely possible that $J$ and $K$ of MnBi$_2$Te$_4$ could be modified due to the $p-d$ coupling. Here, we provide some rough estimation according to previous works. In Ref. \cite{checkelsky2012dirac}, the authors studied the $T_\textrm{c}$ of Mn-doped Bi$_2$(Te/Se)$_3$ as a function of $n$. It was shown (Fig. 2 of that paper) that $T_\textrm{c}$ starts to change as $n$ goes beyond $10^{13}$  cm$^{-2}$). It should be noted that in Mn-doped Bi$_2$(Te/Se)$_3$, Mn ions are dilute and randomly distributed, so the Mn-Mn exchange coupling entirely relies on the RKKY interaction ($p-d$ coupling) through the itinerant electrons. By contrast, in MnBi$_2$Te$_4$, the exchange coupling mainly arises from the Mn-Mn super-exchange. Therefore, it is reasonable to expect that, in order to for MnBi$_2$Te$_4$ to show a strong change of $T_\textrm{N}$ (therefore magnetism), one would need to vary $n$ to an even larger value such as high $10^{13}$  or $10^{13}$ cm$^{-2}$. 
		
		\item Why the magnetoelectric (ME) coupling $\alpha$ is zero above $T_\textrm{N}$, even though MnBi$_2$Te$_4$ in the nonmagnetic state is topologically nontrivial: 
		
		First, as a starting point, this ($\alpha=0$) can be argued by a general symmetry analysis. The magnetoelectric (ME) coupling can be expressed by $M=\alpha E$. We apply time-reversal operation to both sides of the equation, under which $M$ flips sign but $E$ remains invariant. Therefore, the only way for this to make sense is to have $\alpha=0$, i.e., the ME coupling $\alpha$ vanishes with time-reversal symmetry. This symmetry requirement is general, meaning that it is true for any kind of ME coupling (topological or trivial).
		
		Second, we discuss specifically for the case of strong TI with time-reversal symmetry. The strong TI has a nontrivial $\theta$, but it only manifests as a quantized $\alpha$ if we gap out the Dirac surface states. This can be seen from the pioneering theory papers. For example, in the following two theoretical works \cite{qi2008topological,wang2015quantized}, the authors showed the proposed setup to observe quantized $\alpha$ (see Fig. 15a of Ref. \cite{qi2008topological} and Fig. 2b of Ref. \cite{wang2015quantized}). In both works, the strong TI is interfaced with ferromagnetic insulators so that the Dirac surface states are gapped. In this setup, the bulk (i.e., the interior) of the TI can still remain time-reversal symmetric, but the surface is gapped out by interfacial time-reversal breaking. The gapped Dirac surface states have a nonzero Hall conductivity, and the Hall current wraps around the sample, generating the quantized $\alpha$. By contrast, if we have a stand-alone strong TI without the ferromagnetic insulators, then the strong TI has gapless Dirac surface states. The entire sample, which includes the bulk and the surfaces, have time-reversal symmetry. Then the $\alpha$ should vanish because of the time-reversal symmetry of the entire sample, even though the system has nontrivial topology. This fact has also been explicitly stated by another early theory paper \cite{coh2011chern}: Quote: ``Next we analyze the case of a strong $\mathcal{Z}_2$ topological insulator having $\theta=\pi$, or equivalently. We first consider a sample of such a system that has T symmetry conserved at its surfaces, as in Fig. 1(a). Again, since the entire sample is $T$-symmetric, its experimentally measurable magnetoelectric coupling tensor $\beta$ clearly has to vanish”. (Their $\beta$ is our $\alpha$ here).
		
	\end{enumerate}

	\pagebreak
	\newpage
	
	\clearpage

	\bibliographystyle{naturemag}

	\clearpage
	\vspace{0.5cm}

\end{document}